\pdfoutput=1   
\documentclass{article}

\usepackage{PRIMEarxiv}
\usepackage[utf8]{inputenc}
\usepackage[T1]{fontenc}
\usepackage{textcomp}
\usepackage{amsmath}
\usepackage{amssymb}
\usepackage{amsfonts}
\usepackage{amsthm}
\usepackage{bm}
\usepackage{graphicx}
\usepackage{booktabs}
\usepackage{multirow}
\usepackage{array}
\usepackage{algorithm}
\usepackage{algpseudocode}
\usepackage{xcolor}
\usepackage{tikz}
\usetikzlibrary{positioning,arrows.meta,fit,backgrounds,calc}
\usepackage{pgfplots}
\pgfplotsset{compat=1.18}
\usepackage{microtype}
\usepackage{xspace}
\usepackage{fancyhdr}
\usepackage{cite}
\usepackage{xurl}
\usepackage[colorlinks=true,linkcolor=blue!50!black,citecolor=blue!50!black,urlcolor=blue!50!black]{hyperref}
\hypersetup{pdftitle={Aero-VPA: Few-Step Flow-Matching Control over Non-Reconstructive JEPA Latents for Agile Flight on Edge Hardware},
  pdfauthor={Yeonseok Lee}}

\theoremstyle{plain}
\newtheorem{proposition}{Proposition}[section]
\theoremstyle{definition}
\newtheorem{assumption}{Assumption}[section]

\newtheorem{hypothesis}{Hypothesis}
\theoremstyle{remark}
\newtheorem{remark}{Remark}[section]

\newcommand{\R}{\mathbb{R}}
\newcommand{\E}{\mathbb{E}}
\newcommand{\norm}[1]{\left\lVert #1 \right\rVert}
\newcommand{\sg}{\operatorname{sg}}
\DeclareMathOperator*{\argmax}{arg\,max}
\DeclareMathOperator{\Var}{Var}
\DeclareMathOperator{\clip}{clip}

\DeclareMathOperator{\wrap}{wrap}
\DeclareMathOperator{\atantwo}{atan2}
\newcommand{\pr}[1]{\textsc{#1}}      

\newcommand{\ckRone}{\texorpdfstring{D\textsubscript{1}}{D1}\xspace}                         
\newcommand{\ckRtwo}{\texorpdfstring{D\textsubscript{2}}{D2}\xspace}                         
\newcommand{\ckNoact}{\texorpdfstring{D\textsubscript{3$-\bar{a}$}}{D3-a}\xspace}            
\newcommand{\ckNomotion}{\texorpdfstring{D\textsubscript{3$-\bar{a},v$}}{D3-a,v}\xspace}     
\newcommand{\ckSd}{\texorpdfstring{D\textsubscript{3,SD}}{D3,SD}\xspace}                     
\newcommand{\ckKw}{\texorpdfstring{D\textsubscript{3,KW}}{D3,KW}\xspace}                     
\newcommand{\bcVone}{\texorpdfstring{BC\textsubscript{NR}}{BC-NR}\xspace}                    
\newcommand{\bcVtwo}{BC\xspace}                                                              
\newcommand{\bcNoact}{\texorpdfstring{BC\textsubscript{$-\bar{a}$}}{BC-a}\xspace}            
\newcommand{\shieldOne}{shield~v1\xspace}
\newcommand{\shieldTwo}{shield~v2\xspace}

\DeclareRobustCommand{\slashurl}[1]{{\def\UrlBreaks{\do\/}\def\UrlBigBreaks{}\url{#1}}}

\title{Aero-VPA: Few-Step Flow-Matching Control over Non-Reconstructive JEPA Latents for Agile Flight on Edge Hardware}

\author{
  Yeonseok Lee\thanks{Version 2 (3 November 2026). Version 1 (26 October 2026): \url{https://doi.org/10.5281/zenodo.23633375}. Licences: code GPL-3.0-only (release v1.1.0), model weights GPL-3.0, benchmark dataset CC~BY~4.0 (\hyperref[sec:availability]{Code and Data Availability}). The runtime shield of Section~\ref{sec:shield} is \texttt{sim/shield.py}, unchanged from commit \texttt{2aefde1} to release v1.1.0.} \\
  \normalfont SLING AI Inc. \\
  \normalfont Incheon, Republic of Korea \\
  \normalfont\texttt{ylee@sling.ai.kr} \\[6pt]
  \normalfont Code: \url{https://github.com/ylee-sling/Aero-VPA} \\
  \normalfont Trained policy (72.6\,MB checkpoint): \url{https://huggingface.co/ylee-sling/aero-vpa}
}

\begin{document}
\maketitle

\begin{abstract}
Vision-Language-Action (VLA) models generalize across tasks, but their autoregressive decoders need sequential forward passes in proportion to the action horizon, each bounded below on embedded aerial platforms by streaming billions of parameters from memory. We present Aero-VPA, a lightweight Vision-Predictor-Action architecture for 3D navigation on edge hardware. A compact RGB-D Vision Transformer and a five-head ensemble predictor, trained without pixel reconstruction, forecast the next latent state; a Conditional Flow Matching solver generates a velocity and yaw-rate chunk in two Euler steps, so each decision takes five sequential network evaluations; and the ensemble dispersion drives an uncertainty fallback. Version~1 specified the architecture, deployment and protocol without results. This version reports the first closed-loop evaluation: PX4 software-in-the-loop simulation of quadrotor missions in procedurally generated forests. Behaviour cloning failed on all 60 missions despite open-loop errors of 0.02--0.04\,m/s: the policy copies its last command, a copycat form of causal confusion. DAgger with a privileged, safety-gated supervisor gave the first successes but plateaued at 11.1--21.1\% across six checkpoints. A map-free runtime shield with a 15\,s depth memory, barrier constraints, a cap on motion outside the 73\textdegree{} field of view and backoff along the flown path raised the first DAgger policy's success on 180 paired missions from 21.1\% to 52.2\% (95\% CI [45.0, 59.4]; exact McNemar $p = 1.2\times10^{-10}$) and cut its crashes from 72.2\% to 22.2\%, while timeouts rose from 2.8\% to 23.3\%. All six checkpoints gained 30.0--34.4 percentage points, at mean speeds of 1.16--1.36\,m/s. The code (\url{https://github.com/ylee-sling/Aero-VPA}), the trained policy (72.6\,MB, \url{https://huggingface.co/ylee-sling/aero-vpa}) and the full benchmark dataset are public.
\end{abstract}



\section{Introduction}
\label{sec:intro}

Navigating cluttered three-dimensional environments at speed requires a micro aerial vehicle (MAV) to convert onboard sensing into motor commands within a fraction of the time it needs to reach the nearest obstacle. Modular pipelines that chain visual-inertial odometry, volumetric mapping, and trajectory optimization~\cite{qin2018vinsmono,hornung2013octomap,mellinger2011minsnap,zhou2021ego} remain the dominant engineering solution, but each stage adds latency and couples the system to the accuracy of an explicit map; learned sensorimotor policies have accordingly been shown to fly faster and more robustly in the wild~\cite{loquercio2021highspeed} and to reach champion-level performance in drone racing~\cite{kaufmann2023champion,song2023reaching}. In parallel, Vision-Language-Action (VLA) models~\cite{brohan2023rt2,kim2024openvla,black2024pi0} have demonstrated broad semantic generalization by casting control as sequence modeling. Most VLAs, however, emit actions through an autoregressive language-model decoder: an $H$-step chunk of $d_a$-dimensional actions discretized into $q$ tokens per component requires $N = q H d_a$ dependent forward passes. At batch size one, every pass must stream the decoder weights from memory, so the latency is bounded below by $N|W_{\mathrm{LLM}}|b/BW$, where $|W_{\mathrm{LLM}}|$ is the number of parameters, $b$ the bytes per parameter, and $BW$ the memory bandwidth. On an NVIDIA Jetson Orin Nano 8\,GB, whose LPDDR5 interface provides 68\,GB/s in its original configuration and 102\,GB/s in the JetPack~6.2 ``Super'' configurations~\cite{nvidia2025jetpack}, a 3B-parameter model in 16-bit precision (6\,GB, three quarters of the device memory) needs at least 88\,ms per pass at 68\,GB/s. A single four-component velocity command ($q=1$, $H=1$, $d_a=4$) then takes at least 0.35\,s (at most 2.8\,Hz), and a 16-step chunk at least 5.6\,s, during which a vehicle flying at 5\,m/s travels 28\,m. These bounds are optimistic: they ignore compute, attention over the context, and memory shared with perception and middleware.

The Vision-Predictor-Action (VPA) framework~\cite{lee2026vpa} removes the language decoder from the control loop. A compact encoder maps observations into a latent space learned by a Joint-Embedding Predictive Architecture (JEPA)~\cite{lecun2022path,assran2023ijepa,bardes2024vjepa}, which predicts future \emph{embeddings} rather than future pixels and is therefore not required to model task-irrelevant appearance; a predictor forecasts the latent consequence of the selected skill; and a Conditional Flow Matching (CFM) solver~\cite{lipman2023flow,liu2023rectified,albergo2023interpolants} transports Gaussian noise to an entire action chunk by integrating a learned ordinary differential equation (ODE). Denoising diffusion policies~\cite{chi2025diffusion} generate chunks by iterative denoising, commonly with 10 to 100 network evaluations per chunk; the CFM sampler is instead a deterministic ODE map from a sampled initial condition, and straight transport paths admit accurate integration in very few steps; the number of sequential network evaluations per decision is $K+3$ for $K$ integration steps, independent of the horizon $H$~\cite{lee2026vpa}. A reference implementation with fewer than 20M trainable parameters ran a decision step in 10.5--14.2\,ms on an Apple M4 and succeeded in 40.0\% of LIBERO-10 manipulation episodes~\cite{lee2026vpa}. Flight, however, differs from tabletop manipulation in three respects that this paper addresses: the platform is underactuated and cannot pause in free space, so the controller needs a safe action whenever its predictions become unreliable; commands must be streamed to a flight controller at a fixed rate under strict failsafe timing~\cite{px4offboard}; and the compute budget is that of an embedded module rather than a workstation.

Aero-VPA is an autonomous navigation framework incorporating a JEPA-based latent world model. The JEPA predictor serves as the latent dynamics world model predicting forward state transitions, while a flow-matching policy acts as the downstream controller. This paper formalizes Aero-VPA, a specialization of VPA to aerial navigation whose long-term goal is agile flight on edge hardware, specifies the protocol by which it is evaluated, and reports its first closed-loop evaluation. The architecture has three parts: a high-level generative actor, the primitive selector $\pi^h_\phi$ and the CFM solver $v_\theta$; a lightweight latent dynamics world model that conditions the actor, the JEPA encoder $E_\psi$ and the five-head ensemble predictor $P_\omega$, whose ensemble mean enters the actor's conditioning and whose dispersion drives an uncertainty fallback, and which is not used for planning; and, added in this version, a deterministic low-level runtime safety shield.

\paragraph{From offline error to closed-loop failure.} Version~1 of this preprint reported no experiments. In PX4 software-in-the-loop (SITL) simulation, the policy trained by behaviour cloning (BC) on about 5.5 million frames of a privileged expert matched the expert almost perfectly offline---on held-out expert frames, the median error of its first velocity knot was 0.02\,m/s---but succeeded in none of 60 missions. Input-swap tests (Section~\ref{sec:copycat}) show why. The command follows the last dispatched setpoint $\bar{a}_t$, which is part of the proprioceptive input $S_t$: replacing $\bar{a}_t$ by that of another frame changed the command by 0.85\,m/s and moved it to within 0.31\,m/s of the other frame's label (medians; two frames' labels differ by a median of 1.13\,m/s), whereas replacing the image or the goal changed it by 0.01\,m/s. On expert data, $\bar{a}_t$ lies within about 0.02\,m/s of the next label, so repeating it is nearly optimal offline and useless in closed loop, where nothing corrects the drift. This is the copycat problem of behaviour cloning~\cite{wen2020copycat}, a form of causal confusion~\cite{dehaan2019causal}. Withholding $\bar{a}_t$ moved the shortcut to the measured velocity. DAgger~\cite{ross2011dagger}, stage~(iii) of the training procedure (Section~\ref{sec:training}), produced the first successes, but neither a second round nor four round-3 recipes aimed at the shortcut (input removal, state dropout, keyframe weighting) improved on the first round, and the DAgger policies still copied $\bar{a}_t$ on expert frames.

\paragraph{Safety under a narrow field of view.} The front camera sees 73\textdegree{} horizontally, and in simulation the vehicle responded to a velocity setpoint after a dead time of 0.24\,s with a first-order lag of 0.5\,s. Across our evaluations, 26.2--36.4\% of the policies' decisions were blind---the commanded direction lay outside the field of view, or the depth there was invalid---compared with 3.7\% for the expert on its own held-out flights. The braking envelope of the learned safety layer (Section~\ref{sec:envelope}) checks only the current frame, and with the identified response its stopping guarantee does not apply. A runtime safety layer therefore has to remember what the camera has seen, and it must not itself move blindly: in a pilot of our first shield version, 10 of its 11 crashes happened while the shield was flying the vehicle, and 8 of its 10 collisions were with obstacles beside or behind the camera.

\paragraph{Runtime shield.} We keep the learned components, train them with DAgger against a privileged supervisor that both labels the visited states and takes over when a safety gate fires, and add a runtime shield with no map and no learned parameters (Section~\ref{sec:shield}). The shield keeps a 15\,s voxel memory of depth returns, so that obstacles beside and behind the camera stay known; projects the commanded velocity onto barrier constraints in the manner of control barrier functions~\cite{ames2019cbf}; caps horizontal motion outside the field of view; enforces an altitude floor; holds the vehicle when decisions stop; and, when the vehicle comes too close to a remembered point, retraces its own flown path and then nudges it toward the waypoint. The filtered setpoint is written back as the policy's input $\bar{a}_t$.

\paragraph{Contributions.}
\begin{itemize}
  \item \textbf{Aero-VPA formulation.} A goal-conditioned partially observable formulation with chunks of heading-frame velocity and yaw-rate setpoints $A_t = [v_x, v_y, v_z, \dot{\psi}]_{t:t+H-1}$; an RGB-D Vision Transformer encoder trained jointly with a primitive- and command-conditioned JEPA predictor under VICReg regularization~\cite{bardes2022vicreg}; and a CFM solver sampled with $K=2$ Euler steps, giving a sequential depth of five network evaluations per decision. We add a noise-persistence rule and prove a Lipschitz bound on the change between consecutive chunks that it enables (Proposition~\ref{prop:coherence}).
  \item \textbf{Latent-uncertainty safety layer.} An epistemic uncertainty score computed from the normalized dispersion of a predictor ensemble~\cite{lakshminarayanan2017ensembles} without access to future observations, an asymmetric filter, a monotone velocity-attenuation law $\alpha(\bar{\sigma})$, an uncertainty-dependent chunk validity horizon, and a minimum-jerk braking transition to hover. A model-free braking envelope complements the learned signal, and we state the assumptions under which it guarantees a stop within the measured free space (Proposition~\ref{prop:envelope}) and why the evaluated configuration does not meet them (Remark~\ref{rem:envelope-evaluated}).
  \item \textbf{Asynchronous edge deployment.} A two-rate architecture for the Jetson Orin Nano (8\,GB, 15\,W) and a PX4 flight controller connected through micro-XRCE-DDS~\cite{px4uxrce}, comprising latency-compensated chunk indexing, continuous chunk blending, minimum-acceleration setpoint interpolation, an analytic parameter and memory-bandwidth budget, and an adaptation to quadruped robots. The embedded deployment is specified, not yet built; the evaluation runs the same executor logic on a cloud GPU (Section~\ref{sec:implementation}).
  \item \textbf{Benchmark protocol.} A software-in-the-loop (SITL) and hardware-in-the-loop (HITL) protocol with baselines, metric definitions, falsifiable hypotheses with acceptance criteria, a statistical analysis plan, and an ablation matrix, all fixed in version~1 before any experiment was executed (Section~\ref{sec:experiments}). Section~\ref{sec:evaluation} executes a first part of it, and Section~\ref{sec:deviations} lists every deviation.
  \item \textbf{A closed-loop diagnosis of copycat imitation and of DAgger's plateau.} BC failed on all 60 missions despite open-loop errors of 0.02--0.04\,m/s. Swap tests on expert frames and on the policies' own states identify the shortcut---copying $\bar{a}_t$, or the measured velocity once $\bar{a}_t$ is withheld---and show that neither input removal, state dropout nor keyframe weighting removes it. DAgger with a privileged, safety-gated supervisor gave the first successes (21.1\% against 0\% for BC; exact McNemar $p = 0.0005$ on the 60 shared missions), then plateaued: the second round was worse than the first (12.8\%, $p = 0.0315$), no round-3 recipe beat the first round (11.1--15.6\%), and the error on the policies' own states stayed near 0.5\,m/s across rounds (Sections~\ref{sec:copycat}--\ref{sec:plateau}).
  \item \textbf{A map-free runtime shield and its failure analysis.} On 180 paired test missions the shield raised success by 30.0--34.4 percentage points for all six DAgger checkpoints, to 52.2\% [45.0, 59.4] for the first-round policy ($p = 1.2 \times 10^{-10}$), and lowered crashes from 72.2--82.2\% to 22.2--42.8\%; its memory update took 2.9--3.0\,ms per frame at the 95th percentile. The gain held on the 144 missions never used to revise the shield (first-round policy: 52.1\%), and for the first-round policy it was largest in the densest worlds (6.7\% $\to$ 46.7\%). The shield moved failures from collisions to timeouts (2.8\% $\to$ 23.3\% for the first-round policy), and its design iteration moved its own collisions from obstacles beside or behind the camera (8 of 10 in the first version) to obstacles in view (6 of 7 in the second) (Sections~\ref{sec:shielded}--\ref{sec:activity} and~\ref{sec:discussion}).
  \item \textbf{Open release.} The trained policies are public on \href{https://huggingface.co/ylee-sling/aero-vpa}{Hugging Face} (GPL-3.0): the six DAgger checkpoints, each a single file of 72.6--72.7\,MB, the three behaviour-cloning baselines and the two stage-(i) checkpoints. So are the code (GPL-3.0-only, release v1.1.0)~\cite{slingai2026code} and the benchmark dataset of every closed-loop mission record, with the decision traces, the 36 test worlds and the analysis reports (CC~BY~4.0)~\cite{slingai2026benchmark} (\hyperref[sec:availability]{Code and Data Availability}).
\end{itemize}

\paragraph{Changes in version 2.} Version~1 (26 October 2026) specified the architecture, the deployment and the benchmark protocol and reported no results. Version~2 keeps Sections~\ref{sec:related}--\ref{sec:experiments} and changes them only where the evaluated system differs or the evaluation required an addition: the predictor is in residual form (Eq.~\eqref{eq:predictor}); $\sigma_{\mathrm{crit}}$ is calibrated from per-flight maxima (Section~\ref{sec:uncertainty}); Remark~\ref{rem:envelope-evaluated} states that the evaluated braking envelope does not satisfy Assumption~\ref{ass:envelope}; the DAgger stage is specified with the supervisor that was used (Section~\ref{sec:training}); the runtime shield is new (Section~\ref{sec:shield}, Figure~\ref{fig:system}); Section~\ref{sec:implementation} describes the implementation used in the evaluation; and Section~\ref{sec:experiments} keeps the protocol of version~1 with its execution status. Sections~\ref{sec:evaluation}--\ref{sec:conclusion} and the appendices are new; Section~\ref{sec:related} adds work on causal confusion, interactive imitation and safety filters. The networks and their sizes are unchanged (18.13M trainable parameters).

\paragraph{Scope.} All results are from PX4 SITL with Gazebo in static, procedurally generated forests, with one training seed per recipe. The shielded mean speeds were 1.16--1.36\,m/s, far below the agile flight that remains the goal, and timing was measured on an NVIDIA L4 GPU, not on the Jetson Orin Nano of Section~\ref{sec:system}. The embedded deployment, the hardware-in-the-loop stage, the baselines and most ablations of Section~\ref{sec:experiments} have not been executed.


\section{Related Work}
\label{sec:related}

\subsection{Visual Predictive World Models and Non-Reconstructive Latents}
Generative world models learn dynamics by reconstructing observations and have achieved strong results in reinforcement learning~\cite{hafner2025dreamerv3}, but reconstruction obliges the model to account for every pixel, including illumination, texture, and sensor noise that are irrelevant to control. Joint-embedding predictive architectures~\cite{lecun2022path} instead predict the embedding of a target from the embedding of a context: I-JEPA~\cite{assran2023ijepa} predicts masked image regions in representation space, V-JEPA~\cite{bardes2024vjepa} extends this to video, and V-JEPA~2~\cite{assran2025vjepa2} shows that an action-conditioned latent predictor post-trained on robot interaction data supports zero-shot manipulation through planning in representation space, at the cost of iterative optimization at inference time. TD-MPC2~\cite{hansen2024tdmpc2} likewise learns a decoder-free latent model for continuous control. Without a reconstruction target, joint-embedding objectives admit collapsed solutions; VICReg~\cite{bardes2022vicreg} prevents them with variance and covariance penalties on the embedding batch. In robot learning, ACT-JEPA~\cite{vujinovic2026actjepa} uses latent prediction to regularize imitation policies, VLA-JEPA~\cite{sun2026vlajepa} adds a latent world model to a VLA backbone, and VPA~\cite{lee2026vpa,lee2026nsjepa} couples a VICReg-regularized JEPA predictor with a discrete skill selector and a flow-matching action head without a language decoder. Aero-VPA retains this structure and adapts its inputs, conditioning, and safety logic to flight.

\subsection{Continuous Normalizing Flows and Flow Matching for Robot Action Generation}
Continuous normalizing flows define a generative model by integrating a learned vector field~\cite{chen2018neuralode}. Flow matching~\cite{lipman2023flow}, rectified flow~\cite{liu2023rectified}, and stochastic interpolants~\cite{albergo2023interpolants} train such fields by regression onto the velocity of a prescribed probability path, avoiding simulation during training; straight paths permit few-step integration. In robotics, action chunking with transformers (ACT)~\cite{zhao2023act} showed that predicting multi-step chunks mitigates compounding error, Diffusion Policy~\cite{chi2025diffusion} models multimodal chunk distributions by iterative denoising, and consistency distillation reduces the number of denoising steps~\cite{prasad2024consistency}. $\pi_0$~\cite{black2024pi0} attaches a flow-matching action expert to a multi-billion-parameter vision-language model, and Octo~\cite{octo2024} attaches a diffusion head to a transformer backbone of 27M--93M parameters. Real-time chunking~\cite{black2025rtc} addresses the discontinuities that arise when chunks are replanned asynchronously by inpainting the portion of the next chunk that overlaps the current one. Aero-VPA uses a CFM head with $K=2$, conditions it on forecast latents rather than on the hidden states of a language backbone, and addresses inter-chunk consistency by noise persistence and continuous blending (Section~\ref{sec:cfm}).

Closest to our work is Qantara~\cite{rakhimov2026qantara}, which also couples a JEPA latent model with flow matching on actions. Qantara trains its encoder and a shared block-causal transformer predictor jointly from scratch: the next latent is learned with a Brownian-bridge objective anchored at the current latent, the action with a noise-to-data flow-matching objective, and SIGReg serves as the only anti-collapse term. By sampling separate noise levels for the action and latent axes, $(\tau^a, \tau^z)$ in their notation, along the edges and the diagonal of the unit square, a single 21M-parameter checkpoint serves latent planning with the cross-entropy method, behaviour cloning, and a video-inverse path that predicts the next latent and then infers the connecting action. Aero-VPA operates at a similar scale, also samples actions with two Euler steps, and follows the same predict-then-act chain as that last path, but resolves the design trade-off differently: it commits to a single goal- and primitive-conditioned inference path and decouples its components. The VICReg-regularized encoder and ensemble predictor are trained first and frozen, which fixes the latent scale that normalizes the uncertainty signal of Section~\ref{sec:uncertainty}. The CFM head is trained afterwards on velocity chunks, so it can be retrained per embodiment and compiled as an independent inference engine. Qantara is evaluated in simulated navigation, manipulation, and reaching tasks, with timing measured on an H100 GPU. Aero-VPA instead targets agile three-dimensional flight under the fixed-rate setpoint interface of PX4 on a 15\,W embedded module, and couples action generation to an uncertainty-gated safety layer.

\subsection{Edge Autonomous Flight and Computational Constraints in Aerial and Legged Systems}
Classical agile-flight stacks combine visual-inertial state estimation~\cite{qin2018vinsmono,qin2019vinsfusion}, occupancy mapping~\cite{hornung2013octomap}, and polynomial or gradient-based planning~\cite{mellinger2011minsnap,zhou2021ego}, and have been scaled to swarms in unstructured environments~\cite{zhou2022swarm}. Learning-based approaches replace part of this stack: Loquercio~et~al.~\cite{loquercio2021highspeed} trained a depth-based policy by imitating a privileged planner and executed it on an embedded GPU, Kaufmann~et~al.~\cite{kaufmann2023champion} trained racing policies with deep reinforcement learning, and Agilicious~\cite{foehn2022agilicious} provides open hardware and software with onboard GPU compute for vision-based agile flight. These systems flew far faster than the 1.16--1.36\,m/s reported in Section~\ref{sec:evaluation}. Learned residual dynamics address aerodynamic effects such as ground effect~\cite{shi2019neurallander} and strong wind~\cite{oconnell2022neuralfly}. General navigation models such as GNM~\cite{shah2023gnm} and NoMaD~\cite{sridhar2024nomad} transfer across ground robots but target lower speeds. In legged locomotion, hierarchical designs pair a high-level command policy with a low-level controller trained by reinforcement learning~\cite{lee2020quadruped,kumar2021rma,margolis2022wtw}; this separation of time scales motivates the cross-embodiment discussion in Section~\ref{sec:legged}. Across these platforms the binding constraints are size, weight, and power (SWaP), memory bandwidth, and the fixed-rate interfaces of the low-level controller~\cite{meier2015px4,px4offboard}, which together determine the design choices of Section~\ref{sec:system}.

\subsection{Causal Confusion and Interactive Imitation}
\label{sec:related-imitation}
De Haan et al.~\cite{dehaan2019causal} showed that giving an imitation policy more information can hurt it, because it may rely on nuisance correlates of the expert's action. Wen et al.~\cite{wen2020copycat} described the copycat problem of policies that observe their own history and repeat their previous action, and later reweighted training toward keyframes where the action changes~\cite{wen2021keyframe}; our \ckKw recipe follows that idea. Codevilla et al.~\cite{codevilla2019limitations} described the inertia problem of driving policies that condition on their speed, and ChauffeurNet~\cite{bansal2019chauffeurnet} dropped out past motion during training, the precedent of our state-dropout recipe \ckSd. In Aero-VPA the copied quantity is the last dispatched velocity setpoint $\bar{a}_t$, and withholding it moves the shortcut to the measured velocity.

DAgger~\cite{ross2011dagger} labels the states the learner visits. Learning by Cheating~\cite{chen2019lbc} first trains a privileged agent and then distils it into a sensorimotor student, and SafeDAgger~\cite{zhang2017safedagger} and HG-DAgger~\cite{kelly2019hgdagger} let a learned safety policy or a human expert decide when the expert takes over. Our supervisor combines these roles: a privileged planner labels every visited state and takes control when a safety gate fires, softly when it can keep flying its plan.

\subsection{Safety Filters, Shields, and Barrier Functions}
\label{sec:related-shields}
Shields~\cite{alshiekh2018shielding} override a learning agent's actions that would violate a specification, and safety filters~\cite{wabersich2023datadriven,hsu2024safetyfilter} minimally modify a nominal controller's command, often through control barrier functions~\cite{ames2019cbf}. NanoMap~\cite{florence2018nanomap} answers proximity queries against a history of depth frames, so that a vehicle with a narrow field of view can account for obstacles it no longer sees. Our runtime shield combines such a memory with a barrier-function-style velocity constraint enforced by sequential projection and with recovery manoeuvres. It provides no formal guarantee; Section~\ref{sec:shield} states the conditions under which its constraint is invariant.


\section{Problem Formulation and Proposed Architecture}
\label{sec:method}


\begin{figure}[t]
\centering
\begin{tikzpicture}[
  font=\footnotesize,
  blk/.style={draw, rounded corners=2pt, align=center, minimum height=10mm, text width=18.5mm, inner sep=2pt, fill=white},
  io/.style={align=center, text width=16mm, inner sep=1pt},
  arr/.style={-{Stealth[length=1.8mm]}, semithick},
  lbl/.style={font=\scriptsize, inner sep=1pt}
]
\node[io] (obs) at (0,0) {RGB-D $I_t$\\ odometry $S_t$};
\node[blk] (enc) at (2.6,0) {ViT encoder\\ $E_\psi$};
\node[blk] (sel) at (5.2,0) {Primitive\\ selector $\pi^h_\phi$};
\node[blk] (pred) at (7.8,0) {JEPA ensemble\\ $P_\omega$, $N_e{=}5$};
\node[blk] (cfm) at (10.4,0) {CFM solver\\ $v_\theta$, $K{=}2$};
\node[blk] (safe) at (13.0,0) {Safety filter\\ $\alpha(\bar{\sigma})$, $v_{\mathrm{safe}}$};
\node[blk] (buf) at (13.0,-2.3) {Chunk buffer\\ (validity $H_t\Delta$)};
\node[blk] (interp) at (10.4,-2.3) {Interpolate, blend,\\ limit; $\mathcal{H}_t \to$ NED};
\node[blk, fill=orange!8] (rot) at (7.8,-2.3) {Runtime shield\\ (Sec.~\ref{sec:shield})};
\node[blk] (dds) at (5.2,-2.3) {uXRCE-DDS\\ agent};
\node[blk, text width=26mm] (px4) at (5.2,-4.5) {PX4 DDS client,\\ vel./att./rate control};
\node[io] (plant) at (0,-4.5) {Vehicle or\\ PX4 SITL};
\draw[arr] (obs) -- (enc);
\draw[arr] (enc) -- node[lbl, above] {$z_t$} (sel);
\draw[arr] (sel) -- node[lbl, above] {$u_t$} (pred);
\draw[arr] (pred) -- node[lbl, above] {$\hat{z}_{t+1}$} (cfm);
\draw[arr] (cfm) -- node[lbl, above] {$\hat{A}_t$} (safe);
\draw[arr] (pred.north) to[out=35, in=145] node[lbl, above] {$\sigma_{t+1}$} (safe.north);
\draw[arr] (safe) -- node[lbl, right] {$\mathcal{C}_t$} (buf);
\draw[arr] (buf) -- (interp);
\draw[arr] (interp) -- (rot);
\draw[arr] (rot) -- (dds);
\draw[arr] (dds) -- node[lbl, right, pos=0.55] {NED setpoints, 100\,Hz} (px4);
\draw[arr] (px4) -- node[lbl, above] {motor commands} (plant);
\draw[arr] (plant) -- node[lbl, left, align=right] {camera,\\ EKF2} (obs);
\draw[arr, dashed] (rot.north) -- ++(0,0.55) -| node[lbl, pos=0.75, right] {$\bar{a}_t$} ($(obs.south)+(0.45,0)$);
\begin{scope}[on background layer]
  \node[draw=blue!50!black, fill=blue!4, rounded corners=3pt, inner sep=6pt, fit=(enc)(safe)(buf)(dds)] (jet) {};
  \node[draw=orange!60!black, fill=orange!6, rounded corners=3pt, inner sep=4pt, fit=(px4)] (fcu) {};
\end{scope}
\node[lbl, anchor=south west] at (jet.north west) {companion computer (target: Jetson Orin Nano)};
\node[lbl, anchor=west] at ($(fcu.east)+(0.15,0)$) {flight controller (FMU or SITL)};
\node[lbl, anchor=north west] at ($(enc.south west)+(0,-0.15)$) {inference loop, event-driven at $f_{\mathrm{dec}} \ge 30$\,Hz; sequential depth $K{+}3$};
\node[lbl, anchor=south west] at ($(dds.north west)+(0,0.15)$) {executor, fixed rate $f_c = 100$\,Hz};
\end{tikzpicture}
\caption{Aero-VPA deployment architecture. The inference loop (top) runs once per camera frame and produces a velocity and yaw-rate chunk $\mathcal{C}_t$ through five sequential network evaluations; the safety filter applies the uncertainty-dependent attenuation $\alpha(\bar{\sigma})$, the braking envelope $v_{\mathrm{safe}}$, and the validity horizon $H_t$. The executor (bottom) evaluates the newest valid chunk at its chunk-local time, converts it to NED setpoints, and streams them to PX4 over micro-XRCE-DDS; it engages the braking fallback when a fallback is requested or no valid chunk is available. The runtime shield (orange, added in version~2) filters every setpoint against a 15\,s depth memory or flies recovery manoeuvres itself; the setpoint it dispatches is fed back as $\bar{a}_t$ (dashed). In the evaluation of Section~\ref{sec:evaluation} the companion-computer software ran on a cloud GPU against PX4 SITL (Section~\ref{sec:implementation}).}
\label{fig:system}
\end{figure}
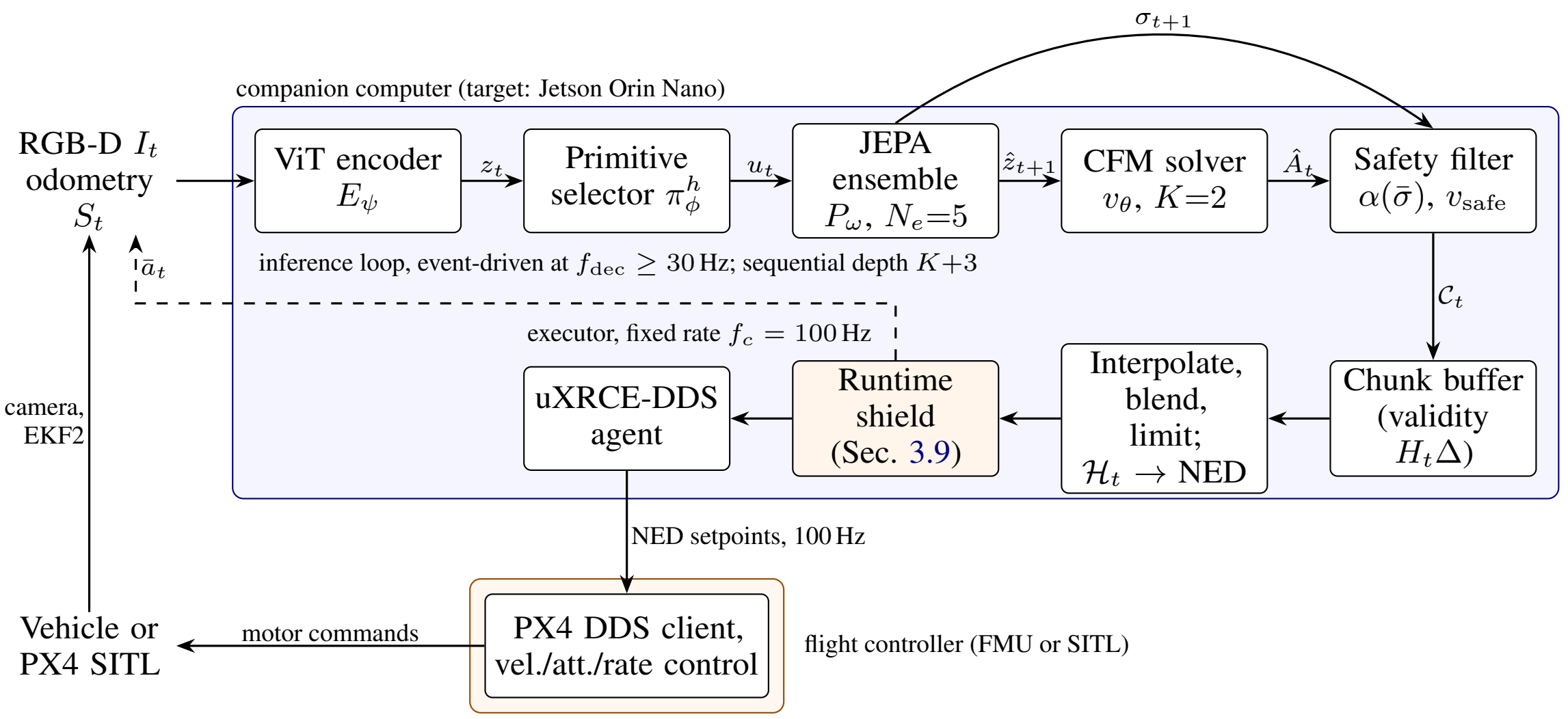

\subsection{Goal-Conditioned POMDP with Velocity Chunks}
\label{sec:pomdp}
We model goal-directed flight as a goal-conditioned partially observable Markov decision process (POMDP)
\begin{equation}
  \mathcal{M} = \left(\mathcal{S}, \mathcal{A}, \mathcal{O}, \mathcal{T}, \Omega, \mathcal{G}, R, \gamma_{\mathrm{disc}}\right),
  \label{eq:pomdp}
\end{equation}
whose components are defined as follows. A state $s = (p, v, \mathbf{R}, \omega, w, \mathcal{X}) \in \mathcal{S}$ comprises the position $p \in \R^3$ and velocity $v \in \R^3$ of the vehicle in a local north-east-down (NED) world frame $\mathcal{W}$, its attitude $\mathbf{R} \in SO(3)$, its body angular rate $\omega \in \R^3$, the ambient wind $w \in \R^3$, and the occupied set $\mathcal{X} \subset \R^3$ of the environment. The vehicle observes $o_t = (I_t, S_t) \in \mathcal{O}$ at decision step $t$, captured at wall-clock time $\tau_t$, where $I_t \in \R^{H_{\mathrm{img}} \times W_{\mathrm{img}} \times 4}$ is an RGB-D frame ($H_{\mathrm{img}} = 128$, $W_{\mathrm{img}} = 160$) and
\begin{equation}
  S_t = \big(\hat{v}^{\mathcal{H}}_t,\, \hat{n}^{\mathcal{B}}_t,\, \hat{\omega}^{\mathcal{B}}_t,\, \bar{a}_t\big) \in \R^{d_s}, \qquad d_s = 13,
  \label{eq:proprio}
\end{equation}
is the proprioceptive state estimated by the PX4 EKF2: the velocity $\hat{v}^{\mathcal{H}}_t \in \R^3$ expressed in the heading frame $\mathcal{H}_t$ (the gravity-aligned forward-right-down frame obtained by rotating $\mathcal{W}$ about its down axis by the estimated yaw $\hat{\psi}_t$), the unit gravity direction $\hat{n}^{\mathcal{B}}_t \in \mathbb{S}^2$ in the body frame $\mathcal{B}$, which encodes roll and pitch, the body rates $\hat{\omega}^{\mathcal{B}}_t \in \R^3$, and the setpoint $\bar{a}_t \in \R^4$ dispatched most recently to the flight controller, expressed in $\mathcal{H}_t$. The observation model $\Omega$ is induced by image formation, $I_t = \mathcal{I}(s(\tau_t), \eta_t)$, where $\eta_t$ collects nuisance factors such as illumination, texture, motion blur, and depth noise.

A mission $G = (x, (w^{(m)})_{m=1}^{M}) \in \mathcal{G}$ consists of an optional natural-language instruction $x$ and an ordered sequence of $M$ waypoints $w^{(m)} \in \R^3$ in $\mathcal{W}$. With the active waypoint index $m_t$ (Section~\ref{sec:selector}) and the estimated position $\hat{p}_t$, the policy receives the goal feature
\begin{equation}
  g_t = \left(\frac{r_t}{\max(\norm{r_t}_2, \epsilon_r)},\; \frac{\min(\norm{r_t}_2, D_{\max})}{D_{\max}}\right) \in \R^4, \qquad r_t = \mathbf{R}_z(\hat{\psi}_t)^{\top}\big(w^{(m_t)} - \hat{p}_t\big),
  \label{eq:goal}
\end{equation}
where $\mathbf{R}_z(\psi) \in SO(3)$ is the rotation about the down axis by $\psi$, $D_{\max}$ is a saturation distance, and $\epsilon_r > 0$ avoids division by zero.

The per-step action $a = (v_x, v_y, v_z, \dot{\psi}) \in \mathcal{A}_0 = [-\bar{v}_h, \bar{v}_h]^2 \times [-\bar{v}_z, \bar{v}_z] \times [-\bar{\dot{\psi}}, \bar{\dot{\psi}}]$ is a velocity setpoint and a yaw-rate setpoint, with speed limits $\bar{v}_h$, $\bar{v}_z$, and $\bar{\dot{\psi}}$. At each decision step the controller outputs an open-loop chunk of $H$ knots spaced $\Delta$ seconds apart,
\begin{equation}
  A_t = [v_x, v_y, v_z, \dot{\psi}]_{t:t+H-1} = \big[a_{t,0};\, a_{t,1};\, \dots;\, a_{t,H-1}\big] \in \mathcal{A}_0^H \subset \R^{H \times d_a}, \qquad d_a = 4,
  \label{eq:chunk}
\end{equation}
where knot $a_{t,k}$ applies at time $\tau_t + k\Delta$ and its velocity components are expressed in the heading frame $\mathcal{H}_t$ frozen at the capture time $\tau_t$. We use $\Delta = 20$\,ms and $H = 16$, so that a chunk spans 0.32\,s. The transition kernel $\mathcal{T}(s' \mid s, a)$ is induced by the PX4 velocity and attitude controllers, the rigid-body and rotor dynamics, and the wind. The reward
\begin{equation}
  R(s, G; m) = -\norm{p - w^{(m)}}_2 - \lambda_{\mathrm{col}}\, \mathbb{1}\!\left[\operatorname{dist}(p, \mathcal{X}) < r_v\right],
  \label{eq:reward}
\end{equation}
with vehicle radius $r_v$ and collision weight $\lambda_{\mathrm{col}} > 0$, and the discount $\gamma_{\mathrm{disc}} \in [0,1)$ define the task for analysis and for the reinforcement-learning baselines of Section~\ref{sec:baselines}. Aero-VPA itself is trained by self-supervised latent prediction and by imitation of a privileged expert (Section~\ref{sec:training}).

\subsection{Non-Reconstructive JEPA World Model}
\label{sec:jepa}
A Vision Transformer~\cite{dosovitskiy2021vit} $E_\psi$ with patch size 16 divides the four-channel frame into $N_p = (128/16)(160/16) = 80$ patches, processes them together with a class token through six blocks of width 384, and projects the class token to the latent
\begin{equation}
  z_t = E_\psi(I_t) \in \mathcal{Z} \subset \R^d, \qquad d = 256.
  \label{eq:encoder}
\end{equation}
The depth channel is clipped to the sensor range and normalized before patch embedding. When an instruction $x$ is given, a frozen CLIP text encoder~\cite{radford2021clip} computes $c_{\mathrm{text}} \in \R^{512}$ once per mission; otherwise $c_{\mathrm{text}}$ is a learned null embedding.

The predictor $P_\omega$ forecasts the latent state $T_\nu = 0.1$\,s ahead, conditioned on the current latent, the active primitive $u_t \in \mathcal{U}$ (Section~\ref{sec:selector}), and the executing setpoint $\bar{a}_t$. It consists of $N_e = 5$ independently initialized heads $P^{(i)}_\omega$ in residual form: head $i$ adds the output of an MLP $f^{(i)}_\omega$ ($292 \to 512 \to 512 \to 256$) to the current latent,
\begin{equation}
  \hat{z}^{(i)}_{t+1} = z_t + f^{(i)}_\omega\big(z_t, \operatorname{Embed}(u_t), \bar{a}_t\big), \qquad \hat{z}_{t+1} = \frac{1}{N_e}\sum_{i=1}^{N_e} \hat{z}^{(i)}_{t+1},
  \label{eq:predictor}
\end{equation}
where $\operatorname{Embed}: \mathcal{U} \to \R^{d_u}$ is a learned primitive embedding ($d_u = 32$) and the subscript $t+1$ abbreviates the time $\tau_t + T_\nu$. Version~1 wrote the heads without the residual connection; every evaluated policy except \bcVone (Table~\ref{tab:names}) uses the residual form. Conditioning on $\bar{a}_t$, which is known before the decision, makes the predictor action-conditioned without placing the solver output on its input, so the dependency chain of a decision step remains acyclic. The predictor is not used for planning, imagined rollouts or model-predictive control: its ensemble mean conditions the solver (Eq.~\eqref{eq:cond}), and its dispersion drives the uncertainty layer (Section~\ref{sec:uncertainty}). The encoder and predictor are trained jointly with
\begin{equation}
  \mathcal{L}_{\mathrm{JEPA}}(\psi, \omega) = \E_t\!\left[\frac{1}{N_e}\sum_{i=1}^{N_e}\norm{\hat{z}^{(i)}_{t+1} - \sg\!\big(E_{\bar{\psi}}(I_{t+1})\big)}_2^2\right] + \lambda_v\, v(Z) + \lambda_c\, c(Z),
  \label{eq:jepa}
\end{equation}
where $I_{t+1}$ is the frame captured closest to $\tau_t + T_\nu$, $\sg(\cdot)$ is the stop-gradient operator, $E_{\bar{\psi}}$ is a momentum copy of the encoder updated as $\bar{\psi} \leftarrow \mu \bar{\psi} + (1-\mu)\psi$ with $\mu \in (0,1)$, and $Z \in \R^{B \times d}$ is the batch of online latents $z_t$. The variance and covariance penalties of VICReg~\cite{bardes2022vicreg} are
\begin{align}
  v(Z) &= \frac{1}{d}\sum_{j=1}^{d} \max\!\Big(0,\; \gamma - \sqrt{\Var(Z_{\cdot,j}) + \epsilon}\Big), \label{eq:var}\\
  c(Z) &= \frac{1}{d}\sum_{i \neq j} \big[C(Z)\big]_{i,j}^2, \qquad C(Z) = \frac{1}{B-1}\sum_{b=1}^{B}\big(Z_{b,\cdot} - \bar{Z}\big)^{\top}\big(Z_{b,\cdot} - \bar{Z}\big), \label{eq:cov}
\end{align}
with margin $\gamma > 0$, numerical constant $0 < \epsilon < \gamma^2$, and batch mean $\bar{Z} \in \R^{1 \times d}$. The variance hinge counteracts collapse to a constant embedding and the covariance term counteracts dimensional collapse; the conditions under which the constant encoder is excluded from the global minimizers, and the limits of this guarantee, are analyzed in~\cite{lee2026vpa}. During training we track the realized latent scale
\begin{equation}
  \bar{\gamma}_n = (1 - \alpha_\gamma)\,\bar{\gamma}_{n-1} + \alpha_\gamma \Big(\frac{1}{d}\sum_{j=1}^{d}\Var\big(Z^{(n)}_{\cdot,j}\big) + \epsilon\Big)^{1/2},
  \label{eq:scale}
\end{equation}
where $Z^{(n)}$ is the batch at iteration $n$ and $\alpha_\gamma \in (0,1]$. After this stage $E_\psi$ and $P_\omega$ are frozen, which fixes the scale $\bar{\gamma}^\star$ and the latent mean $\bar{\mu}_Z$ used below. The evaluated system uses $\lambda_v = 25$, $\lambda_c = 1$ and $\mu = 0.996$ (Appendix~\ref{app:hyper}).

\subsection{Primitive Selection and Waypoint Progression}
\label{sec:selector}
The selector $\pi^h_\phi$ is a multilayer perceptron that chooses one of $N_u = 6$ flight primitives $\mathcal{U} = \{\textsc{Cruise},\allowbreak\ \textsc{LateralAvoid},\allowbreak\ \textsc{VerticalAvoid},\allowbreak\ \textsc{YawScan},\allowbreak\ \textsc{Approach},\allowbreak\ \textsc{BrakeHover}\}$:
\begin{equation}
  u_t = \argmax_{u \in \mathcal{U}} \pi^h_\phi\big(u \mid z_t, g_t, S_t, c_{\mathrm{text}}\big),
  \label{eq:selector}
\end{equation}
trained with the cross-entropy loss
\begin{equation}
  \mathcal{L}_{\mathrm{CE}}(\phi) = -\E_t\big[\log \pi^h_\phi(u^\star_t \mid z_t, g_t, S_t, c_{\mathrm{text}})\big],
  \label{eq:ce}
\end{equation}
where $u^\star_t$ is a label derived from the expert command by fixed rules applied in priority order (\textsc{BrakeHover} if the expert speed is below a hover threshold; \textsc{Approach} within a distance of the final waypoint; \textsc{VerticalAvoid}, \textsc{LateralAvoid}, or \textsc{YawScan} if the corresponding command component exceeds a threshold; \textsc{Cruise} otherwise). These rules are a stand-in for annotated or learned segmentations and are not part of the method. The selector output is restricted to $\mathcal{U}$, and \textsc{BrakeHover} doubles as the target of the safety override (Section~\ref{sec:uncertainty}).

The active waypoint advances by the symbolic rule
\begin{equation}
  m_{t+1} =
  \begin{cases}
    m_t + 1, & \text{if } \norm{\hat{p}_t - w^{(m_t)}}_2 < \tau_w \text{ and } m_t < M,\\
    m_t, & \text{otherwise,}
  \end{cases}
  \label{eq:waypoint}
\end{equation}
with acceptance radius $\tau_w > 0$. In~\cite{lee2026vpa} the corresponding test is applied to latent distances to goal images, and its soundness depends on a lower co-Lipschitz bound of the encoder that cannot be enforced architecturally. Because flight waypoints are geometric, Eq.~\eqref{eq:waypoint} tests the state estimate instead: if $\norm{\hat{p}_t - p_t}_2 \le \epsilon_p$, a transition implies $\norm{p_t - w^{(m_t)}}_2 < \tau_w + \epsilon_p$ by the triangle inequality, without any assumption on $E_\psi$.

\subsection{Conditional Flow-Matching Policy Solver}
\label{sec:cfm}
The solver conditions on the standardized vector
\begin{equation}
  e_t = \operatorname{Concat}\big(\operatorname{Embed}(u_t),\; \tilde{z}_t,\; \tilde{\hat{z}}_{t+1},\; \tilde{S}_t,\; g_t,\; c_{\mathrm{text}}\big) \in \R^{d_e}, \qquad d_e = d_u + 2d + d_s + 4 + 512 = 1073,
  \label{eq:cond}
\end{equation}
where $\tilde{z} = (z - \bar{\mu}_Z)/\bar{\gamma}^\star$ is applied to both latents and $\tilde{S}_t$ is $S_t$ standardized with dataset statistics. Actions are mapped to $[-1,1]$ componentwise by the box bounds of $\mathcal{A}_0$; we keep the symbol $A_t$ for the normalized chunk. The vector field $v_\theta: \R^{H \times d_a} \times [0,1] \times \R^{d_e} \to \R^{H \times d_a}$ is a four-block transformer of width 256 that attends over the $H$ knot tokens and a conditioning token. With integration time $\rho \in [0,1]$ and Gaussian noise $\xi \sim \mathcal{N}(0, I)$, $\xi \in \R^{H \times d_a}$, the linear (optimal-transport) path
\begin{equation}
  A_t^{(\rho)} = \rho A_t + (1 - \rho)\,\xi
  \label{eq:path}
\end{equation}
has constant conditional velocity $A_t - \xi$, and $v_\theta$ is trained by the flow-matching regression
\begin{equation}
  \mathcal{L}_{\mathrm{CFM}}(\theta) = \E_{t,\, \xi,\, \rho \sim \mathcal{U}[0,1]}\left[\norm{v_\theta\big(A_t^{(\rho)}, \rho \mid e_t\big) - (A_t - \xi)}_F^2\right],
  \label{eq:cfm}
\end{equation}
where $\norm{\cdot}_F$ is the Frobenius norm and the expectation over $t$ runs over the training data. At deployment, a chunk is obtained by integrating $\mathrm{d}A/\mathrm{d}\rho = v_\theta(A, \rho \mid e_t)$ from $A^{(0)} = \xi_t$ with $K$ Euler steps,
\begin{equation}
  A^{(\rho_{k+1})} = A^{(\rho_k)} + \frac{1}{K}\, v_\theta\big(A^{(\rho_k)}, \rho_k \mid e_t\big), \qquad \rho_k = \frac{k}{K}, \quad k = 0, \dots, K-1,
  \label{eq:euler}
\end{equation}
which for $K = 2$ reads
\begin{equation}
  \hat{A}_t = \xi_t + \tfrac{1}{2} v_\theta(\xi_t, 0 \mid e_t) + \tfrac{1}{2} v_\theta\Big(\xi_t + \tfrac{1}{2} v_\theta(\xi_t, 0 \mid e_t),\, \tfrac{1}{2} \;\Big|\; e_t\Big).
  \label{eq:euler2}
\end{equation}
The chunk is then denormalized and projected componentwise onto $\mathcal{A}_0$. Although each conditional path in Eq.~\eqref{eq:path} is straight, the learned marginal field can be curved where the action distribution is multimodal, so the accuracy of two-step integration is an empirical property to be measured (ablation A1, Section~\ref{sec:ablation}).

\paragraph{Noise persistence.} In~\cite{lee2026vpa}, frequent replanning with fresh noise reduced manipulation success, plausibly because consecutive chunks switched between distinct but individually valid modes. Because the sampler of Eq.~\eqref{eq:euler} is deterministic given $\xi_t$, we reuse the initial noise across decisions,
\begin{equation}
  \xi_t =
  \begin{cases}
    \xi_{t-1}, & \text{if } u_t = u_{t-1} \text{ and } t \not\equiv 0 \pmod{N_\xi},\\
    \text{a fresh draw from } \mathcal{N}(0, I), & \text{otherwise,}
  \end{cases}
  \label{eq:noise}
\end{equation}
so that the noise is redrawn when the primitive changes and at least every $N_\xi$ decisions. The following proposition shows what persistence buys.

\begin{proposition}[Chunk coherence under noise persistence]
\label{prop:coherence}
Suppose that, on a set containing all Euler iterates, $v_\theta$ is $L_A$-Lipschitz in its first argument and $L_e$-Lipschitz in its conditioning argument with respect to $\norm{\cdot}_F$ and $\norm{\cdot}_2$, respectively, with $L_A > 0$. Let $\Phi_K(\xi; e)$ denote the output of Eq.~\eqref{eq:euler} for initial condition $\xi$ and conditioning $e$. Then, for any $\xi$ and any $e, e'$,
\begin{equation}
  \norm{\Phi_K(\xi; e) - \Phi_K(\xi; e')}_F \le \frac{L_e}{L_A}\Big[\big(1 + \tfrac{L_A}{K}\big)^K - 1\Big]\norm{e - e'}_2 \le \frac{L_e}{L_A}\big(e^{L_A} - 1\big)\norm{e - e'}_2.
  \label{eq:coherence}
\end{equation}
\end{proposition}
\begin{proof}
Let $\delta_k = \norm{A^{(\rho_k)} - A'^{(\rho_k)}}_F$ for the iterates under $e$ and $e'$, so $\delta_0 = 0$. By Eq.~\eqref{eq:euler} and the triangle inequality, $\delta_{k+1} \le (1 + L_A/K)\,\delta_k + (L_e/K)\norm{e - e'}_2$. Unrolling the recursion gives $\delta_K \le (L_e/K)\norm{e-e'}_2\sum_{j=0}^{K-1}(1 + L_A/K)^j = (L_e/L_A)[(1 + L_A/K)^K - 1]\norm{e-e'}_2$, and $(1 + L_A/K)^K \le e^{L_A}$.
\end{proof}
With a shared $\xi$, the change between consecutive chunks is therefore bounded in proportion to the change of the conditioning vector, which is small between decisions 33\,ms apart; with independent draws no such bound exists, since $\xi_t$ and $\xi_{t-1}$ may be transported to different modes. The constants $L_A$ and $L_e$ are not controlled by training, and dot-product attention is not globally Lipschitz~\cite{kim2021lipschitz}, so Eq.~\eqref{eq:coherence} explains the mechanism rather than providing a certified bound; its practical effect is tested in ablation A5.

\subsection{Epistemic Uncertainty and Velocity Attenuation}
\label{sec:uncertainty}
At deployment the future latent is unobserved, so the reliability of $\hat{z}_{t+1}$ must be assessed from quantities available at time $\tau_t$. Following deep ensembles~\cite{lakshminarayanan2017ensembles}, we use the disagreement of the heads in Eq.~\eqref{eq:predictor}, normalized by the frozen latent scale so that it is dimensionless:
\begin{equation}
  \sigma_{t+1} = \frac{1}{\sqrt{d}\,\bar{\gamma}^\star}\left(\frac{1}{N_e}\sum_{i=1}^{N_e}\norm{\hat{z}^{(i)}_{t+1} - \hat{z}_{t+1}}_2^2\right)^{1/2}.
  \label{eq:sigma}
\end{equation}
An asymmetric filter reacts immediately to increases and decays slowly,
\begin{equation}
  \bar{\sigma}_{t+1} = \max\big\{\sigma_{t+1},\; (1 - \alpha_\sigma)\,\bar{\sigma}_t + \alpha_\sigma\,\sigma_{t+1}\big\}, \qquad \alpha_\sigma \in (0, 1],
  \label{eq:filter}
\end{equation}
which prevents rapid fluctuations of $\sigma_{t+1}$ from modulating the command at the decision rate. The filtered score governs three mechanisms. First, the translational components of the chunk are attenuated by the monotone, piecewise-linear law
\begin{equation}
  \alpha(\bar{\sigma}) =
  \begin{cases}
    1, & \bar{\sigma} \le \sigma_{\mathrm{lo}},\\[2pt]
    1 - (1 - \alpha_{\min})\dfrac{\bar{\sigma} - \sigma_{\mathrm{lo}}}{\sigma_{\mathrm{crit}} - \sigma_{\mathrm{lo}}}, & \sigma_{\mathrm{lo}} < \bar{\sigma} < \sigma_{\mathrm{crit}},
  \end{cases}
  \label{eq:alpha}
\end{equation}
with $0 < \alpha_{\min} \le 1$ and $0 \le \sigma_{\mathrm{lo}} < \sigma_{\mathrm{crit}}$; $\alpha$ is non-increasing and Lipschitz with constant $(1 - \alpha_{\min})/(\sigma_{\mathrm{crit}} - \sigma_{\mathrm{lo}})$, and the yaw rate is not attenuated so that the vehicle can keep reorienting its camera. Second, each chunk carries the validity horizon
\begin{equation}
  H_t = \max\Big(H_{\min},\; \big\lfloor H_{\max}\exp(-\beta\,\bar{\sigma}_{t+1})\big\rfloor\Big), \qquad 1 \le H_{\min} \le H_{\max} = H,
  \label{eq:horizon}
\end{equation}
with sensitivity $\beta > 0$: the executor (Section~\ref{sec:algorithm}) may use the chunk only while its age satisfies $\tau - \tau_t \le H_t\Delta$. In the synchronous execution of~\cite{lee2026vpa}, $H_t$ is the number of knots executed before replanning; in the asynchronous design adopted here, newer chunks preempt older ones and $H_t$ bounds how stale a chunk may become, for instance when inference stalls, before the fallback is engaged. Third, if $\bar{\sigma}_{t+1} \ge \sigma_{\mathrm{crit}}$, the primitive is overridden, $u_t \leftarrow \textsc{BrakeHover}$, and the executor performs the braking manoeuvre of Section~\ref{sec:envelope}.

The thresholds are calibrated on in-distribution validation flights: $\sigma_{\mathrm{lo}}$ is the empirical 0.9-quantile of $\bar{\sigma}$, and $\sigma_{\mathrm{crit}}$ is its $(1 - p_{\mathrm{fa}})$-quantile, with $p_{\mathrm{fa}}$ chosen such that the expected number of spurious fallbacks per mission, $p_{\mathrm{fa}} f_{\mathrm{dec}} T_{\mathrm{mis}}$, does not exceed a target $n_{\mathrm{fa}}$, where $f_{\mathrm{dec}}$ is the decision rate and $T_{\mathrm{mis}}$ the mission duration. For $f_{\mathrm{dec}} = 30$\,Hz, $T_{\mathrm{mis}} = 60$\,s, and $n_{\mathrm{fa}} = 0.05$, this gives $p_{\mathrm{fa}} \approx 2.8 \times 10^{-5}$, which requires a large validation set because consecutive values of $\bar{\sigma}$ are strongly correlated. The trade-off between safety and conservatism that these choices induce is mapped by ablation A2. The evaluated system calibrates $\sigma_{\mathrm{crit}}$ differently, from per-flight maxima (the episode-maximum method): with $\alpha_\sigma = 0.1$ and $\sigma_{\mathrm{lo}}$ the 0.9-quantile of $\bar{\sigma}$ on 360 validation flights, $\sigma_{\mathrm{crit}}$ is calibrated per checkpoint from the maxima of $\bar{\sigma}$ of the individual flights with a Poisson model, for about 0.05 spurious fallbacks per 60\,s of flight, and the attenuation factor stays at or above 0.5.

\subsection{Braking Envelope and Hover Fallback}
\label{sec:envelope}
Ensemble disagreement is a heuristic: it can remain low on inputs that the frozen encoder maps into familiar regions of $\mathcal{Z}$. We therefore add a model-free bound derived from the depth channel. Let $d_t$ be the smallest range measured in an image region $\mathcal{R}_t$ whose back-projection covers the volume swept by a sphere of radius $r_s \ge r_v$ moving along the commanded direction $\hat{u}_t = v_{t,0}/\norm{v_{t,0}}_2$ out to the sensor range, where $v_{t,k}$ denotes the translational part of $a_{t,k}$. If $\hat{u}_t$ lies outside the field of view, or if the fraction of invalid depth pixels in $\mathcal{R}_t$ exceeds a threshold, the speed is limited to a constant $v_{\mathrm{blind}}$.

\paragraph{Braking profile.} When the fallback is engaged at speed $v_0$ (taken here with zero initial acceleration), the executor commands the velocity profile that minimizes the integrated squared jerk $\int_0^{T_b}\norm{\ddot{v}(\tau)}_2^2\,\mathrm{d}\tau$ subject to $v(0) = v_0$, $\dot{v}(0) = 0$, $v(T_b) = 0$, and $\dot{v}(T_b) = 0$, with the stopping position left free. The Euler--Lagrange condition $v^{(4)} = 0$ makes the minimizer cubic, and the boundary conditions give
\begin{equation}
  v_{\mathrm{brk}}(\tau) = v_0\left(1 - 3\Big(\frac{\tau}{T_b}\Big)^2 + 2\Big(\frac{\tau}{T_b}\Big)^3\right), \qquad T_b = \frac{3\norm{v_0}_2}{2a_{\max}},
  \label{eq:brake}
\end{equation}
whose peak deceleration, attained at $\tau = T_b/2$, equals $1.5\norm{v_0}_2/T_b = a_{\max}$ and whose stopping distance is $\int_0^{T_b}\norm{v_{\mathrm{brk}}}_2\,\mathrm{d}\tau = \norm{v_0}_2 T_b/2 = 3\norm{v_0}_2^2/(4a_{\max})$. After the stop, the executor holds the stopping point with a position setpoint, and yaw scanning is permitted so that new observations can lower $\bar{\sigma}$ and release the vehicle.

\paragraph{Speed envelope.} Let $T_r$ upper-bound the time from the capture of $I_t$ to the onset of braking if braking becomes necessary; it comprises the worst-case end-to-end latency $T_{\mathrm{e2e}}^{\max}$, the chunk validity $H_t\Delta$ of Eq.~\eqref{eq:horizon}, and the velocity-tracking lag $T_{\mathrm{trk}}$ of the flight controller. With depth error bound $\epsilon_d$, define
\begin{equation}
  v_{\mathrm{safe}}(d) =
  \begin{cases}
    \dfrac{2a_{\max}}{3}\left(-T_r + \sqrt{T_r^2 + \dfrac{3(d - \epsilon_d - r_s)}{a_{\max}}}\right), & d > \epsilon_d + r_s,\\[8pt]
    0, & \text{otherwise.}
  \end{cases}
  \label{eq:vsafe}
\end{equation}
The commanded translational velocity of every knot is scaled by the common factor
\begin{equation}
  s_t = \min\left\{\alpha(\bar{\sigma}_{t+1}),\; \frac{v_{\mathrm{safe}}(d_t)}{\max\big(\max_k \norm{v_{t,k}}_2,\, \epsilon_v\big)}\right\}, \qquad a^{\mathrm{cmd}}_{t,k} = \big(s_t\, v_{t,k},\; \dot{\psi}_{t,k}\big),
  \label{eq:scaling}
\end{equation}
where $\epsilon_v > 0$ guards the division; scaling preserves the direction of every knot.

\begin{assumption}[Braking envelope]
\label{ass:envelope}
Over the interval from $\tau_t$ to the end of a possible braking manoeuvre: (i) the obstacles intersecting the swept volume are static; (ii) every range measured in $\mathcal{R}_t$ exceeds the true range by at most $\epsilon_d$; (iii) the vehicle moves along $\hat{u}_t$ with speed at most $v_{\mathrm{safe}}(d_t)$ until braking begins, and braking begins within $T_r$ of $\tau_t$ with zero acceleration; (iv) the flight controller tracks Eq.~\eqref{eq:brake}, which requires a deceleration of at most $a_{\max}$.
\end{assumption}

\begin{proposition}[Stop within measured free space]
\label{prop:envelope}
Under Assumption~\ref{ass:envelope}, the vehicle comes to rest without its center entering the $r_s$-neighbourhood of any obstacle point imaged in $\mathcal{R}_t$.
\end{proposition}
\begin{proof}
Let $v \le v_{\mathrm{safe}}(d_t)$ be the speed during the reaction interval. The distance travelled before braking is at most $vT_r$ and the braking distance is $3v^2/(4a_{\max})$ by Eq.~\eqref{eq:brake}. For $d_t > \epsilon_d + r_s$, $v_{\mathrm{safe}}(d_t)$ is the positive root of $\frac{3}{4a_{\max}}v^2 + T_r v - (d_t - \epsilon_d - r_s) = 0$, and the left-hand side is increasing in $v \ge 0$, so $vT_r + 3v^2/(4a_{\max}) \le d_t - \epsilon_d - r_s$. By (ii), the true range of every obstacle point in $\mathcal{R}_t$ is at least $d_t - \epsilon_d$; the total travel therefore ends at least $r_s$ short of it. For $d_t \le \epsilon_d + r_s$, $v_{\mathrm{safe}} = 0$ and the vehicle is commanded to hold.
\end{proof}

\begin{remark}
The guarantee is conditional. It does not cover dynamic obstacles, which motivate fast replanning and the uncertainty signal, nor obstacles outside the field of view, which motivate $v_{\mathrm{blind}}$. If braking starts with acceleration $a_0$ along $\hat{u}_t$, the cubic profile matching $\dot{v}(0) = a_0$ stops after $\norm{v_0}_2 T_b/2 + a_0 T_b^2/12$, and the excess can be absorbed into $r_s$. For illustration, with $a_{\max} = 5$\,m/s$^2$, $T_r = 0.4$\,s, and $d_t - \epsilon_d - r_s = 6$\,m, Eq.~\eqref{eq:vsafe} gives $v_{\mathrm{safe}} \approx 5.1$\,m/s. The envelope is a simplified braking-feasibility constraint in the spirit of control barrier functions~\cite{ames2019cbf}.
\end{remark}

\begin{remark}[Evaluated configuration]
\label{rem:envelope-evaluated}
The evaluated system uses $r_s = 0.5$\,m, $a_{\max} = 5$\,m/s$^2$, $\epsilon_d = 0.2$\,m, $v_{\mathrm{blind}} = 1$\,m/s and $T_r = 0.06 + 0.02\,H_t + 0.15$\,s, which is at most 0.53\,s. This reaction time is shorter than the response identified in simulation, a dead time of 0.24\,s followed by a first-order lag of 0.5\,s (Section~\ref{sec:implementation}), so Assumption~\ref{ass:envelope}(iii)--(iv) does not hold for the evaluated system, Proposition~\ref{prop:envelope} does not apply, and we treat the envelope as a heuristic. The envelope also checks only the current frame; obstacles that have left the field of view are handled by the runtime shield of Section~\ref{sec:shield}.
\end{remark}

\subsection{Receding-Horizon Control Algorithm}
\label{sec:algorithm}
Algorithm~\ref{alg:aerovpa} summarizes the two loops of Aero-VPA. The inference loop runs once per camera frame whenever the previous decision has finished; the executor runs at a fixed rate $f_c = 100$\,Hz and evaluates the newest valid chunk at its chunk-local time $\tau - \tau_t$. Because the chunk is indexed by elapsed time since capture rather than by arrival, the knots that correspond to the inference and transport latency are skipped automatically, an approach related to the frozen prefix of real-time chunking~\cite{black2025rtc}.

The sequential structure of a decision step is that of~\cite{lee2026vpa}: the chain $I_t \to z_t \to u_t \to (\hat{z}_{t+1}, \sigma_{t+1}) \to e_t \to A^{(\rho_1)} \to \dots \to A^{(\rho_K)}$ contains one evaluation each of $E_\psi$, $\pi^h_\phi$, and $P_\omega$ (whose heads are independent and can be batched) and $K$ evaluations of $v_\theta$. The sequential depth is therefore $D_{\mathrm{VPA}} = K + 3 = 5$ for $K = 2$, independent of $H$ and $d_a$, and the network time of a decision step is
\begin{equation}
  T_{\mathrm{VPA}} = t_E + t_h + t_P + K\, t_v(H),
  \label{eq:tvpa}
\end{equation}
where $t_E$, $t_h$, $t_P$, and $t_v(H)$ are the wall-clock times of the encoder, selector, predictor, and one solver evaluation. The arithmetic work of $t_v(H)$ grows with $H d_a$; approximate independence of $H$ holds only while one solver evaluation fits within the available parallelism, which on an Apple M4 was the case up to $H = 64$~\cite{lee2026vpa} and must be re-measured on the target hardware (Section~\ref{sec:hitl}). The safety computations of Eqs.~\eqref{eq:sigma}--\eqref{eq:scaling} add only vector and image-region reductions and no network evaluation. In version~2, the runtime shield of Section~\ref{sec:shield} filters every setpoint of the executor before it is sent to PX4, and the setpoint it dispatches is the $\bar{a}_t$ of the next decision.


\begin{algorithm}[t]
\caption{Aero-VPA asynchronous receding-horizon control with safety filter}
\label{alg:aerovpa}
\begin{algorithmic}[1]
\Require frozen $E_\psi$, $P_\omega$, $\bar{\gamma}^\star$, $\bar{\mu}_Z$; trained $\pi^h_\phi$, $v_\theta$; mission $G$; $c_{\mathrm{text}}$ (computed once); $K = 2$
\Statex \textit{Inference loop (event-driven, one decision per camera frame)}
\Procedure{Decide}{$I_t$, $S_t$, $\tau_t$}
  \State $z_t \gets E_\psi(I_t)$; update $m_t$ by Eq.~\eqref{eq:waypoint}; compute $g_t$ by Eq.~\eqref{eq:goal}
  \State $u_t \gets \argmax_u \pi^h_\phi(u \mid z_t, g_t, S_t, c_{\mathrm{text}})$ \Comment{Eq.~\eqref{eq:selector}}
  \State $\{\hat{z}^{(i)}_{t+1}\}_{i=1}^{N_e} \gets P_\omega(z_t, \operatorname{Embed}(u_t), \bar{a}_t)$; $\hat{z}_{t+1} \gets$ ensemble mean \Comment{Eq.~\eqref{eq:predictor}}
  \State compute $\sigma_{t+1}$ and $\bar{\sigma}_{t+1}$ \Comment{Eqs.~\eqref{eq:sigma}, \eqref{eq:filter}}
  \If{$\bar{\sigma}_{t+1} \ge \sigma_{\mathrm{crit}}$}
    \State $u_t \gets \textsc{BrakeHover}$; \Call{RequestFallback}{} \textbf{and return}
  \EndIf
  \State update $\xi_t$ by Eq.~\eqref{eq:noise}; form $e_t$ by Eq.~\eqref{eq:cond}
  \State $A \gets \xi_t$
  \For{$k = 0, \dots, K-1$} \Comment{Eq.~\eqref{eq:euler}}
    \State $A \gets A + \frac{1}{K}\, v_\theta(A, k/K \mid e_t)$
  \EndFor
  \State $\hat{A}_t \gets \Pi_{\mathcal{A}_0}(\text{denormalize}(A))$
  \State $d_t \gets$ minimum valid range in $\mathcal{R}_t$; compute $s_t$ by Eq.~\eqref{eq:scaling}; $H_t$ by Eq.~\eqref{eq:horizon}
  \State publish chunk $\mathcal{C}_t = (\tau_t,\, \hat{\psi}_t,\, \{(s_t v_{t,k}, \dot{\psi}_{t,k})\}_{k=0}^{H-1},\, H_t)$ to the buffer
\EndProcedure
\Statex \textit{Executor (fixed rate $f_c = 100$\,Hz)}
\Procedure{Tick}{$\tau$}
  \State $\mathcal{C} \gets$ newest chunk in buffer; $\Delta\tau \gets \tau - \tau_{\mathcal{C}}$
  \If{fallback requested \textbf{or} no chunk \textbf{or} $\Delta\tau > H_{\mathcal{C}}\Delta$ \textbf{or} odometry stale}
    \State $(v^{\mathcal{W}}_{\mathrm{ref}}, \dot{v}^{\mathcal{W}}_{\mathrm{ref}}) \gets$ braking profile Eq.~\eqref{eq:brake}, then hold position
  \Else
    \State interpolate knots at $\Delta\tau$ (Eq.~\eqref{eq:interp}); blend with previous chunk (Eq.~\eqref{eq:blend}); rate-limit (Eq.~\eqref{eq:ratelimit})
    \State rotate to NED: $v^{\mathcal{W}}_{\mathrm{ref}} \gets \mathbf{R}_z(\hat{\psi}_{\mathcal{C}})\, v^{\mathcal{H}}_{\mathrm{ref}}$ \Comment{Eq.~\eqref{eq:rotation}}
  \EndIf
  \State pass the setpoint through the runtime shield, which filters it or replaces it by its own (Algorithm~\ref{alg:shield})
  \State publish \texttt{OffboardControlMode} and \texttt{TrajectorySetpoint}$(v^{\mathcal{W}}_{\mathrm{ref}}, \dot{v}^{\mathcal{W}}_{\mathrm{ref}}, \dot{\psi}_{\mathrm{ref}})$
  \State $\bar{a} \gets$ dispatched setpoint, stored for the next decision
\EndProcedure
\end{algorithmic}
\end{algorithm}

\subsection{Training Procedure and DAgger Supervisor}
\label{sec:training}
Training follows the staged procedure of~\cite{lee2026vpa}, adapted to flight. (i) \emph{Representation learning:} $E_\psi$ and $P_\omega$ are trained with Eq.~\eqref{eq:jepa} on flight logs, then frozen, fixing $\bar{\gamma}^\star$ and $\bar{\mu}_Z$. (ii) \emph{Policy learning:} with latents cached, the selector and solver minimize
\begin{equation}
  \mathcal{L}_{\mathrm{pol}}(\phi, \theta) = \mathcal{L}_{\mathrm{CFM}}(\theta) + \lambda_h\,\mathcal{L}_{\mathrm{CE}}(\phi),
  \label{eq:policy}
\end{equation}
with weight $\lambda_h > 0$, on expert chunks expressed in the frozen heading frame of their capture time. (iii) \emph{Dataset aggregation:} because imitation from expert states alone suffers from covariate shift, Aero-VPA is rolled out in simulation under a mixture of its own and the expert's commands, the privileged expert relabels the visited states, and stage (ii), optionally also stage (i), is repeated on the aggregated data, following DAgger~\cite{ross2011dagger}. The expert is a planner with access to the ground-truth occupied set $\mathcal{X}$, in the manner of~\cite{loquercio2021highspeed}.

\paragraph{Evaluated training runs.} In the evaluated system, one stage-(ii) run took 153,528 steps at batch size 1024, about 2\,h\,40\,min on one NVIDIA L4 GPU. Inputs that a recipe withholds (Table~\ref{tab:names}) enter $\pi^h_\phi$ and $e_t$ at their training mean, while $P_\omega$ always receives $\bar{a}_t$. Stage~(iii) was realized with the supervisor described next.

\paragraph{Expert and demonstrations.} The expert plans with A* on the ground-truth occupancy (1\,m inflation) and a minimum-snap trajectory~\cite{mellinger2011minsnap}. Demonstrations come from 200 training worlds (seeds 0--199), with the speed limit of each mission drawn from $\{3, 5, 7\}$\,m/s: about 5.5 million RGB-D frames. Another 360 expert flights on calibration worlds (seeds 1000--1011) serve validation and the calibration of $\sigma_{\mathrm{crit}}$. The label of knot $k$ follows the expert's tracking law from the path point $\tau^\star$ nearest the vehicle,
\begin{equation}
a^\star_{k} = w(e)\; v^\star(\tau^\star + k\Delta) + k_p\, e_p\, e^{-k_p k\Delta}, \qquad k_p = 1.5\,\mathrm{s}^{-1},
\label{eq:label}
\end{equation}
where $v^\star$ is the velocity of the expert's reference trajectory, $e_p$ the vector from the vehicle to the path point and $e = \norm{e_p}$; on the path ($e \le 0.3$\,m) $w = 1$, further off $w \to 0$ and the horizontal speed limit falls to 2\,m/s at $e \ge 1$\,m, so that the expert rejoins its path first.

\paragraph{Gate and takeover.} The policy flies until one of six triggers fires: an obstacle within $0.5$\,m $+$ $0.06$\,m per m/s of speed of the straight look-ahead along the current velocity (0.3\,s of reaction plus the braking distance, at least 1.5\,m, sampled every 0.1\,m); an altitude below 0.8\,m; more than 1.5\,m off the path; a heading more than 75\textdegree{} from the path; a missed waypoint; or 5\,s without progress. The expert then takes over. In a soft takeover, which keeps recording frames, it chooses between flying on, slowing down along its path (a time-scaled reference) and a straight stop by the clearance predicted in a rollout of the delayed closed loop (0.24\,s dead time, 0.5\,s first-order lag). A hard takeover brakes, backs off and re-plans. Control returns to the policy after 2\,s on track. Appendix~\ref{app:supervisor} gives the details.

\paragraph{Rounds and recipes.} Each round flies 201 missions over the training worlds, and the expert labels every visited state. In stage~(ii), DAgger frames are weighted $\times3$ relative to demonstrations. Round~1 was flown by the behaviour-cloning policy \bcVtwo, round~2 by \ckRone and round~3 by \ckRtwo (Table~\ref{tab:rounds}). Four round-3 recipes target the copycat shortcut of Section~\ref{sec:copycat} (Table~\ref{tab:names}): \ckNoact withholds $\bar{a}_t$ from $\pi^h_\phi$ and $e_t$; \ckNomotion also withholds $v^{\mathcal{H}}_t$; \ckSd replaces $v^{\mathcal{H}}_t$ and $\bar{a}_t$ by their training mean in 50\% of training frames (state dropout); and \ckKw weights each frame in proportion to $\min\big(1 + \norm{A_t[0] - \bar{a}_t}/0.1,\, 10\big)$, the error of repeating the last command (keyframe weighting), with DAgger weight $\times1$.

\subsection{Runtime Safety Shield}
\label{sec:shield}
The runtime shield, added in version~2 after the evaluation of Section~\ref{sec:evaluation} exposed the limits of the learned safety layer, sits between the policy's executor and the flight controller (Figure~\ref{fig:system}). It is deterministic, has no learned parameters and uses no map or privileged information: its inputs are the depth image, the odometry estimate (position $\hat{p}$, velocity $\hat{v}$, yaw $\hat\psi$, attitude $\hat{R}$), the active waypoint and the decision timestamps. It runs at every executor tick.

\begin{figure}[t]
\centering
\begin{tikzpicture}[
  font=\footnotesize,
  st/.style={draw, rounded corners=4pt, thick, minimum height=10mm, align=center, fill=white, inner sep=3pt},
  arr/.style={-{Stealth[length=1.8mm]}, semithick},
  lbl/.style={font=\scriptsize, align=center, inner sep=1.5pt}
]
\node[st, fill=blue!6, text width=40mm]   (fly)  at (0,0)      {\textsc{Fly}\\[1pt]{\scriptsize policy flies; every setpoint is filtered (altitude floor $\to$ barrier $\to$ blind cap) and written back as $\bar{a}_t$}};
\node[st, fill=orange!8, text width=27mm] (back) at (6.7,2.1)  {\textsc{Backoff}\\[1pt]{\scriptsize retrace the breadcrumbs at 0.3\,m/s}};
\node[st, fill=orange!8, text width=27mm] (nud)  at (6.7,-2.1) {\textsc{Nudge}\\[1pt]{\scriptsize face the waypoint, then 0.6\,m/s for 1\,s}};
\node[st, fill=orange!8, text width=17mm] (hold) at (-5.3,0)   {\textsc{Hold}\\[1pt]{\scriptsize zero velocity}};
\draw[arr] ($(fly.north east)+(-0.2,0)$) -- node[lbl, above, sloped, pos=0.45] {nearest remembered point $< 0.45$\,m} (back.south west);
\draw[arr] (back) -- node[lbl, right, text width=27mm, align=left] {retraced 1\,m, clearance $\ge r + 0.4$\,m, or blocked ($\norm{u} < 0.075$\,m/s)} (nud);
\draw[arr] ($(fly.east)+(0,-0.25)$) to[bend left=6] node[lbl, above, sloped, pos=0.58] {5\,s without 0.3\,m of progress} (nud.north west);
\draw[arr] (nud.west) to[bend left=18] node[lbl, below, sloped] {nudge over} ($(fly.south)+(0.6,0)$);
\draw[arr] (fly.west) to[bend right=22] node[lbl, above] {no decision for 0.3\,s} (hold.east);
\draw[arr] (hold.south east) to[bend right=22] node[lbl, below] {next decision} ($(fly.south west)+(0.2,0)$);
\end{tikzpicture}
\caption{Mode automaton of the runtime shield. In \textsc{Fly} the policy's executor runs and the shield filters each setpoint. In \textsc{Backoff}, \textsc{Nudge} and \textsc{Hold} (orange) the executor is paused, and the shield's own setpoint, passed through the same filters, is dispatched and fed back as $\bar{a}_t$. The radius $r = 0.55$\,m is the barrier's zero; a backoff that ends without an active waypoint returns to \textsc{Fly}.}
\label{fig:automaton}
\end{figure}
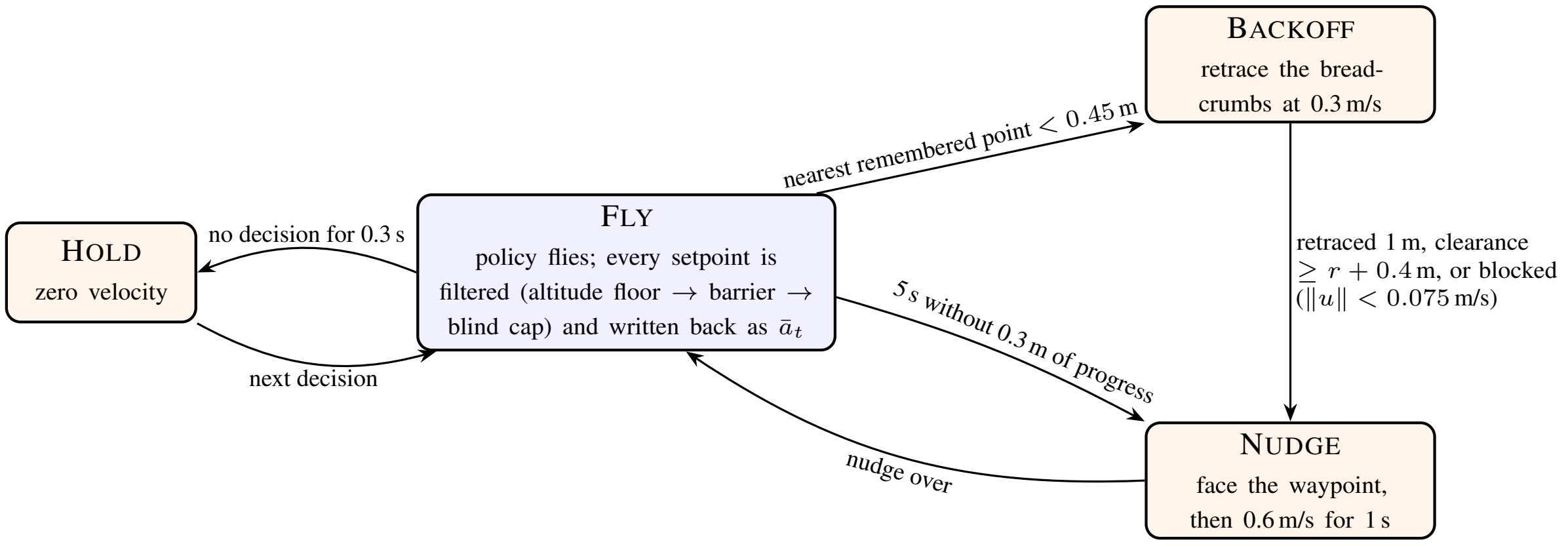

\begin{algorithm}[t]
\caption{One step of the runtime shield (every executor tick, 100\,Hz). The memory $\mathcal{M}$ and the local set $\mathcal{N}$ are updated per frame (Eqs.~\eqref{eq:backproject}--\eqref{eq:local}).}
\label{alg:shield}
\begin{algorithmic}[1]
\Require estimate $(\hat{p}, \hat{v}, \hat\psi)$; active waypoint $w$; time $t$; time $t_f$ of the last frame; breadcrumbs $\mathcal{B}$
\State add $\hat{p}$ to $\mathcal{B}$ if it is $\ge 0.25$\,m from the last breadcrumb (not while backing off); update the progress toward $w$
\State $c \gets \min_{x \in \mathcal{N}} \norm{x - \hat{p}}$ \Comment{clearance to the memory}
\If{mode $=$ \textsc{Fly} \textbf{and} $c < r_{\mathrm{hard}}$}
  \State mode $\gets$ \textsc{Backoff} \Comment{retrace at most 1\,m}
\ElsIf{mode $=$ \textsc{Fly} \textbf{and} no 0.3\,m of progress for 5\,s}
  \State mode $\gets$ \textsc{Nudge}
\EndIf
\If{mode $=$ \textsc{Backoff}}
  \State $u \gets \Call{Filter}{0.3\,\mathrm{m/s}\text{ toward the previous breadcrumb}}$
  \If{retraced 1\,m \textbf{or} $c \ge r + 0.4$\,m \textbf{or} $\norm{u} < 0.075$\,m/s}
    \State mode $\gets$ \textsc{Nudge}
  \Else
    \State \Return $u$ as an external setpoint \Comment{becomes $\bar{a}_t$}
  \EndIf
\EndIf
\If{mode $=$ \textsc{Nudge}}
  \If{the nudge is over}
    \State mode $\gets$ \textsc{Fly}
  \Else
    \State $\dot\psi \gets \clip(1.5\,\Delta\psi, \pm0.8)$; $u \gets 0$ while $|\Delta\psi| > 36.5^\circ$, else $0.6\cos(\Delta\psi)$\,m/s toward $w$ for 1\,s
    \State \Return $(\Call{Filter}{u}, \dot\psi)$ as an external setpoint \Comment{becomes $\bar{a}_t$}
  \EndIf
\EndIf
\If{$t - t_f > 0.3$\,s}
  \State mode $\gets$ \textsc{Hold}; \Return zero velocity as an external setpoint
\EndIf
\State mode $\gets$ \textsc{Fly}; $(u, a_{\mathrm{ff}}, \dot\psi) \gets$ one executor tick of the policy
\State $(u, a_{\mathrm{ff}}) \gets \Call{Filter}{u, a_{\mathrm{ff}}}$; if it changed, write $(u, \dot\psi)$ back as $\bar{a}_t$
\State \Return $(u, a_{\mathrm{ff}}, \dot\psi)$
\Statex
\Function{Filter}{$u, a_{\mathrm{ff}}$}
  \State altitude floor; barrier: two sweeps of Eq.~\eqref{eq:projection} over the 24 points of $\mathcal{N}$ nearest to $q = \hat{p} + 0.4\,\hat{v}$; blind cap, Eq.~\eqref{eq:blind}
  \State \Return $(u, a_{\mathrm{ff}})$
\EndFunction
\end{algorithmic}
\end{algorithm}

\paragraph{Depth memory.} For each new frame, the shield takes the nearest valid return $z_b \in [0.2, 8]$\,m of every $4\times4$ pixel block $b$ and back-projects it,
\begin{equation}
x_b = \hat{p}_t + \hat{R}_t\,(z_b\, r_b),
\label{eq:backproject}
\end{equation}
where $r_b$ is the ray through the block centre in the body frame, scaled to unit forward component (depth is measured along the optical axis). Points less than 0.25\,m above the ground are dropped. The memory $\mathcal{M}_t$ is the union of all points no older than $T_M = 15$\,s and within 10\,m horizontally, voxelised at 0.15\,m with the newest point kept per voxel, so that obstacles beside and behind the camera remain known after they leave the field of view. The barrier uses the local set
\begin{equation}
\mathcal{N}_t = \{x \in \mathcal{M}_t : \norm{x - \hat{p}_t} \le 3.5\,\mathrm{m}\}.
\label{eq:local}
\end{equation}

\paragraph{Barrier.} Let $u \in \R^3$ be the commanded velocity, $a_{\mathrm{ff}}$ its acceleration feed-forward, and $q = \hat{p} + \hat{v}\,T_{\mathrm{la}}$ the position $T_{\mathrm{la}} = 0.4$\,s ahead at the flown velocity, a horizon that covers the 0.24\,s dead time. For each of the 24 points $x_i \in \mathcal{N}_t$ nearest to $q$, let $n_i = (x_i - q)/\norm{x_i - q}$ and $h_i = \norm{x_i - q} - r$ with $r = 0.55$\,m. The shield enforces
\begin{equation}
n_i^\top u \le \beta(h_i), \qquad
\beta(h) = \begin{cases} \min\big(k h,\; \sqrt{2 a_b h}\big), & h > 0,\\[2pt] \max\big(k h,\; -v_{\mathrm{push}}\big), & h \le 0, \end{cases}
\label{eq:barrier}
\end{equation}
with $k = 1.5$\,s$^{-1}$, $a_b = 3$\,m/s$^2$ and $v_{\mathrm{push}} = 0.5$\,m/s, by two sequential projection sweeps over the 24 points: whenever $n_i^\top u > \beta(h_i)$,
\begin{equation}
u \leftarrow u - \big(n_i^\top u - \beta(h_i)\big)\, n_i, \qquad a_{\mathrm{ff}} \leftarrow a_{\mathrm{ff}} - \max\big(0,\, n_i^\top a_{\mathrm{ff}}\big)\, n_i,
\label{eq:projection}
\end{equation}
which removes the approaching components of the velocity and of the feed-forward and keeps tangential motion past the obstacle. With $h(p) = \norm{x - p} - r$ and $\dot{p} = u$, the constraint reads $\dot{h} \ge -\beta(h)$, a control-barrier-function condition~\cite{ames2019cbf}; $\beta$ is continuous and non-decreasing with $\beta(0) = 0$. The square-root term caps the approach speed at the speed from which a deceleration $a_b$ stops within the distance $h$; the linear term is the smaller one for $h < 2a_b/k^2 \approx 2.7$\,m; and for $h \le 0$ the constraint pushes the vehicle away at up to $v_{\mathrm{push}}$. The projection is sequential, not a quadratic program: it is neither optimal nor guaranteed to satisfy all constraints jointly.

\begin{proposition}[Invariance under exact tracking]
\label{prop:barrier}
Let $x \in \R^3$ be a static point, and on an interval $[0, T]$ let the vehicle position $p$ be absolutely continuous with $p(t) \neq x$ and $\dot{p}(t) = u(t)$, where $u(t)$ satisfies $n(t)^\top u(t) \le \beta(h(t))$ for $n = (x - p)/\norm{x - p}$ and $h = \norm{x - p} - r$ (Eq.~\eqref{eq:barrier} evaluated at $q = p$). Then (i) if $h(0) \ge 0$, $h(t) \ge h(0)\,e^{-kt} \ge 0$ for all $t \in [0, T]$; and (ii) if $h(0) < 0$, $h$ is strictly increasing while it is negative, with $\dot{h} \ge \min(k\,|h|, v_{\mathrm{push}})$ almost everywhere.
\end{proposition}
\begin{proof}
Wherever $p \ne x$, $\tfrac{d}{dt}\norm{x - p} = -n^\top \dot{p}$, so almost everywhere $\dot{h} = -n^\top u \ge -\beta(h)$. For $h \ge 0$, $\beta(h) \le kh$, hence $\dot{h} \ge -kh$; for $h < 0$, $\beta(h) = \max(kh, -v_{\mathrm{push}}) < 0$, hence $\dot{h} \ge \min(k|h|, v_{\mathrm{push}}) > 0$, which is (ii). For (i), $h$ cannot become negative: at a first time $t_0$ with $h(t_0) = 0$ after which $h < 0$, $h$ would have to decrease on an interval where (ii) makes it increase. On $[0, T]$ we therefore have $h \ge 0$ and $\tfrac{d}{dt}\big(h\, e^{kt}\big) = e^{kt}(\dot{h} + kh) \ge 0$ almost everywhere, so $h(t)\,e^{kt} \ge h(0)$.
\end{proof}

\begin{remark}
\label{rem:not-guaranteed}
Proposition~\ref{prop:barrier} concerns one remembered point, exact velocity tracking and the constraint evaluated at the current position. The implemented filter departs from it in four ways, so the shield carries no safety guarantee. (1) It evaluates the constraint at the predicted position $q$, a heuristic for the dead time, while the vehicle tracks $u$ with a 0.24\,s dead time and a 0.5\,s lag. (2) When several constraints are active, the sequential projection may leave some of them violated, and the blind cap (below), applied after the projection, scales the horizontal command without re-checking them. (3) Only remembered points are constrained: never-observed obstacles, points forgotten after 15\,s or beyond 10\,m, points outside the 3.5\,m local set or beyond the 24 nearest, and depth errors are not. (4) Obstacles are assumed static.
\end{remark}

\paragraph{Blind cap, altitude floor and hold.} After the barrier, the horizontal command is capped where the camera cannot see:
\begin{equation}
u_{xy} \leftarrow \frac{v_b}{\norm{u_{xy}}}\, u_{xy} \quad \text{if} \quad \norm{u_{xy}} > v_b \;\;\text{and}\;\; \big|\wrap\big(\atantwo(u_y, u_x) - \hat\psi\big)\big| > \tfrac{1}{2}\,\mathrm{HFOV},
\label{eq:blind}
\end{equation}
with $v_b = 0.7$\,m/s and $\tfrac{1}{2}\mathrm{HFOV} = 36.5$\textdegree, unless $u_{xy}$ retraces the flown path (it points within 35\textdegree{} of a breadcrumb 0.3--2.5\,m away); the feed-forward is scaled alike. Because the cap follows the barrier, a slide around a trunk that leaves the view is capped as well. Before the barrier, an altitude floor sets the upward speed to at least 0.5\,m/s below 0.8\,m of height and removes downward feed-forward. If no decision has arrived for 0.3\,s, the shield commands zero velocity until the next one. An uncertainty cap (speed limit while $\bar\sigma > \sigma_{\mathrm{lo}}$) exists but is disabled ($v_\sigma = \infty$): in the first shield version it held the vehicle at 1\,m/s for 18.0\,s per mission and turned flights into timeouts. The filter order is altitude floor, barrier, blind cap.

\paragraph{Recovery: backoff and nudge.} The shield drops a breadcrumb every 0.25\,m along the flown path, which is known free space. When the nearest remembered point is closer than $r_{\mathrm{hard}} = 0.45$\,m, it backs off: it retraces the breadcrumbs at 0.3\,m/s, with the barrier, the blind cap and the altitude floor still applied, until it has retraced 1\,m, the clearance to the memory reaches $r + 0.4$\,m, or the filtered command falls below 0.075\,m/s because the way back is blocked. It then nudges the vehicle toward the active waypoint: it yaws toward the waypoint at $\dot\psi = \clip(1.5\,\Delta\psi, \pm 0.8\,\mathrm{rad/s})$, where $\Delta\psi$ is the bearing of the waypoint relative to the heading, without translating while the waypoint is outside the field of view, and then moves toward it at $0.6\cos(\Delta\psi)$\,m/s for 1\,s, filtered like any other command. A nudge also follows a stall: 5\,s without 0.3\,m of progress toward the waypoint. The nudge ensures that a policy that copies its last command does not remain at rest after an intervention.

\paragraph{Coupling to the policy.} While the shield only filters (mode \textsc{Fly}), the filtered setpoint is dispatched and written back as $\bar{a}_t$, so the policy conditions on what was actually sent. While it flies itself (\textsc{Backoff}, \textsc{Nudge}, \textsc{Hold}), the executor is paused, the shield's setpoint is dispatched as an external setpoint, and it becomes $\bar{a}_t$. Figure~\ref{fig:automaton} shows the mode automaton and Algorithm~\ref{alg:shield} one shield step.


\section{Asynchronous System Integration and Edge Deployment}
\label{sec:system}

\subsection{Compute and Memory Budget on the Jetson Orin Nano}
\label{sec:budget}
The target module is the Jetson Orin Nano 8\,GB with a 1024-core Ampere GPU, 8\,GB of LPDDR5 shared by CPU and GPU, and 68\,GB/s of memory bandwidth in its original configuration (102\,GB/s in the JetPack~6.2 Super configurations)~\cite{nvidia2025jetpack}; we use the 15\,W power mode. Unlike the Orin NX and AGX Orin, the Orin Nano has no Deep Learning Accelerator, so all networks run on the GPU as TensorRT engines. Table~\ref{tab:budget} lists the configuration and its parameter count, computed analytically from the layer dimensions. Since every evaluation at batch size one reads the weights of the evaluated network, a decision step streams at least
\begin{equation}
  M_{\mathrm{dec}} = b\left(|W_E| + |W_h| + |W_P| + K|W_v|\right)
  \label{eq:membound}
\end{equation}
bytes, where $|W_\cdot|$ are the parameter counts of the four networks and $b = 2$ for FP16. For the configuration of Table~\ref{tab:budget}, $M_{\mathrm{dec}} = 43.4$\,MB, a lower bound of 0.64\,ms at 68\,GB/s, compared with 88\,ms for a single pass of the 3B-parameter decoder discussed in Section~\ref{sec:intro}. At this scale the latency is expected to be dominated by kernel launch overhead and arithmetic (about 1.0 giga multiply-accumulates per decision) rather than by weight streaming, which is why the latency targets of Section~\ref{sec:hypotheses} are hypotheses to be measured rather than consequences of Eq.~\eqref{eq:membound}. The released policy checkpoints are single files of 72.6--72.7\,MB, under 1\% of the module's 8\,GB of memory; their timing on the module has not been measured (Section~\ref{sec:implementation}).


\begin{table}[t]
\centering
\caption{Aero-VPA network configuration and parameter budget (analytic counts for $H = 16$, $d = 256$, $N_u = 6$; FP16 weights). The frozen text encoder is evaluated once per mission and lies outside the per-decision dependency chain.}
\label{tab:budget}
\small
\begin{tabular}{@{}l>{\raggedright\arraybackslash}p{5.4cm}rcr@{}}
\toprule
Module & Configuration & Params & Evals./decision & FP16 weights \\
\midrule
Encoder $E_\psi$ & ViT, $4{\times}128{\times}160$ input, patch 16, width 384, depth 6, 6 heads & 11.17M & 1 & 22.3\,MB \\
Selector $\pi^h_\phi$ & MLP $785\!\to\!512\!\to\!512\!\to\!6$ & 0.67M & 1 & 1.3\,MB \\
Predictor $P_\omega$ & 5 heads, MLP $292\!\to\!512\!\to\!512\!\to\!256$ & 2.72M & 1 (heads batched) & 5.4\,MB \\
Solver $v_\theta$ & Transformer, width 256, depth 4, 4 heads, $H{+}1 = 17$ tokens & 3.57M & $K = 2$ & 7.1\,MB \\
\midrule
Trainable total & & 18.13M & $K{+}3 = 5$ sequential & 36.3\,MB \\
Text encoder (frozen) & CLIP ViT-B/32 text tower & 63.4M & once per mission & -- \\
\bottomrule
\end{tabular}
\end{table}

\subsection{Hierarchical Software Stack}
\label{sec:stack}
The high level runs on the Jetson under ROS~2~\cite{macenski2022ros2}. A camera driver publishes time-stamped, hardware-synchronized RGB and aligned depth frames; a preprocessing node resizes and normalizes them on the GPU; the inference node executes $E_\psi$, $\pi^h_\phi$, $P_\omega$, and $v_\theta$ as TensorRT FP16 engines on a single CUDA stream with preallocated buffers, so that no memory is allocated on the critical path; and the executor node runs on a real-time-prioritized timer at $f_c = 100$\,Hz. The two loops communicate only through a lock-free single-slot chunk buffer. Each chunk carries the capture stamp $\tau_t$, which the executor uses for latency compensation, so the decision rate $f_{\mathrm{dec}} = \min\{f_{\mathrm{cam}}, 1/T_{\mathrm{inf}}\}$, with camera rate $f_{\mathrm{cam}}$ and inference time $T_{\mathrm{inf}}$, can vary without affecting the setpoint rate. A watchdog monitors chunk age, odometry age, and link status and requests the fallback when any of them exceeds its bound.

\subsection{PX4 Interface and Minimum-Acceleration Setpoint Interpolation}
\label{sec:px4}
The low level is a PX4 flight controller~\cite{meier2015px4} (e.g., an FMUv6X-class Pixhawk) connected to the Jetson through the uXRCE-DDS bridge, in which a client on PX4 exchanges uORB topics with an agent on the companion computer over a serial link (e.g., 921\,600\,baud) or a dedicated Ethernet link~\cite{px4uxrce}. In offboard mode, PX4 requires a continuous proof-of-life stream of at least 2\,Hz, provided by \texttt{OffboardControlMode}, and triggers its configured failsafe if the stream is interrupted for longer than \texttt{COM\_OF\_LOSS\_T} (default 1.0\,s)~\cite{px4offboard}; we publish both this message and \texttt{TrajectorySetpoint} at $f_c = 100$\,Hz, so that PX4's failsafe acts only as the last of three layers behind the executor fallback and the watchdog. \texttt{TrajectorySetpoint} is expressed in the local NED frame, with unused fields set to NaN, so the heading-frame velocities of a chunk are rotated with the yaw frozen at its capture time,
\begin{equation}
  v^{\mathcal{W}}_{\mathrm{ref}}(\tau) = \mathbf{R}_z(\hat{\psi}_t)\, v^{\mathcal{H}_t}_{\mathrm{ref}}(\tau), \qquad \dot{\psi}_{\mathrm{ref}}(\tau) = \dot{\psi}^{\mathcal{H}_t}_{\mathrm{ref}}(\tau),
  \label{eq:rotation}
\end{equation}
where the yaw rate maps directly onto the NED \texttt{yawspeed} field because the down axes of $\mathcal{H}_t$ and $\mathcal{W}$ coincide.

\paragraph{Interpolation.} The executor runs faster than the knot rate $1/\Delta$ and also supplies an acceleration feedforward, which PX4 expects to be kinematically consistent with the velocity setpoint. Among all absolutely continuous velocity references that pass through the knots, $v_{\mathrm{ref}}(k\Delta) = v_{t,k}$, the one minimizing the integrated squared acceleration $\int\norm{\dot{v}_{\mathrm{ref}}(\tau)}_2^2\,\mathrm{d}\tau$ satisfies $\ddot{v}_{\mathrm{ref}} = 0$ between knots by the Euler--Lagrange equation and is therefore piecewise linear:
\begin{equation}
  v_{\mathrm{ref}}(\tau) = v_{t,k} + \frac{\tau - k\Delta}{\Delta}\big(v_{t,k+1} - v_{t,k}\big), \qquad \dot{v}_{\mathrm{ref}}(\tau) = \frac{v_{t,k+1} - v_{t,k}}{\Delta}, \qquad \tau \in [k\Delta, (k+1)\Delta),
  \label{eq:interp}
\end{equation}
with $\tau$ measured from the capture time $\tau_t$; the yaw rate is interpolated in the same way. The acceleration feedforward is piecewise constant and exactly consistent with the velocity reference. To keep it feasible, consecutive knots are rate-limited before interpolation,
\begin{equation}
  v_{t,k+1} \leftarrow v_{t,k} + \clip_{a_{\max}\Delta}\big(v_{t,k+1} - v_{t,k}\big), \qquad \clip_{c}(y) = y\,\min\!\big(1, c/\norm{y}_2\big),
  \label{eq:ratelimit}
\end{equation}
which bounds $\norm{\dot{v}_{\mathrm{ref}}}_2$ by $a_{\max}$.

\paragraph{Chunk blending.} When a new chunk $\mathcal{C}_{t'}$ replaces $\mathcal{C}_t$ at time $\tau_s$, the executor cross-fades the two references over a window $T_{\mathrm{bl}}$,
\begin{equation}
  v_{\mathrm{ref}}(\tau) = \lambda(\tau)\, v^{(t')}_{\mathrm{ref}}(\tau) + \big(1 - \lambda(\tau)\big)\, v^{(t)}_{\mathrm{ref}}(\tau), \qquad \lambda(\tau) = \min\!\Big(1, \frac{\tau - \tau_s}{T_{\mathrm{bl}}}\Big),
  \label{eq:blend}
\end{equation}
where both references are expressed in $\mathcal{W}$ and evaluated at their own chunk-local times. Since $\lambda(\tau_s) = 0$ and both references are continuous, $v_{\mathrm{ref}}$ is continuous at $\tau_s$; if $\mathcal{C}_t$ expires within the window, its last knot is held. The acceleration feedforward of the blended reference is the corresponding derivative of Eq.~\eqref{eq:blend}. Inpainting-based real-time chunking~\cite{black2025rtc} is an alternative evaluated in ablation A8.

\paragraph{Link budget.} A \texttt{TrajectorySetpoint} carries 64 bytes of payload. Assuming at most 128 bytes per serialized message including framing, the setpoint stream and an odometry stream of equal rate occupy about 28\% of the 92\,160\,B/s capacity of a 921\,600-baud 8N1 link at 100\,Hz each; the dedicated Ethernet link removes this constraint.

\subsection{Latency Budget}
\label{sec:latency}
We decompose the end-to-end latency from photon capture to the setpoint received by PX4 as
\begin{equation}
  T_{\mathrm{e2e}} = t_{\mathrm{cap}} + t_{\mathrm{pre}} + \underbrace{t_E + t_h + t_P + K\,t_v(H)}_{T_{\mathrm{VPA}}} + t_{\mathrm{safe}} + t_{\mathrm{exec}} + t_{\mathrm{dds}},
  \label{eq:e2e}
\end{equation}
where $t_{\mathrm{cap}}$ is the exposure, readout, and driver delay, $t_{\mathrm{pre}}$ the preprocessing time, $t_{\mathrm{safe}}$ the time of the safety computations, $t_{\mathrm{exec}} \le 1/f_c$ the wait until the next executor tick, and $t_{\mathrm{dds}}$ the transport time to the PX4 client. We do not adopt a numerical value for $t_{\mathrm{dds}}$ from the literature; every term of Eq.~\eqref{eq:e2e} is measured in the HITL protocol (Section~\ref{sec:hitl}), and the measured worst case $T_{\mathrm{e2e}}^{\max}$ enters $T_r$ in Eq.~\eqref{eq:vsafe}.

\subsection{Cross-Embodiment Adaptation to Quadrupeds}
\label{sec:legged}
The architecture transfers to legged robots through the same separation of time scales. For a quadruped (Legged-VPA), the chunk consists of planar base twists, elements of $\mathfrak{se}(2)$,
\begin{equation}
  A^{\mathrm{leg}}_t = [v_x, v_y, \omega_z]_{t:t+H-1} \in \R^{H \times 3},
  \label{eq:legchunk}
\end{equation}
where $v_x, v_y$ are base-frame linear velocities and $\omega_z$ is the yaw rate. Because leg phase and contact state make single-step proprioception non-Markovian, the proprioceptive input is a history of $L = 30$ steps of joint positions $q_j \in \R^{12}$, joint velocities $\dot{q}_j \in \R^{12}$, and projected gravity $n^{\mathcal{B}} \in \R^3$,
\begin{equation}
  S^{\mathrm{leg}}_t = \big(q_{j,t-\ell},\, \dot{q}_{j,t-\ell},\, n^{\mathcal{B}}_{t-\ell}\big)_{\ell=0}^{L-1} \in \R^{L(12+12+3)} = \R^{810},
  \label{eq:legstate}
\end{equation}
which a linear layer projects before it enters $e_t$. The executor samples Eq.~\eqref{eq:interp} at 50\,Hz and passes the twist to a low-level locomotion policy trained by reinforcement learning, such as Walk These Ways~\cite{margolis2022wtw}, whose joint targets are tracked by a proportional-derivative controller at 200\,Hz. The safety layer carries over with $\alpha(\bar{\sigma})$ applied to $(v_x, v_y)$, the braking envelope recomputed with the deceleration limit of the gait, and \textsc{BrakeHover} replaced by a stop-and-stand primitive. Whether $E_\psi$ and $P_\omega$ can be shared across aerial and legged embodiments is an open question: it requires pretraining data from both, and camera height, ego-motion, and motion blur differ substantially between them. We treat sharing as a hypothesis, whereas $v_\theta$ and the executor are embodiment-specific by construction.

\subsection{Implementation Used in the Evaluation}
\label{sec:implementation}
The evaluation of Section~\ref{sec:evaluation} does not use the Jetson stack of Sections~\ref{sec:budget}--\ref{sec:px4}. The networks run in PyTorch on an NVIDIA L4 GPU of a cloud virtual machine, and one ROS~2 node per simulated vehicle runs the inference loop, the executor and the runtime shield against PX4 SITL over uXRCE-DDS~\cite{px4uxrce}. Each camera frame triggers one decision, a chain of $K + 3 = 5$ sequential network evaluations. A 100\,Hz executor evaluates the newest chunk with latency compensation and a cross-fade of $T_{\mathrm{bl}} = 0.04$\,s (Eq.~\eqref{eq:blend}) and sends offboard velocity setpoints with acceleration feed-forward; the shield filters every setpoint. Unlike the two-rate design of Section~\ref{sec:stack}, decisions and executor ticks share one single-threaded ROS~2 executor, so a slow decision delays the setpoints. With the shield, the 95th percentile of a decision's compute time was 32.0--33.1\,ms and that of its age from capture to commit 43.5--44.6\,ms (medians over missions), and the longest setpoint gap of a mission had a median of 0.10\,s and a maximum of 0.13--0.18\,s. These values were measured on the L4 and do not bear on Hypothesis~\ref{hyp:latency}, which concerns the Jetson. A fit of the simulated vehicle's horizontal response to its setpoints gave a dead time of 0.24\,s, a first-order lag of 0.5\,s and a feed-forward share of 1.0 (RMSE 0.273\,m/s).


\section{Experimental Design and Benchmark Protocol}
\label{sec:experiments}

\textbf{Status.} Version~1 of this preprint fixed this protocol before any experiment was executed: it specifies the environments, data, hardware, baselines, metrics, hypotheses, and statistical analysis, so that results can be judged against criteria fixed in advance. Version~2 executes a first part of it in software-in-the-loop simulation (Section~\ref{sec:evaluation}) and lists every deviation in Section~\ref{sec:deviations}; none of Hypotheses~\ref{hyp:latency}--\ref{hyp:coherence} has been tested yet. The rest of this section is unchanged from version~1 and remains the plan for the experiments not yet executed. Target values in Tables~\ref{tab:schema} and~\ref{tab:ablation} are acceptance criteria, not results.

\subsection{Simulation Setup}
\label{sec:sim}
\paragraph{Platforms.} The primary environment is PX4 SITL coupled to Gazebo~\cite{koenig2004gazebo} in its Harmonic long-term-support release (Gazebo Garden reached end of life in November 2024 and is not supported by current PX4 releases), using the PX4 quadrotor model with a front-facing depth camera (\texttt{gz\_x500\_depth}) augmented with an aligned RGB stream. For photorealistic tests of visual robustness, the same vehicle is simulated in NVIDIA Isaac Sim~\cite{nvidia2026isaacsim} with PX4 SITL attached through a bridge such as the Pegasus Simulator~\cite{jacinto2024pegasus}. A single PX4 release is pinned for all runs and recorded with the results.

\paragraph{Environment classes.} Worlds are generated procedurally from integer seeds. (E1) \emph{Dense forest}: trees placed by Poisson-disc sampling at three densities, with trunk diameters and canopy heights drawn from fixed distributions. (E2) \emph{Collapsed indoor structure}: rooms and corridors with rubble, tilted slabs, hanging cables, and openings whose minimum width ranges from 1.5 to 3 times the vehicle diameter, under low and non-uniform illumination. (E3) \emph{Dynamic clutter}: E1 and E2 with moving obstacles crossing the nominal path at speeds of 0.5--2\,m/s. Missions consist of 3--6 waypoints with straight-line spacing of 10--40\,m. Each episode runs at a commanded speed limit $\bar{v}_h \in \{3, 5, 7\}$\,m/s.

\paragraph{Perturbations.} Physical perturbations comprise steady wind and gusts up to 5\,m/s injected mid-flight through the simulator's wind model. Visual perturbations comprise illumination changes, unseen textures, moving background elements, motion blur, and depth noise and dropout following a range-dependent stereo error model. Training uses domain randomization~\cite{tobin2017domain} over a subset of these factors; held-out factors define the out-of-distribution (OOD) test set.

\paragraph{Splits.} Training, validation, and test worlds use disjoint seed ranges. The test set comprises 10 held-out worlds per environment class and difficulty level, an in-distribution (ID) condition, and an OOD condition with held-out visual factors and wind.

\paragraph{Data.} Demonstrations are produced by a privileged expert that plans on the ground-truth occupancy and tracks the resulting minimum-snap trajectories~\cite{mellinger2011minsnap}; expert chunks are recorded at the knot rate $1/\Delta = 50$\,Hz together with the RGB-D stream and EKF2 estimates. We plan 2{,}000 expert flights over 200 training worlds followed by three DAgger rounds of 500 flights each (Section~\ref{sec:training}); dataset sizes and the expert's own success rate are reported.

\subsection{Hardware-in-the-Loop Specification}
\label{sec:hitl}
\paragraph{Configuration.} The trained pipeline is exported to ONNX and compiled with TensorRT in FP16 (INT8 with a calibration cache as ablation A9) on a Jetson Orin Nano 8\,GB developer kit in the 15\,W mode, with clock settings fixed and reported. A simulation host with a discrete GPU runs PX4 SITL and the simulator; rendered RGB-D frames are streamed to the Jetson over a dedicated Gigabit Ethernet link, and the uXRCE-DDS agent on the Jetson connects to the SITL client over UDP. In a second stage, PX4 runs on a physical flight controller in PX4's HITL mode, connected to the Jetson over a serial link at 921\,600\,baud, so that the serial transport and flight-controller scheduling are exercised.

\paragraph{Timing validity.} Host and Jetson clocks are synchronized with the Precision Time Protocol. The simulator runs in real time, and episodes whose real-time factor falls below 0.98 at any point are discarded and reported. The image transport from host to Jetson is not present on a vehicle; its latency is measured separately and reported both included in and excluded from $T_{\mathrm{e2e}}$.

\paragraph{Instrumentation.} Every term of Eq.~\eqref{eq:e2e} is logged per decision from synchronized timestamps: capture, preprocessing, the four network times (CUDA events), safety filter, executor wait, and DDS transport (round trip halved and cross-checked against synchronized one-way stamps). Power and utilization are logged with \texttt{tegrastats} at 100\,ms intervals, including module input power (VDD\_IN), the combined CPU--GPU--CV rail (VDD\_CPU\_GPU\_CV), GPU and memory-controller utilization, and temperatures. On this module the GPU is not metered on a separate rail, so the combined rail is reported as the GPU-power metric.

\subsection{Baselines}
\label{sec:baselines}
All baselines receive the same observations (RGB-D, EKF2 state, mission waypoints), run on the same Jetson in the same power mode, and command PX4 through the same interface with the same speed limits; imitation-learning baselines use the same expert data and DAgger budget, and hyperparameters are tuned with equal search budgets on validation worlds.
\begin{itemize}
  \item \textbf{B1, classical:} VINS-Fusion~\cite{qin2019vinsfusion} stereo-inertial odometry, an OctoMap~\cite{hornung2013octomap} at 0.1\,m resolution, A$^\ast$ search on the occupancy grid, and minimum-snap trajectory generation~\cite{mellinger2011minsnap} tracked through PX4 position setpoints.
  \item \textbf{B2, end-to-end reinforcement learning:} vision-based PPO~\cite{schulman2017ppo} and SAC~\cite{haarnoja2018sac} policies with an encoder of the same size as $E_\psi$ and an asymmetric critic with privileged state, trained on the reward of Eq.~\eqref{eq:reward} with a progress term in a GPU-parallel simulator~\cite{mittal2023orbit} with matched vehicle parameters and depth rendering, and evaluated in the common SITL protocol. The training budget is reported in environment steps.
  \item \textbf{B3, generative policies:} a downscaled Diffusion Policy~\cite{chi2025diffusion} with the encoder of Table~\ref{tab:budget}, a solver of equal size, and DDIM sampling with 10 and 100 denoising steps; Octo-Small (27M parameters)~\cite{octo2024}, fine-tuned with a four-dimensional velocity action head; and Qantara~\cite{rakhimov2026qantara}, trained from scratch on the same data with its published objective, with RGB-D input and the waypoint feature $g_t$ of Eq.~\eqref{eq:goal} added to its conditioning so that it receives the same mission information as the other methods. Qantara is evaluated through its behaviour-cloning and video-inverse paths with two Euler steps; its latent-planning path is excluded from the real-time protocol because its reported planning time of 0.4--1.2\,s per step on an H100 GPU exceeds the control period by more than an order of magnitude.
\end{itemize}

\subsection{Evaluation Metrics}
\label{sec:metrics}
For $N$ episodes, let $\mathcal{E}^{\mathrm{succ}}_i$ denote the event that episode $i$ reaches the final waypoint within $\tau_w = 1$\,m before the time limit $T_{\max}$ without collision, and $\mathcal{E}^{\mathrm{col}}_i$ the event that $\operatorname{dist}(p(\tau), \mathcal{X}) < r_v$ at some time $\tau$ during the episode. The primary metrics are
\begin{align}
  \mathrm{SR} &= \frac{1}{N}\sum_{i=1}^{N}\mathbb{1}\big[\mathcal{E}^{\mathrm{succ}}_i\big], \qquad
  \mathrm{CR} = \frac{1}{N}\sum_{i=1}^{N}\mathbb{1}\big[\mathcal{E}^{\mathrm{col}}_i\big], \label{eq:srcr}\\
  \bar{v} &= \frac{1}{|\mathcal{I}_{\mathrm{succ}}|}\sum_{i \in \mathcal{I}_{\mathrm{succ}}}\frac{\ell_i}{T_i}, \label{eq:speed}\\
  T_{\mathrm{e2e}}^{(q)} &= \text{the } q\text{-quantile of } \{T_{\mathrm{e2e},t}\}_t, \quad q \in \{0.5, 0.95, 0.99\}, \label{eq:latq}\\
  \bar{P}_{\mathrm{rail}} &= \frac{1}{T}\int_0^T P_{\mathrm{rail}}(\tau)\,\mathrm{d}\tau, \qquad E_{\mathrm{dec}} = \frac{\bar{P}_{\mathrm{VDD\_IN}}}{f_{\mathrm{dec}}}, \label{eq:power}
\end{align}
where $\mathcal{I}_{\mathrm{succ}}$ is the set of successful episodes, $\ell_i$ and $T_i$ are the flown path length and duration of episode $i$, $T_{\mathrm{e2e},t}$ is the latency of Eq.~\eqref{eq:e2e} for decision $t$, $P_{\mathrm{rail}}$ is the power of the rail VDD\_CPU\_GPU\_CV or VDD\_IN over a logging window of length $T$, and $E_{\mathrm{dec}}$ is the module energy per decision. Secondary metrics are the decision rate $f_{\mathrm{dec}}$; the deadline-miss rate, i.e., the fraction of executor ticks served by the fallback because no valid chunk was available; the root-mean-square jerk of the flown trajectory; the magnitude of velocity-reference discontinuities at chunk switches before blending; and, for the safety layer, the area under the receiver operating characteristic curve
\begin{equation}
  \mathrm{AUROC} = \Pr\big(\bar{\sigma}_{t^+} > \bar{\sigma}_{t^-}\big),
  \label{eq:auroc}
\end{equation}
where $t^+$ and $t^-$ are decision steps drawn uniformly from those followed and not followed, respectively, by a collision or a near-miss within $T_h = 1$\,s, together with the rates of true and spurious fallbacks and the mean speed reduction caused by $s_t$ in perturbation-free segments, which quantifies conservatism.

\subsection{Hypotheses and Statistical Protocol}
\label{sec:hypotheses}
\begin{hypothesis}[Latency]\label{hyp:latency}
On the Jetson Orin Nano in the 15\,W mode with TensorRT FP16, $T_{\mathrm{VPA}}$ for $K = 2$ and $H = 16$ has a 99th percentile of at most 20\,ms, and $T_{\mathrm{e2e}}^{(0.99)}$ excluding the simulator image transport is at most 50\,ms; $t_v(H)$ remains within 10\% of its value at $H = 8$ for $H \le 64$.
\end{hypothesis}
\begin{hypothesis}[Task performance]\label{hyp:task}
At each speed limit, the SR of Aero-VPA is non-inferior to that of the best B3 baseline with a margin of 5 percentage points, while its $T_{\mathrm{e2e}}^{(0.95)}$ is lower; and its SR decreases less than that of B1 between $\bar{v}_h = 3$ and $7$\,m/s.
\end{hypothesis}
\begin{hypothesis}[Uncertainty and safety]\label{hyp:safety}
Under OOD conditions, $\bar{\sigma}$ predicts collisions and near-misses within 1\,s with $\mathrm{AUROC} \ge 0.75$, and the full safety layer reduces CR relative to the unfiltered policy while increasing the mean mission time by at most 15\%.
\end{hypothesis}
\begin{hypothesis}[Representation]\label{hyp:repr}
Under held-out visual perturbations, the relative SR decrease of Aero-VPA is smaller than that of an otherwise identical policy whose encoder is trained with a pixel-reconstruction objective.
\end{hypothesis}
\begin{hypothesis}[Chunk coherence]\label{hyp:coherence}
Noise persistence (Eq.~\eqref{eq:noise}) reduces the median velocity-reference discontinuity at chunk switches and does not reduce SR in environment class E3.
\end{hypothesis}

\paragraph{Statistical analysis.} Each method is trained with three seeds. For each seed, environment class, condition, and speed limit, 500 episodes are run (10 test worlds $\times$ 50 start states), with identical world seeds and start states across methods. Proportions are reported with 95\% Wilson intervals and continuous metrics with bootstrap intervals over episodes, together with seed-level means and standard deviations. Paired binary outcomes are compared with McNemar's exact test, and the non-inferiority test of Hypothesis~\ref{hyp:task} uses the lower bound of a one-sided 95\% interval on the paired SR difference; $p$-values within each hypothesis family are corrected with the Holm--Bonferroni procedure. For an unpaired two-sided two-proportion $z$-test at significance level $\alpha_s$ and power $1 - \beta_s$, the number of episodes per method needed to detect a difference between success probabilities $p_1$ and $p_2$ is
\begin{equation}
  n = \frac{\Big(z_{1-\alpha_s/2}\sqrt{2\bar{p}(1 - \bar{p})} + z_{1-\beta_s}\sqrt{p_1(1-p_1) + p_2(1-p_2)}\Big)^2}{(p_1 - p_2)^2}, \qquad \bar{p} = \frac{p_1 + p_2}{2},
  \label{eq:samplesize}
\end{equation}
where $z_q$ is the standard normal $q$-quantile. For $p_1 = 0.6$, $p_2 = 0.7$, $\alpha_s = 0.05$, and $1 - \beta_s = 0.8$, Eq.~\eqref{eq:samplesize} gives $n \approx 356$; with 500 episodes the power is about 0.91, and the paired design is expected to increase it further.

\subsection{Target Comparison Schema}
\label{sec:schema}
Table~\ref{tab:schema} is the template in which the main comparison will be reported, for each environment class, condition, and speed limit. All measured cells are empty by design.


\begin{table}[t]
\centering
\caption{Target comparison schema (template; no results). Each block is reported per environment class (E1--E3), condition (ID/OOD), and speed limit $\bar{v}_h \in \{3, 5, 7\}$\,m/s. ``--'' denotes a quantity to be measured under the protocol of Section~\ref{sec:experiments}; $\bar{P}_{\mathrm{CGC}}$ is the mean power of the VDD\_CPU\_GPU\_CV rail. The last row states the acceptance criteria for Aero-VPA (V1) and the hypothesis each criterion tests.}
\label{tab:schema}
\footnotesize
\setlength{\tabcolsep}{5pt}
\begin{tabular}{@{}llrcccccc@{}}
\toprule
 & & & SR & CR & $\bar{v}$ & $T_{\mathrm{e2e}}$ p50 / p95 & $\bar{P}_{\mathrm{CGC}}$ & $E_{\mathrm{dec}}$ \\
ID & Method & Params & (\%) & (\%) & (m/s) & (ms) & (W) & (mJ) \\
\midrule
B1 & VINS-Fusion + OctoMap + min-snap & -- & -- & -- & -- & -- / -- & -- & -- \\
B2a & Vision PPO & matched & -- & -- & -- & -- / -- & -- & -- \\
B2b & Vision SAC & matched & -- & -- & -- & -- / -- & -- & -- \\
B3a & Diffusion Policy (DDIM 10 / 100) & matched & -- & -- & -- & -- / -- & -- & -- \\
B3b & Octo-Small, fine-tuned & 27M & -- & -- & -- & -- / -- & -- & -- \\
B3c & Qantara (behaviour cloning / video-inverse) & $\approx$22M & -- & -- & -- & -- / -- & -- & -- \\
\midrule
V0 & Aero-VPA without safety layer & 18.13M & -- & -- & -- & -- / -- & -- & -- \\
V1 & Aero-VPA, full ($K = 2$) & 18.13M & -- & -- & -- & -- / -- & -- & -- \\
\midrule
\multicolumn{3}{@{}l}{Acceptance criteria for V1} & $\ge$ B3$^\ast$ $-$ 5 & $<$ V0 & report & p99 $\le$ 50 & 15\,W & report \\
\multicolumn{3}{@{}l}{(hypothesis tested)} & (H\ref{hyp:task}) & (H\ref{hyp:safety}) & vs.\ B1 & (H\ref{hyp:latency}) & mode & \\
\bottomrule
\multicolumn{9}{@{}l}{\scriptsize B3$^\ast$: best B3 baseline at the same speed limit (non-inferiority margin 5 percentage points). Network-only p99 target: $T_{\mathrm{VPA}} \le 20$\,ms.}
\end{tabular}
\end{table}

\subsection{Ablation Matrix}
\label{sec:ablation}
Ablations vary one factor at a time around the default configuration (V1), except A2, which is a full grid because its purpose is to trace the safety--conservatism Pareto front in the plane of collision rate and normalized mission time. Table~\ref{tab:ablation} lists the factors, levels, readouts, and the hypothesis each ablation addresses.


\begin{table}[t]
\centering
\caption{Planned ablation matrix. Defaults are underlined; all other factors are held at their defaults. ``Pareto'' denotes the front of collision rate versus mission time normalized by that of V0.}
\label{tab:ablation}
\small
\setlength{\tabcolsep}{4pt}
\begin{tabular}{@{}l>{\raggedright\arraybackslash}p{3.0cm}>{\raggedright\arraybackslash}p{5.0cm}>{\raggedright\arraybackslash}p{4.3cm}c@{}}
\toprule
ID & Factor & Levels & Primary readouts & Tests \\
\midrule
A1 & CFM integration steps & $K \in \{1, \underline{2}, 4, 8\}$ & $K t_v$, $T_{\mathrm{VPA}}$, SR, CR, chunk error vs.\ $K = 64$ reference & H\ref{hyp:latency}, H\ref{hyp:task} \\
A2 & Uncertainty threshold $\times$ attenuation floor & $p_{\mathrm{fa}} \in \{10^{-2}, 10^{-3}, \underline{2.8{\times}10^{-5}}\}$ or fallback off; $\times$ $\alpha_{\min} \in \{0.25, \underline{0.5}, 1\}$ & Pareto, spurious and true fallback rates & H\ref{hyp:safety} \\
A3 & Attenuation law & none, \underline{piecewise linear (Eq.~\eqref{eq:alpha})}, exponential & CR, $\bar{v}$, jerk & H\ref{hyp:safety} \\
A4 & Braking envelope & off, \underline{on} & CR in E1/E2, $\bar{v}$, conservatism & H\ref{hyp:safety} \\
A5 & Noise persistence & fresh noise, \underline{persistent, $N_\xi = 30$}, persistent, $N_\xi = 10$ & discontinuity at switches, SR in E3 & H\ref{hyp:coherence} \\
A6 & Latent predictor & removed (reactive CFM), \underline{JEPA + VICReg}, JEPA without $v(Z), c(Z)$ & SR, effective rank of $C(Z)$, AUROC & H\ref{hyp:repr} \\
A7 & Encoder objective & \underline{JEPA}, pixel reconstruction, frozen pretrained DINOv2-S~\cite{oquab2024dinov2} & relative SR drop under visual OOD & H\ref{hyp:repr} \\
A8 & Execution mode & \underline{asynchronous + blending}, asynchronous + inpainting~\cite{black2025rtc}, synchronous prefix with fixed $H_t \in \{4, 8, 16\}$, synchronous adaptive $H_t$ & SR, deadline-miss rate, discontinuity & H\ref{hyp:coherence} \\
A9 & Numerical precision & FP32, \underline{FP16}, INT8 & $T_{\mathrm{VPA}}$, $E_{\mathrm{dec}}$, chunk deviation from FP32 & H\ref{hyp:latency} \\
A10 & Chunk horizon & $H \in \{8, \underline{16}, 32, 64\}$ & $t_v(H)$, SR, jerk & H\ref{hyp:latency} \\
\bottomrule
\end{tabular}
\end{table}


\section{Closed-Loop Evaluation in Simulation}
\label{sec:evaluation}

This section reports the first executed part of the protocol of Section~\ref{sec:experiments}: closed-loop waypoint missions through procedurally generated forests, environment class E1 of Section~\ref{sec:sim}, in PX4 SITL with Gazebo Harmonic. It evaluates behaviour cloning, six DAgger checkpoints and the runtime shield of Section~\ref{sec:shield}; Section~\ref{sec:deviations} lists how the evaluation departs from the protocol, and the appendices give the full results.

\subsection{Setup}
\label{sec:setup}

\paragraph{Vehicle.} The vehicle is a PX4 x500 quadrotor model with a front RGB-D camera of $160 \times 128$ pixels, a horizontal field of view of 73\textdegree{} and a depth range of 0.1--10\,m, which provides the frame $I_t$ of Section~\ref{sec:pomdp}; PX4 runs in SITL, and the companion-computer software of Section~\ref{sec:implementation} commands it in offboard mode.

\paragraph{Worlds and missions.} The test worlds are 36 procedurally generated forests, seeds 2000--2011 at three densities (Poisson-disc trunk spacing of 7, 5 and 3.5\,m for sparse, medium and dense), disjoint from the training (0--199) and calibration (1000--1011) worlds. Each world has five missions, 180 per policy; mission $k$ of a world is seeded by (world seed, $k$), so every policy flies identical missions and all comparisons are paired.\footnote{Every closed-loop mission record behind the tables of this section and of the appendices is released, with the decision traces, the 36 test worlds and their obstacle maps, and the analysis reports, in the benchmark dataset~\cite{slingai2026benchmark}, \slashurl{https://huggingface.co/datasets/ylee-sling/aero-vpa-benchmark} (CC~BY~4.0).} A mission takes off to 2.5\,m and visits 3--6 waypoints placed 10--40\,m apart within a time limit of $1.5\,T_{\mathrm{ref}} + 15$\,s, where $T_{\mathrm{ref}}$ is the duration of the privileged expert's reference plan at 3\,m/s. The policies are not speed-conditioned.

\paragraph{Endings.} A mission ends in \emph{success} when the final waypoint is reached within $\tau_w = 1$\,m before the time limit; in a \emph{collision} when the privileged distance from the vehicle centre to a trunk surface (trunks are vertical cylinders in the world file) falls below 0.35\,m; in \emph{ground} contact below 0.3\,m of altitude; in a \emph{timeout} at the limit; in a \emph{camera timeout} after 0.5\,s without a decision; or \emph{out of bounds}. A \emph{crash} is a collision or a ground contact.

\paragraph{Policies.} Table~\ref{tab:names} lists the nine policies. The three BC policies flew 60 missions (12 of the 36 test worlds) before a start-up warm-up fix; the six DAgger checkpoints flew all 180 missions without and with the shield.

\begin{table}[tb]
\centering
\caption{Evaluated policies. Except \bcVone, all use the frozen residual encoder $E_\psi$ and predictor $P_\omega$; all are calibrated with the episode-maximum method of Section~\ref{sec:uncertainty}. Withheld inputs enter $\pi^h_\phi$ and $e_t$ at their training mean; $P_\omega$ always receives $\bar{a}_t$. ``Demos'': demonstrations from the 200 training worlds.}
\label{tab:names}
\small
\begin{tabular}{@{}l>{\raggedright\arraybackslash}p{5.0cm}>{\raggedright\arraybackslash}p{7.2cm}@{}}
\toprule
Policy & Training data & Inputs and recipe \\
\midrule
\bcVone  & demos & non-residual stage~(i); all inputs \\
\bcVtwo  & demos & residual stage~(i); all inputs \\
\bcNoact & demos & as \bcVtwo, with $\bar{a}_t$ withheld \\
\midrule
\ckRone     & demos + DAgger round 1 (flown by \bcVtwo) & all inputs \\
\ckRtwo     & demos + rounds 1--2 (round 2 flown by \ckRone) & all inputs \\
\ckNoact    & demos + rounds 1--3 (round 3 flown by \ckRtwo) & $\bar{a}_t$ withheld \\
\ckNomotion & as \ckNoact & $\bar{a}_t$ and $v^{\mathcal{H}}_t$ withheld \\
\ckSd       & as \ckNoact & all inputs; state dropout: $v^{\mathcal{H}}_t$ and $\bar{a}_t$ replaced by their training mean in 50\% of training frames \\
\ckKw       & as \ckNoact & all inputs; keyframe-weighted sampling, frames drawn $\propto \min\!\big(1 + \norm{A_t[0] - \bar{a}_t}/0.1,\; 10\big)$, the error of repeating the last command; DAgger weight $\times1$ \\
\bottomrule
\end{tabular}
\end{table}

\paragraph{Compute.} Three PX4 SITL and Gazebo Harmonic~\cite{koenig2004gazebo} instances ran headless in parallel on one virtual machine with 16 vCPUs and an NVIDIA L4 GPU.

\paragraph{Metrics.} Besides the endings we report the share of all mission waypoints reached; the median duration and path length; the ratio of duration to the expert's reference time and of path length to mission length (medians over successful missions); the mean speed (mean over missions of the per-mission mean); the smallest clearance (median over missions of the per-mission minimum privileged distance); the share of blind decisions, whose commanded direction lay outside the field of view or whose depth was invalid; fallback onsets per mission; and, for the shield, interventions per mission, seconds per mission with each rule active, and the clearance to the memory (median over missions of the minimum distance from the estimated position to a remembered return). These metrics differ from those planned in Section~\ref{sec:metrics}: the mean speed averages over all missions rather than over successful ones as in Eq.~\eqref{eq:speed}, the latency quantiles of Eq.~\eqref{eq:latq} and the power and energy of Eq.~\eqref{eq:power} were not measured, and Section~\ref{sec:implementation} reports the decision timing on the L4 instead.

\paragraph{Statistics.} Success and crash rates carry Wilson 95\% intervals~\cite{wilson1927}. Paired comparisons use the two-sided exact McNemar test on the discordant pairs~\cite{mcnemar1947}; tables give the counts of missions in which both, only one or neither of the two policies succeeded. $N$ is stated in every caption. The primary family of tests is the six shielded-against-unshielded success comparisons; with Holm's correction~\cite{holm1979} its largest adjusted $p$ is $2.4\times10^{-10}$. All other tests are secondary and unadjusted. Each recipe was trained once, so the tests capture variation between missions, not between training runs.

\subsection{Behaviour Cloning Fails: The Copycat Shortcut}
\label{sec:copycat}
BC succeeded in none of the 60 missions it flew (Appendix~\ref{app:bc}): \bcVone crashed in 83\% of them, \bcVtwo in 51.7\% with timeouts in the rest, and \bcNoact in 70.0\%, mostly into the ground. Offline, the same policies looked nearly perfect (Table~\ref{tab:copycat}, top): on 5000 held-out expert frames, the median error of \bcVtwo's first knot was 0.02\,m/s (\bcNoact: 0.04\,m/s), and its selector agreed with the expert's primitive in 98.0\% of frames (93.6\%).

The swap test explains the gap. It replaces one input by that of another frame and measures the change of the first knot; the labels of two frames differ by a median of 1.13\,m/s. For \bcVtwo, swapping $\bar{a}_t$ changed the command by 0.85\,m/s, moved it to within 0.31\,m/s of the donor frame's label, and changed the primitive in 35.5\% of frames, whereas swapping the image or the goal changed the command by 0.01\,m/s. With $\bar{a}_t$ withheld, \bcNoact copied the measured velocity instead (1.17\,m/s, 0.21\,m/s from the donor's label). On expert data the last command is within about 0.02\,m/s of the next label, so this shortcut fits the demonstrations almost exactly and no offline metric reveals it.

DAgger did not remove the shortcut on expert frames. \ckRone and \ckRtwo followed the donor's $\bar{a}_t$ (0.96 and 0.93\,m/s; 0.15 and 0.27\,m/s from its label) and ignored the image and the goal ($\le 0.02$\,m/s); their velocity swaps changed the command by medians of 0.17 and 0.31\,m/s but by 1.24 and 1.62\,m/s at the 90th percentile, a growing dependence rather than a copy. With $\bar{a}_t = 0$, \ckRone's first knot was still 1.42\,m/s (\bcVtwo: 0.57\,m/s). The round-3 recipes moved the shortcut rather than removing it. \ckNoact copied velocity (1.10\,m/s, 0.28\,m/s from the donor's label). \ckNomotion copied nothing---image and goal swaps moved its command by 0.43 and 0.42\,m/s, with donor errors of at least 1.04\,m/s---but underfitted (open-loop error 0.31\,m/s median and 1.19\,m/s at the 90th percentile; selector agreement 78.3\%), and $\bar{a}_t$ reached its command through the predictor's forecast (0.33\,m/s, 0.86\,m/s from the donor's label against 0.57\,m/s from its own). State dropout (\ckSd) reduced the velocity dependence (0.16\,m/s) but not the $\bar{a}_t$ dependence (0.83\,m/s, 0.28\,m/s from the donor's label against 0.84\,m/s from its own), and keyframe weighting (\ckKw) left both (0.49 and 0.88\,m/s; Appendix~\ref{app:swap}).

On the policies' own states the picture differs (Table~\ref{tab:copycat}, bottom). On the states \ckRone flew in round 2, labelled by the expert and held out for both policies shown, \bcVtwo and \ckRone erred by 0.50 and 0.49\,m/s, pointed more than 45\textdegree{} away from the label in 22.2\% and 24.2\% of frames, and agreed with the expert's primitive in only 56.8\% and 61.2\%; there \ckRone did use the image (swap 0.17\,m/s median, 1.41\,m/s at the 90th percentile). On \ckRtwo's round-3 states, \ckRone and \ckRtwo erred by 0.51 and 0.47\,m/s, and swapping $\bar{a}_t$ still changed their commands by 1.38 and 1.19\,m/s. More rounds with $\bar{a}_t$ available did not converge. The expert itself was blind in 3.7\% of its held-out frames but in 34\% of the frames on \ckRone's round-2 states.

\begin{table}[t]
\centering
\caption{Copycat diagnostics. \emph{Top:} 5000 held-out expert frames (validation flights). A swap replaces one input by that of another frame; entries are the median change of the first knot $v_0$ (m/s), with the error to the donor frame's label in parentheses (the median label difference between frames is 1.13\,m/s). Open loop: median $|v_0 - v_0^\star|$ and selector agreement with the expert's primitive. \emph{Bottom:} the policies' own DAgger states, labelled by the expert and held out for the policies shown; ``$>45^\circ$'': share of frames whose command points more than 45\textdegree{} from the label. ``--'': withheld input, or not measured.}
\label{tab:copycat}
\small
\setlength{\tabcolsep}{4.5pt}
\begin{tabular}{@{}lcccccccc@{}}
\toprule
\multicolumn{9}{@{}l}{\emph{Expert frames}} \\
Policy & $\bar{a}_t$ swap & $v^{\mathcal{H}}_t$ swap & image & goal & all of $S_t$ & $\bar{a}_t$ to $P_\omega$ only & $|v_0 - v_0^\star|$ & selector \\
\midrule
\bcVtwo     & 0.85 (0.31) & 0.31 (0.86) & 0.01 & 0.01 & 1.14 & 0.00 & 0.02 & 98.0\% \\
\bcNoact    & --          & 1.17 (0.21) & 0.03 & 0.02 & 1.14 & 0.00 & 0.04 & 93.6\% \\
\ckRone     & 0.96 (0.15) & 0.17 (1.02) & 0.02 & 0.01 & --   & --   & --   & --     \\
\ckRtwo     & 0.93 (0.27) & 0.31 (0.99) & 0.02 & 0.01 & --   & --   & --   & --     \\
\ckNoact    & --          & 1.10 (0.28) & --   & --   & --   & --   & --   & --     \\
\ckNomotion & --          & --          & 0.43 & 0.42 & --   & 0.33 (0.86) & 0.31 & 78.3\% \\
\ckSd       & 0.83 (0.28) & 0.16 (0.98) & 0.04 & 0.03 & 1.09 & 0.02 & 0.04 & 91.8\% \\
\ckKw       & 0.88 (0.43) & 0.49 (0.97) & 0.03 & 0.02 & 1.14 & 0.00 & 0.03 & 92.9\% \\
\midrule
\multicolumn{9}{@{}l}{\emph{Policies' own states}} \\
States & Policy & error to label & $>45^\circ$ & selector & \multicolumn{4}{l}{swap on these states} \\
\midrule
round 2 (flown by \ckRone) & \bcVtwo & 0.50 & 22.2\% & 56.8\% & \multicolumn{4}{l}{--} \\
                           & \ckRone & 0.49 & 24.2\% & 61.2\% & \multicolumn{4}{l}{image 0.17 (p90 1.41)} \\
round 3 (flown by \ckRtwo) & \ckRone & 0.51 & 22.0\% & 62.7\% & \multicolumn{4}{l}{$\bar{a}_t$ 1.38} \\
                           & \ckRtwo & 0.47 & 21.8\% & 62.9\% & \multicolumn{4}{l}{$\bar{a}_t$ 1.19} \\
\bottomrule
\end{tabular}
\end{table}

\subsection{DAgger Gives the First Successes, Then Plateaus}
\label{sec:plateau}
Under the supervisor, success rose from 73.6\% in round~1 to 94.0\% in round~3 (Table~\ref{tab:rounds}), but these flights were gated and partly flown by the expert. Autonomously, without the shield, \ckRone succeeded in 38 of the 180 test missions (21.1\%, 95\% CI [15.8, 27.6]; Table~\ref{tab:main}); on the 60 missions shared with BC, \ckRone alone succeeded in 12 and \bcVtwo alone in none ($p = 0.0005$). No later checkpoint improved on it (Table~\ref{tab:unshielded-pairs}): \ckRtwo, trained on two DAgger rounds, succeeded in 12.8\% [8.7, 18.4] ($p = 0.0315$ against \ckRone), and the round-3 recipes in 11.1--15.6\%, with \ckNoact significantly worse than \ckRone ($p = 0.0051$) and the others not significantly different from it ($p = 0.0704$--$0.184$) or from \ckRtwo ($p = 0.473$--$1$). Crashes ended 72.2--82.2\% of the unshielded missions, and the median smallest clearance was 0.35\,m for all six checkpoints, the collision distance, because most missions ended in a collision. Mission outcomes varied strongly between checkpoints: \ckRone and \ckKw shared only 12 of their 38 and 28 successes, where independent draws at those rates would share about 6. Together with the errors of about 0.5\,m/s on the policies' own states (Section~\ref{sec:copycat}), we read this as a plateau: the unshielded DAgger policies stayed at 11.1--21.1\% success, and neither input removal, state dropout nor keyframe weighting beat the first round.

\begin{table}[t]
\centering
\caption{DAgger rounds under the privileged supervisor ($N = 201$ missions per round over the training worlds). These flights are gated and partly flown by the expert, so their success and crash rates measure the supervised system, not autonomous performance.}
\label{tab:rounds}
\small
\begin{tabular}{@{}cccccc@{}}
\toprule
Round & Flown by & Missions & Success & Crash & Policy's share of the flight \\
\midrule
1 & \bcVtwo & 201 & 148 (73.6\%) & 28 (13.9\%) & $\approx$50\% \\
2 & \ckRone & 201 & 166 (82.6\%) & 5 (2.5\%)   & 63\% \\
3 & \ckRtwo & 201 & 189 (94.0\%) & 10 (5.0\%)  & 64\% \\
\bottomrule
\end{tabular}
\end{table}

\begin{table}[t]
\centering
\caption{Paired success comparisons without the shield ($N = 180$ paired test missions; last row $N = 60$, the missions the BC policies flew). Discordant pairs and the two-sided exact McNemar $p$.}
\label{tab:unshielded-pairs}
\small
\begin{tabular}{@{}lccc@{}}
\toprule
Comparison (A vs B) & Only A succeeded & Only B succeeded & $p$ \\
\midrule
\ckRone{} vs \ckRtwo      & 29 & 14 & 0.0315 \\
\ckRone{} vs \ckNoact     & 28 & 10 & 0.0051 \\
\ckRone{} vs \ckNomotion  & 33 & 19 & 0.0704 \\
\ckRone{} vs \ckSd        & 28 & 18 & 0.184 \\
\ckRone{} vs \ckKw        & 26 & 16 & 0.164 \\
\ckRtwo{} vs \ckNoact     & 19 & 16 & 0.736 \\
\ckRtwo{} vs \ckNomotion  & 17 & 18 & 1 \\
\ckRtwo{} vs \ckSd        & 14 & 19 & 0.487 \\
\ckRtwo{} vs \ckKw        & 13 & 18 & 0.473 \\
\midrule
\ckRone{} vs \bcVtwo{} (60 missions) & 12 & 0 & 0.0005 \\
\bottomrule
\end{tabular}
\end{table}

\subsection{Shielded Benchmark}
\label{sec:shielded}
The shield raised success for every checkpoint (Table~\ref{tab:main}, Figure~\ref{fig:endings}): for \ckRone by 31.1 percentage points, from 21.1\% to 52.2\% [45.0, 59.4] (only shielded 68, only unshielded 12; $p = 1.2\times10^{-10}$), and for the other five by 30.0--34.4 points, to 42.8--50.0\%. All six paired comparisons gave $p \le 1.33\times10^{-10}$. Crashes fell from 72.2--82.2\% to 22.2--42.8\%; for \ckRone from 130 to 40 missions (only unshielded 98, only shielded 8; $p = 8.08\times10^{-21}$; Appendix~\ref{app:endings}). Every shielded checkpoint also succeeded more often than the best unshielded one, \ckRone, on the same missions ($p \le 1.69\times10^{-5}$). Paired with \ckRone{} + shield, however, no shielded checkpoint differed significantly ($p = 0.0712$--$0.724$; Table~\ref{tab:pairs-shielded}), so we do not rank them.

Failures moved from crashes to timeouts: \ckRone's timeouts rose from 2.8\% to 23.3\% and \ckKw's from 1.7\% to 21.7\%. The flight itself slowed only slightly: \ckRone's mean speed went from 1.24 to 1.18\,m/s, while its median smallest clearance rose from 0.35 to 0.52\,m; its successful missions took 1.26 times the expert's reference time and flew 1.13 times the mission length (medians). The learned fallback started 1.04--1.53 times per shielded mission, in 30.0--36.7\% of missions, and braked or held for 0.1--0.3\,s per mission (Appendix~\ref{app:endings}).

\begin{table}[t]
\centering
\caption{Main result: the six DAgger checkpoints without and with the runtime shield on the same $N = 180$ paired test missions. Success counts with Wilson 95\% intervals; $\Delta$: change in percentage points; paired: missions in which both, only the unshielded (U), only the shielded (S), or neither run succeeded, with the two-sided exact McNemar $p$. Rows follow the training lineage, not a ranking: no shielded checkpoint differs significantly from \ckRone{} + shield (Table~\ref{tab:pairs-shielded}).}
\label{tab:main}
\footnotesize
\setlength{\tabcolsep}{3.5pt}
\begin{tabular}{@{}lccccccc@{}}
\toprule
& \multicolumn{2}{c}{Success} & & \multicolumn{2}{c}{Paired success} & Crash (\%) & Timeout (\%) \\
\cmidrule(lr){2-3}\cmidrule(lr){5-6}
Checkpoint & U & S [95\% CI] & $\Delta$ (pp) & both / U / S / neither & $p$ & U $\to$ S & U $\to$ S \\
\midrule
\ckRone     & 38 (21.1\%) & 94 (52.2\%) [45.0, 59.4] & $+31.1$ & 26 / 12 / 68 / 74 & $1.20\times10^{-10}$ & 72.2 $\to$ 22.2 & 2.8 $\to$ 23.3 \\
\ckRtwo     & 23 (12.8\%) & 77 (42.8\%) [35.8, 50.1] & $+30.0$ & 14 / 9 / 63 / 94  & $4.18\times10^{-11}$ & 80.0 $\to$ 38.3 & 1.1 $\to$ 16.7 \\
\ckNoact    & 20 (11.1\%) & 79 (43.9\%) [36.8, 51.2] & $+32.8$ & 11 / 9 / 68 / 92  & $2.45\times10^{-12}$ & 82.2 $\to$ 36.1 & 2.8 $\to$ 14.4 \\
\ckNomotion & 24 (13.3\%) & 81 (45.0\%) [37.9, 52.3] & $+31.7$ & 11 / 13 / 70 / 86 & $1.33\times10^{-10}$ & 77.2 $\to$ 42.8 & 3.3 $\to$ 10.0 \\
\ckSd       & 28 (15.6\%) & 82 (45.6\%) [38.4, 52.8] & $+30.0$ & 19 / 9 / 63 / 89  & $4.18\times10^{-11}$ & 76.1 $\to$ 35.0 & 3.3 $\to$ 16.1 \\
\ckKw       & 28 (15.6\%) & 90 (50.0\%) [42.8, 57.2] & $+34.4$ & 16 / 12 / 74 / 78 & $4.69\times10^{-12}$ & 80.6 $\to$ 26.1 & 1.7 $\to$ 21.7 \\
\bottomrule
\end{tabular}
\end{table}

\definecolor{endSuccess}{HTML}{2A78D6}
\definecolor{endCollision}{HTML}{EB6834}
\definecolor{endGround}{HTML}{1BAF7A}
\definecolor{endTimeout}{HTML}{EDA100}
\definecolor{endOther}{HTML}{E87BA4}
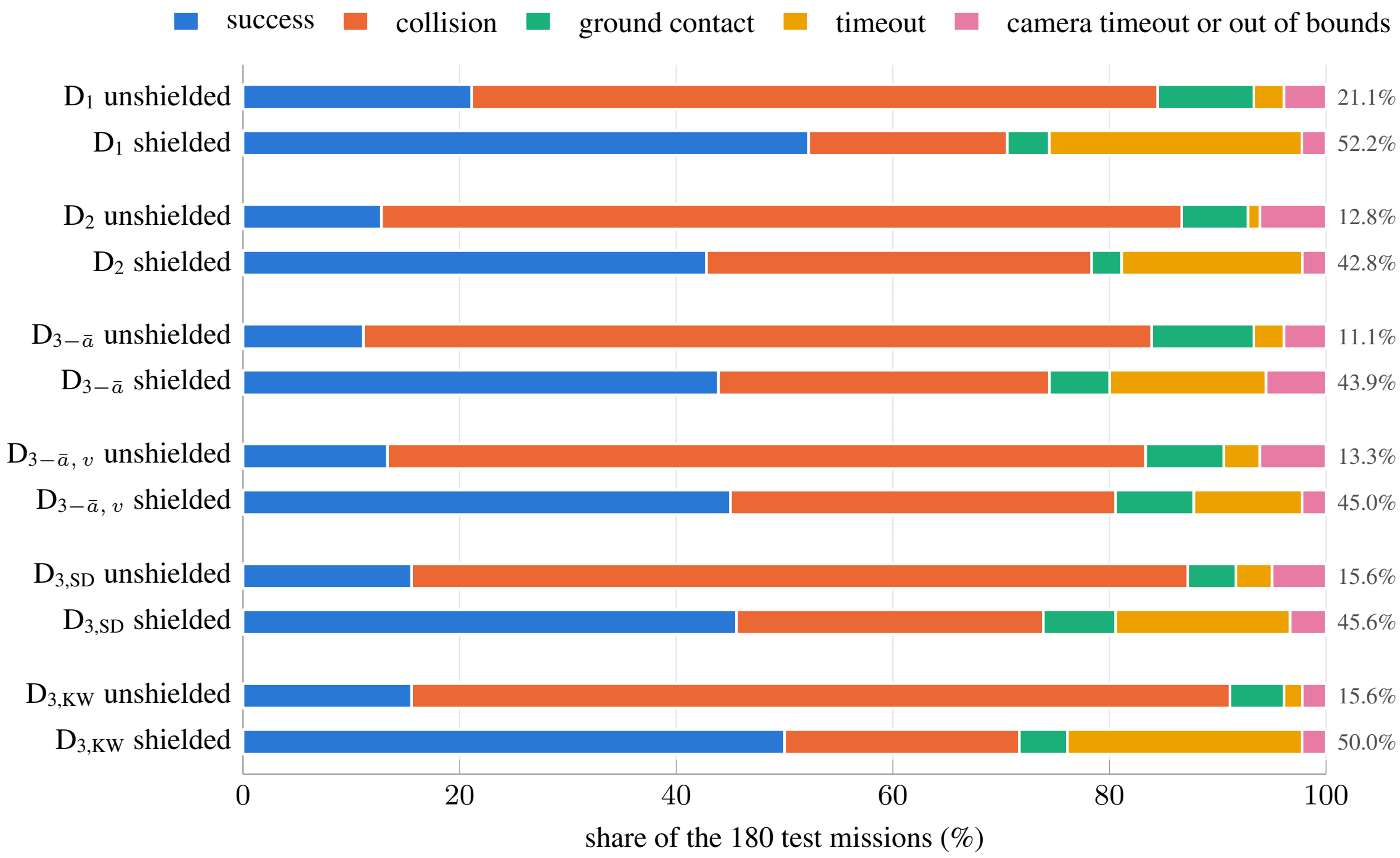
\begin{figure}[t]
\centering
\begin{tikzpicture}
\begin{axis}[
  xbar stacked,
  width=0.80\linewidth, height=9.2cm,
  bar width=7.5pt,
  xmin=0, xmax=100,
  xlabel={share of the 180 test missions (\%)},
  ytick={1,2,3.6,4.6,6.2,7.2,8.8,9.8,11.4,12.4,14,15},
  yticklabels={{\ckRone unshielded}, {\ckRone shielded}, {\ckRtwo unshielded}, {\ckRtwo shielded}, {\ckNoact unshielded}, {\ckNoact shielded}, {\ckNomotion unshielded}, {\ckNomotion shielded}, {\ckSd unshielded}, {\ckSd shielded}, {\ckKw unshielded}, {\ckKw shielded}},
  y dir=reverse,
  ymin=0.3, ymax=15.7,
  tick label style={font=\footnotesize},
  label style={font=\footnotesize},
  xtick={0,20,40,60,80,100},
  xmajorgrids, grid style={black!10},
  axis line style={black!40},
  tick style={black!40},
  axis x line*=bottom, axis y line*=left,
  legend style={at={(0.5,1.02)}, anchor=south, legend columns=5, draw=none, font=\footnotesize, column sep=6pt},
  legend cell align=left,
  clip=false,
  every axis plot/.append style={draw=white, line width=0.8pt},
]
\addplot[fill=endSuccess] coordinates {(21.11,1) (52.22,2) (12.78,3.6) (42.78,4.6) (11.11,6.2) (43.89,7.2) (13.33,8.8) (45.00,9.8) (15.56,11.4) (45.56,12.4) (15.56,14) (50.00,15)};
\addplot[fill=endCollision] coordinates {(63.33,1) (18.33,2) (73.89,3.6) (35.56,4.6) (72.78,6.2) (30.56,7.2) (70.00,8.8) (35.56,9.8) (71.67,11.4) (28.33,12.4) (75.56,14) (21.67,15)};
\addplot[fill=endGround] coordinates {(8.89,1) (3.89,2) (6.11,3.6) (2.78,4.6) (9.44,6.2) (5.56,7.2) (7.22,8.8) (7.22,9.8) (4.44,11.4) (6.67,12.4) (5.00,14) (4.44,15)};
\addplot[fill=endTimeout] coordinates {(2.78,1) (23.33,2) (1.11,3.6) (16.67,4.6) (2.78,6.2) (14.44,7.2) (3.33,8.8) (10.00,9.8) (3.33,11.4) (16.11,12.4) (1.67,14) (21.67,15)};
\addplot[fill=endOther] coordinates {(3.89,1) (2.22,2) (6.11,3.6) (2.22,4.6) (3.89,6.2) (5.56,7.2) (6.11,8.8) (2.22,9.8) (5.00,11.4) (3.33,12.4) (2.22,14) (2.22,15)};
\legend{success, collision, ground contact, timeout, camera timeout or out of bounds}
\node[anchor=west, font=\scriptsize, text=black!70] at (axis cs:100,1) {21.1\%};
\node[anchor=west, font=\scriptsize, text=black!70] at (axis cs:100,2) {52.2\%};
\node[anchor=west, font=\scriptsize, text=black!70] at (axis cs:100,3.6) {12.8\%};
\node[anchor=west, font=\scriptsize, text=black!70] at (axis cs:100,4.6) {42.8\%};
\node[anchor=west, font=\scriptsize, text=black!70] at (axis cs:100,6.2) {11.1\%};
\node[anchor=west, font=\scriptsize, text=black!70] at (axis cs:100,7.2) {43.9\%};
\node[anchor=west, font=\scriptsize, text=black!70] at (axis cs:100,8.8) {13.3\%};
\node[anchor=west, font=\scriptsize, text=black!70] at (axis cs:100,9.8) {45.0\%};
\node[anchor=west, font=\scriptsize, text=black!70] at (axis cs:100,11.4) {15.6\%};
\node[anchor=west, font=\scriptsize, text=black!70] at (axis cs:100,12.4) {45.6\%};
\node[anchor=west, font=\scriptsize, text=black!70] at (axis cs:100,14) {15.6\%};
\node[anchor=west, font=\scriptsize, text=black!70] at (axis cs:100,15) {50.0\%};
\end{axis}
\end{tikzpicture}
\caption{Mission endings of the six DAgger checkpoints without and with the runtime shield ($N = 180$ paired test missions each); the number at the right of each bar is its success rate. The shield turns most collisions into successes or timeouts. Exact counts are in Table~\ref{tab:endings-full}. Six first-person-view (FPV) videos on the \href{https://huggingface.co/ylee-sling/aero-vpa}{model card} (\hyperref[sec:availability]{Code and Data Availability}) show \ckRone{} + shield and \ckKw{} + shield on one mission per density that both completed in the benchmark; they are re-flights for illustration, and the shares shown come from the benchmark runs.}
\label{fig:endings}
\end{figure}

\subsection{Success by Obstacle Density}
\label{sec:density}
For \ckRone the shield's gain was largest in the dense worlds, where success rose from 4 to 28 of 60 missions (6.7\% $\to$ 46.7\%; Table~\ref{tab:density}). Without the shield every checkpoint succeeded in at most 4 of the 60 dense missions; with it, in 17--28. In percentage points, the largest gain fell in the dense worlds for \ckRone and \ckNoact, in the medium worlds for \ckRtwo, \ckNomotion and \ckKw, and in the sparse worlds for \ckSd. Shielded crashes were most frequent in the dense worlds for every checkpoint except \ckNoact (Appendix~\ref{app:endings}).

\begin{table}[t]
\centering
\caption{Success by obstacle density, unshielded (U) and shielded (S), as successes of $N = 60$ missions per density (12 worlds $\times$ 5 missions), with Wilson 95\% intervals (\%) for the shielded runs. Trunk spacing: dense 3.5\,m, medium 5\,m, sparse 7\,m.}
\label{tab:density}
\small
\setlength{\tabcolsep}{4.5pt}
\begin{tabular}{@{}lcccccc@{}}
\toprule
& \multicolumn{2}{c}{Dense} & \multicolumn{2}{c}{Medium} & \multicolumn{2}{c}{Sparse} \\
\cmidrule(lr){2-3}\cmidrule(lr){4-5}\cmidrule(lr){6-7}
Checkpoint & U & S [95\% CI] & U & S [95\% CI] & U & S [95\% CI] \\
\midrule
\ckRone     & 4 (6.7\%) & 28 (46.7\%) [35, 59] & 17 (28.3\%) & 29 (48.3\%) [36, 61] & 17 (28.3\%) & 37 (61.7\%) [49, 73] \\
\ckRtwo     & 0 (0.0\%) & 21 (35.0\%) [24, 48] & 9 (15.0\%)  & 32 (53.3\%) [41, 65] & 14 (23.3\%) & 24 (40.0\%) [29, 53] \\
\ckNoact    & 2 (3.3\%) & 23 (38.3\%) [27, 51] & 6 (10.0\%)  & 25 (41.7\%) [30, 54] & 12 (20.0\%) & 31 (51.7\%) [39, 64] \\
\ckNomotion & 1 (1.7\%) & 17 (28.3\%) [19, 41] & 6 (10.0\%)  & 28 (46.7\%) [35, 59] & 17 (28.3\%) & 36 (60.0\%) [47, 71] \\
\ckSd       & 2 (3.3\%) & 17 (28.3\%) [19, 41] & 9 (15.0\%)  & 26 (43.3\%) [32, 56] & 17 (28.3\%) & 39 (65.0\%) [52, 76] \\
\ckKw       & 1 (1.7\%) & 22 (36.7\%) [26, 49] & 10 (16.7\%) & 33 (55.0\%) [42, 67] & 17 (28.3\%) & 35 (58.3\%) [46, 70] \\
\bottomrule
\end{tabular}
\end{table}

\subsection{Missions Never Used to Revise the Shield}
\label{sec:heldout}
Mission 0 of every test world was the shield pilots' set (Section~\ref{sec:iteration}), so missions 1--4 of each world, 144 per checkpoint, were never used to revise the shield. On them the result is essentially unchanged (Table~\ref{tab:heldout}): \ckRone succeeded in 52.1\% [44.0, 60.1], against 52.2\% on all 180 missions, \ckKw in 54.2\% [46.0, 62.1], and the other four in 43.1--46.5\%; every paired comparison with the unshielded flights of the same missions gave $p \le 5.26\times10^{-8}$. These are the same worlds, so this is a partial check, not a test on fresh worlds; the planned fresh-world development set was not flown.

\begin{table}[t]
\centering
\caption{Missions never used to revise the shield: missions 1--4 of every test world ($N = 144$ paired missions per checkpoint; mission 0 of every world was the shield pilots' set). Same worlds as the full benchmark, so this is a partial check, not a fresh-world test. Paired success counts and exact McNemar $p$; crashes unshielded $\to$ shielded with the exact McNemar $p$ of the paired crash comparison.}
\label{tab:heldout}
\small
\setlength{\tabcolsep}{4.5pt}
\begin{tabular}{@{}lcccccc@{}}
\toprule
& \multicolumn{2}{c}{Success} & \multicolumn{2}{c}{Paired success} & \multicolumn{2}{c}{Crashes} \\
\cmidrule(lr){2-3}\cmidrule(lr){4-5}\cmidrule(lr){6-7}
Checkpoint & unshielded & shielded [95\% CI] & both / only U / only S / neither & $p$ & U $\to$ S & $p$ \\
\midrule
\ckRone     & 31 & 75 (52.1\%) [44.0, 60.1] & 21 / 10 / 54 / 59 & $2.00\times10^{-8}$ & 103 $\to$ 33 & $1.52\times10^{-15}$ \\
\ckRtwo     & 20 & 64 (44.4\%) [36.6, 52.6] & 13 / 7 / 51 / 73  & $2.40\times10^{-9}$ & 112 $\to$ 53 & $3.85\times10^{-13}$ \\
\ckNoact    & 15 & 62 (43.1\%) [35.3, 51.2] & 8 / 7 / 54 / 75   & $4.32\times10^{-10}$ & 120 $\to$ 51 & $9.43\times10^{-16}$ \\
\ckNomotion & 22 & 67 (46.5\%) [38.6, 54.7] & 10 / 12 / 57 / 65 & $3.74\times10^{-8}$ & 109 $\to$ 62 & $3.81\times10^{-8}$ \\
\ckSd       & 24 & 65 (45.1\%) [37.2, 53.3] & 15 / 9 / 50 / 70  & $5.26\times10^{-8}$ & 107 $\to$ 50 & $1.26\times10^{-12}$ \\
\ckKw       & 22 & 78 (54.2\%) [46.0, 62.1] & 15 / 7 / 63 / 59  & $2.28\times10^{-12}$ & 117 $\to$ 32 & $3.09\times10^{-21}$ \\
\bottomrule
\end{tabular}
\end{table}

\subsection{Shield Design Iteration and Failure Locations}
\label{sec:iteration}
The shield was developed with \ckRone on mission 0 of the 36 test worlds (Table~\ref{tab:pilot}). Shield~v1 raised success from 7 to 12 of 36 and lowered crashes from 27 to 11, but its own manoeuvres failed: 10 of its 11 crashes happened while the shield was flying its own setpoint (backing off, nudging or holding); of its 10 collisions, 4 were with obstacles behind the camera, 4 with obstacles beside it and 2 with obstacles in view; and in all 8 crashes with a measurable direction of motion the vehicle was moving more than 36\textdegree{} off the camera axis, outside the field of view. Its uncertainty cap was active for 18.0\,s per mission and turned flights into timeouts. Shield~v2 therefore passes the backoff through the barrier and the altitude floor at 0.3\,m/s (\shieldOne: 0.5\,m/s), ends it in a nudge when the way back is blocked, applies the blind cap after the barrier, and disables the uncertainty cap. On the same 36 missions \shieldTwo succeeded in 20 (both 5, only unshielded 2, only shielded 15; $p = 0.0023$) and crashed in 9, 7 of them while the shield was flying. Its failure pattern moved into the field of view: 6 of its 7 collisions were with obstacles in view, and the vehicle was moving outside the view in 2 of 7. Its 2 ground contacts descended at 2.8--3.3\,m/s, above the 2\,m/s vertical action bound, at total speeds of 4.0--4.4\,m/s, too fast for the 0.8\,m altitude floor.

A separately traced flight of the same 36 missions without the shield (6 successes, 25 collisions, 2 ground contacts, 2 timeouts, 1 camera timeout) gives the unshielded policy's crash locations. Of its 25 collisions, 13 were with obstacles in view and 12 with obstacles beside the camera, none behind; the vehicle was moving outside the view in 5 of the 24 collisions with a measurable direction; the median speed at the crash was 0.95\,m/s; \pr{LateralAvoid} was the active primitive in 14 of the 27 crashes; and the learned fallback was active in the last second before none of the 27. The unshielded policy thus crashed mostly while moving within the field of view, into obstacles in view or beside the camera---consistent with a policy that follows its last command more than its image---whereas the first shield's own failures came from motion outside the field of view (all 8 cases with a measurable direction), which the second version reduced to 2 of its 7 collisions.

Mission 0 was flown twice for \ckRone in both conditions: without the shield it succeeded in 7 of 36 missions in the evaluation and in 6 of 36 in the traced flight; with \shieldTwo in 20 in the pilot and 19 in the benchmark (94 successes on all missions minus 75 on missions 1--4). The aggregate counts agreed within one mission; we did not compute agreement mission by mission.

\begin{table}[t]
\centering
\caption{Shield design iteration on the pilot missions: mission 0 of every test world flown by \ckRone ($N = 36$). Paired: missions in which both, only the unshielded or only the shielded run succeeded, with the exact McNemar $p$. Crash locations give the bearing of the nearest trunk at the collision relative to the camera axis: in view within 36.5\textdegree, beside up to 100\textdegree, behind beyond. ``Moving outside the view'': horizontal motion more than 36.5\textdegree{} off the camera axis, among collisions with a measurable direction (horizontal speed above 0.2\,m/s). $^\dagger$From a separate traced flight of the same 36 missions without the shield (6 successes, 25 collisions, 2 ground contacts, 2 timeouts, 1 camera timeout).}
\label{tab:pilot}
\footnotesize
\setlength{\tabcolsep}{5pt}
\begin{tabular}{@{}lccc@{}}
\toprule
& \ckRone, unshielded & + \shieldOne & + \shieldTwo \\
\midrule
Success [95\% CI]                                  & 7 (19.4\%) [10, 35] & 12 (33.3\%) [20, 50] & 20 (55.6\%) [40, 70] \\
Crashes (collision / ground)                       & 27 (23 / 4)  & 11 (10 / 1) & 9 (7 / 2) \\
Timeouts / camera timeouts                         & 0 / 2        & 12 / 1      & 7 / 0 \\
Paired with the unshielded run (both / only U / only S) & --      & 2 / 5 / 10, $p = 0.30$ & 5 / 2 / 15, $p = 0.0023$ \\
\midrule
Crashes while the shield flew                      & --           & 10 of 11    & 7 of 9 \\
Collisions in view / beside / behind               & 13 / 12 / 0 of 25$^\dagger$ & 2 / 4 / 4 of 10 & 6 / 1 / 0 of 7 \\
Moving outside the view                            & 5 of 24$^\dagger$ & 8 of 8 & 2 of 7 \\
Median speed at the crash (m/s)                    & 0.95$^\dagger$ & 0.77 & 0.69 \\
\pr{LateralAvoid} active at the crash              & 14 of 27$^\dagger$ & 5 of 11 & 3 of 9 \\
Learned fallback active in the last second         & 0 of 27$^\dagger$ & 1 of 11 & 2 of 9 \\
\midrule
Per mission: backoffs / nudges / stalls / holds    & --           & 0.78 / 4.03 / 3.44 / 4.61 & 0.89 / 2.53 / 1.83 / 3.39 \\
Seconds per mission: $\sigma$ cap / blind / barrier & --          & 18.0 / 17.1 / 12.4 & 0.0 / 14.4 / 13.6 \\
Seconds per mission: nudge / backoff               & --           & 5.6 / 2.4   & 3.5 / 3.8 \\
\bottomrule
\end{tabular}
\end{table}

\subsection{Shield Activity, Cost, and What the Timeouts Did}
\label{sec:activity}
Per \ckRone mission (Table~\ref{tab:activity}), the barrier modified the command for 12.9\,s and the blind cap for 13.6\,s; the shield backed off 0.87 times and nudged 2.05 times, 1.35 of them after a stall, for 3.7 and 3.0\,s; it held 2.42 times, for 0.1\,s in total; and the altitude floor acted for 0.3\,s. The six checkpoints drew similar interventions: barrier 11.1--13.5\,s and blind cap 13.6--17.5\,s per mission. The memory update cost 2.9--3.0\,ms at the 95th percentile, and the 95th percentile of decision compute was 2.6--3.7\,ms higher than in the unshielded runs (separate runs, so this difference is approximate). The clearance to the memory, from the estimated position to the nearest remembered depth return (one per 0.15\,m voxel), was 0.32--0.42\,m, for each checkpoint 0.08--0.11\,m below the privileged clearance to the modelled trunk surfaces (0.42--0.52\,m); the two are measured to different targets (remembered returns from the estimated position, modelled cylinders from the true position) and are not directly comparable.

The timeouts were not hovering (Table~\ref{tab:byending}). When the time limit ended them, \ckRone's 42 timed-out missions had reached a median 71\% of their waypoints, at a median mean speed of 0.85\,m/s, with 3.10 stalls, 4.26 nudges and 25.0\,s under the blind cap per mission, against 0.86 stalls, 1.36 nudges and 11.4\,s for its successes. \ckKw's 39 timeouts were similar (67\% of waypoints, 0.87\,m/s, 2.72 stalls, 4.26 nudges, 26.8\,s blind). They ended without collision at the time limit, after slow progress that the shield repeatedly interrupted. Collisions came earlier (median 50.3\,s for \ckRone, 46.6\,s for \ckKw) and after more backoffs (1.52 and 1.49 per mission).

\begin{table}[t]
\centering
\caption{Shield activity in the shielded benchmark ($N = 180$ missions per checkpoint): interventions per mission (means), seconds per mission with each rule active (means), clearance to the memory (median over missions of the per-mission minimum) and the memory update's 95th percentile (median over missions). The uncertainty cap was disabled (0.0\,s for every checkpoint).}
\label{tab:activity}
\small
\setlength{\tabcolsep}{3.8pt}
\begin{tabular}{@{}lcccccccccccc@{}}
\toprule
& \multicolumn{4}{c}{Per mission} & \multicolumn{6}{c}{Seconds per mission} & Memory & Update \\
\cmidrule(lr){2-5}\cmidrule(lr){6-11}
Checkpoint & backoffs & nudges & stalls & holds & barrier & blind & backoff & nudge & altitude & stale & clear.\ (m) & p95 (ms) \\
\midrule
\ckRone     & 0.87 & 2.05 & 1.35 & 2.42 & 12.9 & 13.6 & 3.7 & 3.0 & 0.3 & 0.1 & 0.42 & 2.9 \\
\ckRtwo     & 1.03 & 1.98 & 1.26 & 1.59 & 11.8 & 14.1 & 3.7 & 3.1 & 0.1 & 0.1 & 0.32 & 2.9 \\
\ckNoact    & 1.06 & 2.07 & 1.32 & 1.66 & 12.6 & 17.5 & 4.1 & 3.1 & 0.1 & 0.1 & 0.37 & 2.9 \\
\ckNomotion & 0.91 & 1.72 & 1.11 & 1.35 & 11.1 & 14.4 & 2.6 & 2.6 & 0.0 & 0.0 & 0.33 & 3.0 \\
\ckSd       & 0.96 & 2.22 & 1.49 & 1.19 & 11.9 & 16.5 & 4.0 & 3.2 & 0.0 & 0.0 & 0.36 & 2.9 \\
\ckKw       & 0.98 & 2.20 & 1.42 & 1.59 & 13.5 & 16.9 & 4.5 & 3.2 & 0.0 & 0.1 & 0.39 & 2.9 \\
\bottomrule
\end{tabular}
\end{table}

\begin{table}[t]
\centering
\caption{What the shielded missions did, by ending, for \ckRone and \ckKw ($N = 180$ missions each; $n$: missions with that ending). Waypoint fraction, duration (time), mean speed and smallest privileged clearance are medians over the missions; interventions and the seconds with the barrier and the blind cap active are means per mission.}
\label{tab:byending}
\footnotesize
\setlength{\tabcolsep}{3.2pt}
\begin{tabular}{@{}llrcccccccc@{}}
\toprule
& & & \multicolumn{4}{c}{Medians} & \multicolumn{3}{c}{Per mission} & \multicolumn{1}{c}{Seconds} \\
\cmidrule(lr){4-7}\cmidrule(lr){8-10}\cmidrule(lr){11-11}
& Ending & $n$ & waypoints & time (s) & speed (m/s) & clear.\ (m) & backoffs & nudges & stalls & barrier / blind \\
\midrule
\multirow{5}{*}{\ckRone}
& success        & 94 & 1.00 & 83.1  & 1.24 & 0.60 & 0.50 & 1.36 & 0.86 & 11.7 / 11.4 \\
& timeout        & 42 & 0.71 & 120.3 & 0.85 & 0.53 & 1.17 & 4.26 & 3.10 & 17.7 / 25.0 \\
& collision      & 33 & 0.25 & 50.3  & 1.13 & 0.35 & 1.52 & 1.27 & 0.67 & 11.4 / 7.7 \\
& ground         & 7  & 0.50 & 56.0  & 1.08 & 0.40 & 1.43 & 2.57 & 1.14 & 12.8 / 8.1 \\
& camera timeout & 4  & 0.17 & 28.2  & 1.26 & 1.50 & 0.00 & 0.50 & 0.50 & 1.7 / 5.0 \\
\midrule
\multirow{6}{*}{\ckKw}
& success        & 90 & 1.00 & 81.5  & 1.28 & 0.63 & 0.51 & 1.50 & 0.99 & 12.7 / 16.1 \\
& timeout        & 39 & 0.67 & 112.5 & 0.87 & 0.48 & 1.56 & 4.26 & 2.72 & 21.0 / 26.8 \\
& collision      & 39 & 0.33 & 46.6  & 1.09 & 0.35 & 1.49 & 1.79 & 1.21 & 9.9 / 11.5 \\
& ground         & 8  & 0.12 & 24.4  & 1.25 & 0.39 & 1.25 & 2.12 & 0.88 & 7.7 / 6.9 \\
& camera timeout & 2  & 0.08 & 39.4  & 1.05 & 0.96 & 0.00 & 0.50 & 0.50 & 3.2 / 6.3 \\
& out of bounds  & 2  & 0.53 & 46.6  & 1.68 & 0.50 & 0.50 & 3.50 & 3.00 & 3.0 / 13.8 \\
\bottomrule
\end{tabular}
\end{table}


\section{Discussion}
\label{sec:discussion}

\subsection{What the Evaluation Shows}
\label{sec:findings}
\paragraph{The gain is uniform across recipes.} The shield raised every checkpoint by a similar margin, 30.0--34.4 percentage points, although the recipes differ in what they copy (Section~\ref{sec:copycat}) and in their unshielded success (11.1--21.1\%). It removed the failure that dominated all of them, collision with a trunk, whatever the recipe. The remaining differences between shielded checkpoints are within mission-level noise: none differed significantly from \ckRone{} + shield, and even \ckRone and \ckKw, both shielded, disagreed on 72 of the 180 missions (38 succeeded only with \ckRone, 34 only with \ckKw). In this study it was the runtime shield, not further imitation, that raised closed-loop success.

\paragraph{Failures move from collisions to timeouts.} The shield is conservative by design: per \ckRone mission the barrier acted for 12.9\,s and the blind cap for 13.6\,s, and the timeouts rose from 2.8\% to 23.3\%. The timeouts were slow, repeatedly interrupted progress rather than hovering: they reached a median 71\% of their waypoints at 0.85\,m/s with 4.26 nudges per mission, and spent 25.0\,s per mission under the blind cap, more than twice as long as the successes (11.4\,s). This is the price of map-free safety under a narrow field of view: the shield slows the vehicle wherever it cannot see where it is going. A policy that yawed toward its direction of travel, or a shield that turned the camera before moving sideways, may convert some timeouts into successes; we have not tested either.

\paragraph{Does the shield steer the copycat?} Because the dispatched setpoint is written back as $\bar{a}_t$, a policy that repeats $\bar{a}_t$ continues whatever the shield made it do: after a projection, its next command already points past the obstacle. We hypothesised that this write-back drives part of the gain. The data do not require it: \ckNoact and \ckNomotion, whose actors do not receive $\bar{a}_t$, gained as much (32.8 and 31.7 points) as the checkpoints that do. \ckNoact still copies the measured velocity, which the shield also changes, but \ckNomotion's actor receives neither quantity. Disabling the write-back (experiment F3, Section~\ref{sec:future}) would settle the question.

\paragraph{Remaining failure modes.} With the shield, 22.2--42.8\% of missions still ended in a crash, so the shield mitigates the narrow field of view and the actuation latency; it does not resolve them. In the \shieldTwo pilot, 7 of the 9 crashes happened while the shield was flying its own setpoint (backoff, nudge or hold) and 6 of the 7 collisions were with obstacles in view: the nudge moves toward the waypoint once the camera faces it, and the barrier is evaluated at the position 0.4\,s ahead, which covers the dead time but not the 0.5\,s lag. The two ground contacts descended at 2.8--3.3\,m/s, faster than the vertical action bound, so the altitude floor engaged too late; in the benchmark, 5--13 of each checkpoint's 180 shielded missions ended in ground contact. An altitude floor on the predicted height and a cap on the descent speed near the ground are the next changes.

\paragraph{Latent conditioning and actor sensitivity.} In every evaluated policy the JEPA predictor has two operational roles: its ensemble mean $\hat{z}_{t+1}$ conditions the solver (Eq.~\eqref{eq:cond}), and its ensemble dispersion drives the uncertainty fallback (Section~\ref{sec:uncertainty}); it is not used for planning. The swap tests show how the actor uses these signals. Its command is far less sensitive to the predictor's action conditioning than to direct proprioception: replacing $\bar{a}_t$ in the predictor's input moved the command by 0.00--0.02\,m/s, whereas replacing it in $S_t$ alone moved it by 0.74 and 0.88\,m/s in \ckSd and \ckKw. The predictor's action pathway is nonetheless active: in \ckNomotion, whose actor receives neither $\bar{a}_t$ nor $v^{\mathcal{H}}_t$, $\bar{a}_t$ reached the command through the forecast (0.33\,m/s). The swap test thus places the copycat shortcut in the actor's proprioceptive input, where the round-3 recipes intervened, rather than in the latent forecast. The fallback is calibrated for about 0.05 spurious triggers per 60\,s; at that threshold it braked or held for 0.1--0.3\,s per shielded mission and was inactive in the last second before each of the 27 traced unshielded crashes, so the gain of Section~\ref{sec:shielded} came from the runtime shield. Ablation A6 (Section~\ref{sec:future}) is planned to measure the predictor's contribution to closed-loop success.

\paragraph{Implications for the design.} The proprioceptive input $S_t$ of Eq.~\eqref{eq:proprio} contains the last dispatched setpoint $\bar{a}_t$ so that the predictor is action-conditioned without placing the solver output on its input (Section~\ref{sec:jepa}). The evaluation shows that the same input gives the actor a shortcut: withholding $\bar{a}_t$ from the actor moved the shortcut to the measured velocity, and withholding both left an underfitted actor (Section~\ref{sec:copycat}). DAgger on the policy's own states gave the first successes without removing the shortcut (Section~\ref{sec:plateau}). How the actor can use proprioception without copying it therefore remains a design question for Aero-VPA, separate from the shield, which only constrains what the actor commands.

\subsection{Expected Engineering Hurdles}
\paragraph{Sim-to-real gap in depth and appearance.} Active stereo depth degrades outdoors when sunlight overwhelms the projector, on thin structures such as branches and cables, and at range, where the error grows roughly quadratically with distance. These effects bias both the encoder input and the braking envelope through $\epsilon_d$. We plan range-dependent noise and dropout models in training, a conservative range-dependent $\epsilon_d$ in Eq.~\eqref{eq:vsafe}, and fine-tuning of the stage-(ii) modules on a small amount of real flight data while keeping $E_\psi$ frozen.

\paragraph{Aerodynamic effects.} Ground effect near the floor and rubble~\cite{shi2019neurallander}, wall and ceiling effects in narrow openings, rotor downwash recirculating in confined rooms, and gusts~\cite{oconnell2022neuralfly} all perturb velocity tracking. They enlarge $T_{\mathrm{trk}}$ and may violate Assumption~\ref{ass:envelope}(iv), so the tracking lag and the attainable deceleration will be identified on the physical airframe and entered into $T_r$ and $a_{\max}$ with margins, and the flight envelope will be restricted near surfaces.

\paragraph{Embedded execution.} Converting attention layers to FP16 can overflow or reorder reductions; we will verify the parity of the TensorRT engines with the PyTorch reference to a fixed tolerance on recorded inputs before any flight. In an enclosed airframe, the module may throttle thermally even within the 15\,W mode, which changes $T_{\mathrm{VPA}}$ over a mission; temperature and clock logging during long HITL runs will determine whether active cooling is required. Rolling-shutter distortion and motion blur at speed alter the input distribution in ways that simulation reproduces only approximately.

\subsection{Planned Physical Flight Trials}
Physical trials will proceed only if the HITL stage meets Hypothesis~\ref{hyp:latency} and the safety layer meets Hypothesis~\ref{hyp:safety} in simulation. They are staged as follows: (i) tethered and then free flight in a netted indoor arena with motion-capture ground truth, with the speed limit raised from 1 to 3 and then 5\,m/s only after each level meets its criteria; (ii) static and moving foam obstacles in the arena to test the braking envelope and the fallback, including deliberately induced inference stalls and link interruptions; and (iii) outdoor flight in sparse and then dense forest with RTK-GNSS ground truth. Throughout, a safety pilot holds a radio override, the arena or field is geofenced, PX4's offboard-loss failsafe remains active, and every flight is logged with the instrumentation of Section~\ref{sec:hitl}. Neither condition is met yet: Hypotheses~\ref{hyp:latency} and~\ref{hyp:safety} have not been tested, and 22.2--42.8\% of the shielded missions in simulation still ended in a crash (Section~\ref{sec:shielded}). The runtime shield of Section~\ref{sec:shield}, or a successor, would be part of any physical trial; on hardware its depth memory also depends on the drift of the state estimate over the 15\,s horizon.


\section{Limitations, Threats to Validity, and Protocol Deviations}
\label{sec:limitations}

\subsection{Limitations of the Design}
\label{sec:design-limitations}
The ensemble of Eq.~\eqref{eq:sigma} measures disagreement among predictor heads that share a frozen encoder. It therefore captures predictor uncertainty but can miss failures of the encoder itself: an unfamiliar image mapped to a familiar latent yields confident, agreeing heads. In the evaluation, the fallback braked or held for 0.1--0.3\,s per shielded mission and was active in the last second before none of the 27 traced unshielded crashes (Sections~\ref{sec:shielded} and~\ref{sec:iteration}). The braking envelope partially mitigates this risk for static obstacles; latent-density scores and input-space OOD detectors are candidate complements to be compared within ablation A6. The guarantees stated here are conditional: Proposition~\ref{prop:coherence} depends on Lipschitz constants that are not controlled, and Proposition~\ref{prop:envelope} depends on static obstacles, bounded depth error, and accurate tracking, which the evaluated system does not provide (Remark~\ref{rem:envelope-evaluated}); Proposition~\ref{prop:barrier} concerns one remembered point under exact tracking, and the runtime shield carries no safety guarantee (Remark~\ref{rem:not-guaranteed}). The primitive dictionary bounds the repertoire of behaviours, and the rule-based labels of Section~\ref{sec:selector} may segment expert behaviour imperfectly. Finally, the uncertainty-adaptive horizon did not improve success in the manipulation experiments of~\cite{lee2026vpa}; whether it is beneficial in dynamic flight, where it acts as a staleness bound rather than an execution prefix, is an open question that ablation A8 is designed to answer. The runtime shield constrains the command only against remembered, static points, and it is conservative by design: it turned part of the crashes into timeouts (Section~\ref{sec:findings}).

\subsection{Threats to Validity}
\label{sec:threats}
\begin{enumerate}
  \item \textbf{The shield was revised on test missions.} Shield~v2 was revised after the \shieldOne pilot flew mission 0 of every test world, 36 of the 180 test missions. The 144 missions never used for revision give essentially the same result (\ckRone 52.1\%, \ckKw 54.2\%; Section~\ref{sec:heldout}), but they come from the same worlds. The fresh-world development set was not flown, so we make no claim of generalisation to unseen worlds.
  \item \textbf{Selection among six checkpoints.} All six were evaluated on the same missions. We report \ckRone throughout because it is the first DAgger checkpoint and the best unshielded one; it also had the highest shielded success, so its shielded number is subject to selection, although the paired tests against \ckRone{} + shield show that no checkpoint differs significantly from it.
  \item \textbf{One training seed, and chance.} Each recipe was trained once and each evaluation flown once. The flow-matching noise is not seeded per mission and simulator timing varies, so mission outcomes have a large chance component (Section~\ref{sec:plateau}). Two repeated flights of mission 0 agreed in aggregate within one success (Section~\ref{sec:iteration}), but we did not measure test--retest agreement mission by mission.
  \item \textbf{A narrow scenario.} Simulation only (PX4 SITL with Gazebo Harmonic); static, procedurally generated forests of vertical trunks; no wind, no moving obstacles and no visual distribution shift. The mean speeds of 1.16--1.36\,m/s are far below the demonstrations' speed limits of 3--7\,m/s.
  \item \textbf{Privileged definitions.} A collision is a privileged centre-to-surface distance below 0.35\,m, and a ground contact an altitude below 0.3\,m; neither is a physical contact model.
  \item \textbf{Compute.} Timing was measured on an NVIDIA L4 GPU, not on the Jetson Orin Nano targeted in Section~\ref{sec:system}.
  \item \textbf{The BC baselines.} The BC policies flew 12 of the 36 test worlds (60 missions), before the start-up warm-up fix, and \bcNoact's camera timeouts happened while four simulators shared the machine.
\end{enumerate}

\subsection{Deviations from the Protocol of Section~\ref{sec:experiments}}
\label{sec:deviations}
The evaluation departs from the protocol of Section~\ref{sec:experiments}, fixed in version~1, in eight ways. (i) Only the forest environment class was flown; the collapsed-indoor and dynamic-clutter classes, the out-of-distribution conditions and wind were not. (ii) Each policy flew 180 missions with one training seed, instead of 500 episodes per condition with three seeds. (iii) Results are not stratified by speed limit: the policies are not speed-conditioned, and the time limit uses a 3\,m/s reference. (iv) The classical, reinforcement-learning and generative baselines B1--B3 were not run. (v) The embedded and hardware-in-the-loop latency of Hypothesis~\ref{hyp:latency} was not measured; timing was measured on a cloud GPU (Section~\ref{sec:implementation}). (vi) The runtime shield is a post-hoc addition outside the pre-registered design, revised on test missions. (vii) Holm--Bonferroni correction across hypothesis families was not applied; instead we name one primary family of tests (Section~\ref{sec:setup}) and report all others as secondary and unadjusted. (viii) Each DAgger round flew 201 missions under the safety-gated supervisor of Section~\ref{sec:training}, instead of the 500 flights per round planned in Section~\ref{sec:sim}. Consequently none of Hypotheses~\ref{hyp:latency}--\ref{hyp:coherence} has been tested: the evaluation has no Jetson timing (Hypothesis~\ref{hyp:latency}), no baselines and no stratification by speed limit (Hypothesis~\ref{hyp:task}), no out-of-distribution condition and no AUROC of $\bar{\sigma}$ (Hypothesis~\ref{hyp:safety}), no reconstruction-trained encoder (Hypothesis~\ref{hyp:repr}) and no noise-persistence ablation (Hypothesis~\ref{hyp:coherence}).

\subsection{Experiments for Future Work}
\label{sec:future}
Experiments F1--F6 address the main threats above.
\begin{itemize}
  \item \textbf{F1:} the shield with a non-learned goal-seeking controller (straight-to-waypoint velocity at about 1.2\,m/s, yawing to face the waypoint). It isolates the learned actor's contribution, which matters because all six shielded checkpoints landed within 42.8--52.2\%.
  \item \textbf{F2:} the privileged expert alone on the 180 test missions, as an upper bound.
  \item \textbf{F3:} shield ablations: memory off (current frame only), blind cap off, backoff off, nudge off, and the write-back of the dispatched setpoint into $\bar{a}_t$ off.
  \item \textbf{F4:} the classical map-based stack and the Diffusion Policy baseline of Section~\ref{sec:baselines} (B1 and B3a).
  \item \textbf{F5:} the world-model ablation A6 of Table~\ref{tab:ablation} and the area under the ROC curve of $\bar\sigma$ for collisions (Hypothesis~\ref{hyp:safety}).
  \item \textbf{F6:} several training seeds per recipe, and the fresh-world development set.
\end{itemize}


\section{Conclusion}
\label{sec:conclusion}

Aero-VPA replaces autoregressive action decoding with a short, fixed chain of small networks: a non-reconstructive JEPA world model, a discrete primitive selector, and a two-step flow-matching solver that produces a full velocity chunk in five sequential evaluations. It couples this chain to a safety layer driven by latent ensemble disagreement and by a model-free braking envelope, and to an asynchronous executor designed for the timing contract of the PX4 offboard interface. We have stated which properties follow from the architecture and which depend on assumptions about the learned components or the environment.

The first closed-loop evaluation, in simulation, showed that behaviour cloning learned a copycat shortcut, repeating its last command, that offline metrics could not detect, and that DAgger with a privileged, safety-gated supervisor produced the first successes but plateaued at 11.1--21.1\%. A deterministic, map-free runtime shield raised success by 30.0--34.4 percentage points for all six checkpoints, to 52.2\% for the first-round DAgger policy, and lowered its crashes from 72.2\% to 22.2\%, at the cost of more timeouts; the gain held on the missions never used to revise the shield. Whether the learned actor contributes under the shield, how the system performs in fresh worlds, at higher speeds and on the embedded module, and whether it transfers to hardware remain open; Section~\ref{sec:experiments} and the experiments of Section~\ref{sec:future} state how each will be tested, and their results, positive or negative, will determine whether the latency advantage of the approach translates into safer and faster flight.


\section*{Code and Data Availability}
\phantomsection\label{sec:availability}
The trained policy, the code and the benchmark dataset are public. All released results are from simulation (PX4 SITL with Gazebo Harmonic).
\begin{itemize}
  \item \textbf{Trained policy and model weights} (GPL-3.0):\\
    \url{https://huggingface.co/ylee-sling/aero-vpa}\\
    The six DAgger checkpoints of Table~\ref{tab:names}, each a single file of 72.6--72.7\,MB (the networks have 18.13M trainable parameters, Table~\ref{tab:budget}); the three behaviour-cloning baselines; the two stage-(i) checkpoints; and six FPV videos of re-flights for illustration.
  \item \textbf{Code} (GPL-3.0-only), release v1.1.0~\cite{slingai2026code}:\\
    \url{https://github.com/ylee-sling/Aero-VPA}\\
    release page: \url{https://github.com/ylee-sling/Aero-VPA/releases/tag/v1.1.0}\\
    The runtime shield evaluated here, \texttt{sim/shield.py}, is unchanged from commit \texttt{2aefde1} to this release.
  \item \textbf{Benchmark dataset} (CC~BY~4.0)~\cite{slingai2026benchmark}:\\
    \url{https://huggingface.co/datasets/ylee-sling/aero-vpa-benchmark}\\
    Every closed-loop mission record (a Parquet table and the raw records), the decision traces, the 36 test worlds with their obstacle maps, and the analysis reports.
  \item \textbf{Version~1 of this preprint} (method and pre-registered protocol):\\
    \url{https://doi.org/10.5281/zenodo.23633375}
\end{itemize}
With the dataset downloaded into \texttt{bench/}, the command
\begin{center}
\texttt{python3 sim/paired\_stats.py -{}-root bench/records}
\end{center}
run in the repository recomputes the paired comparisons of the shielded benchmark: Tables~\ref{tab:main}, \ref{tab:heldout}, \ref{tab:byending}, \ref{tab:endings-full}, \ref{tab:crash-pairs} and~\ref{tab:density-endings}, and the comparisons with \ckRone{} + shield in Table~\ref{tab:pairs-shielded}. The other numbers of Section~\ref{sec:evaluation} and of the appendices are in the analysis reports of the dataset. The expert demonstrations and validation flights are not released, so the swap tests of Tables~\ref{tab:copycat} and~\ref{tab:swap-full} can be checked only against those reports.

\section*{Acknowledgments}
The author used an AI assistant (Claude, Anthropic) to help draft and edit the text and to check the reported statistics against the flight logs; the author reviewed all content and takes full responsibility for it.


\appendix
\section{Hyperparameters}
\label{app:hyper}
Table~\ref{tab:hyper} lists every value used in the reported flights.

\begin{table}[p]
\centering
\caption{Values used in every reported flight.}
\label{tab:hyper}
\small
\begin{tabular}{@{}l>{\raggedright\arraybackslash}p{4.6cm}>{\raggedright\arraybackslash}p{6.6cm}@{}}
\toprule
Component & Parameter & Value \\
\midrule
Observation & RGB-D frame; camera & $4\times128\times160$; HFOV 1.274\,rad (73\textdegree), depth 0.1--10\,m \\
            & proprioception $S_t$; goal $g_t$ & $\R^{13}$; $\R^4$, distance saturated at 20\,m \\
            & waypoint acceptance $\tau_w$; instruction & 1\,m; CLIP ViT-B/32 text, 512-d (null embedding) \\
Action      & chunk $H$, knot spacing $\Delta$ & 16 knots, 0.02\,s (0.32\,s) \\
            & bounds & 7\,m/s horizontal, 2\,m/s vertical, 1.5\,rad/s \\
\midrule
$E_\psi$ (11.17M) & ViT & patch 16, 80 tokens + CLS, width 384, depth 6, 6 heads, $z \in \R^{256}$ \\
$P_\omega$ (2.72M) & ensemble & 5 MLP heads $292\to512\to512\to256$, residual, $\operatorname{Embed}(u) \in \R^{32}$, 0.1\,s ahead \\
$\pi^h_\phi$ (0.67M) & selector & MLP $785\to512\to512\to6$ \\
$v_\theta$ (3.57M) & solver & pre-LN Transformer, 4 layers, width 256, 4 heads, 17 tokens; $e_t \in \R^{1073}$ \\
            & sampling & linear path, $K = 2$ Euler steps, noise persistence $N_\xi = 30$ \\
\midrule
Stage (i)   & VICReg weights; EMA momentum & $\lambda_v = 25$, $\lambda_c = 1$; 0.996 \\
Stage (ii)  & steps, batch, time & 153,528 steps, batch 1024, about 2\,h\,40\,min on one L4 \\
            & DAgger frame weight & $\times3$ ($\times1$ for \ckKw) \\
            & state dropout (\ckSd) & $v^{\mathcal{H}}$ and $\bar{a}$ set to their training mean in 50\% of frames \\
            & keyframe weighting (\ckKw) & $\propto \min(1 + \norm{A_t[0] - \bar{a}_t}/0.1, 10)$ \\
\midrule
Learned safety & uncertainty filter; thresholds & $\alpha_\sigma = 0.1$; $\sigma_{\mathrm{lo}}$: 0.9-quantile; $\sigma_{\mathrm{crit}}$: per-flight maxima on 360 validation flights, about 0.05 spurious fallbacks per 60\,s \\
            & attenuation; braking envelope & factor $\ge 0.5$; $r_s = 0.5$\,m, $a_{\max} = 5$\,m/s$^2$, $\epsilon_d = 0.2$\,m, $T_r = 0.06 + 0.02\,H_t + 0.15$\,s, $v_{\mathrm{blind}} = 1$\,m/s \\
Execution   & executor; cross-fade & 100\,Hz; 0.04\,s \\
\midrule
Shield memory & blocks; returns; ground & $4\times4$ pixels, nearest return; 0.2--8\,m; points below 0.25\,m dropped \\
            & voxel; age; extent; local set & 0.15\,m (newest point kept); 15\,s; 10\,m horizontally; 3.5\,m \\
Barrier     & look-ahead; points; radius & 0.4\,s at the flown velocity; 24 nearest; $r = 0.55$\,m \\
            & $\beta$; enforcement & $k = 1.5$\,s$^{-1}$, $a_b = 3$\,m/s$^2$, $v_{\mathrm{push}} = 0.5$\,m/s; two projection sweeps \\
Caps and holds & blind cap & 0.7\,m/s beyond 36.5\textdegree{} of the camera axis, unless within 35\textdegree{} of a breadcrumb 0.3--2.5\,m away \\
            & altitude floor; stale hold; uncertainty cap & climb at $\ge 0.5$\,m/s below 0.8\,m; zero velocity after 0.3\,s without a decision; off ($v_\sigma = \infty$; \shieldOne: 1\,m/s) \\
Recovery    & breadcrumbs; backoff & every 0.25\,m; below 0.45\,m clearance, retrace up to 1\,m at 0.3\,m/s, end if blocked ($\norm{u} < 0.075$\,m/s) or at clearance $\ge r + 0.4$\,m \\
            & nudge; stall & $\dot\psi = \clip(1.5\,\Delta\psi, \pm0.8$\,rad/s$)$, then $0.6\cos\Delta\psi$\,m/s for 1\,s; 5\,s without 0.3\,m of progress \\
\midrule
DAgger      & label law & $k_p = 1.5$\,s$^{-1}$; on path within 0.3\,m; 2\,m/s horizontal limit at $\ge 1$\,m off \\
            & gate & look-ahead 0.3\,s + braking distance ($\ge 1.5$\,m, every 0.1\,m), 0.5\,m + 0.06\,m per m/s; altitude 0.8\,m; off path 1.5\,m; heading 75\textdegree; stall 5\,s \\
            & rollout; hand-back; round size & 0.24\,s dead time, 0.5\,s lag; 2\,s on track; 201 missions \\
\midrule
Missions    & take-off; waypoints & 2.5\,m; 3--6, 10--40\,m apart \\
            & time limit & $1.5\,T_{\mathrm{ref}} + 15$\,s, $T_{\mathrm{ref}}$ at 3\,m/s \\
            & endings & collision $< 0.35$\,m (centre to trunk surface); ground $< 0.3$\,m; camera timeout 0.5\,s \\
\bottomrule
\end{tabular}
\end{table}

\section{DAgger Supervisor Details}
\label{app:supervisor}
\paragraph{Labels.} Every frame of a DAgger round, flown by the policy or by the expert, is labelled with the expert's tracking law, Eq.~\eqref{eq:label}, on its minimum-snap plan of the mission, from the path point $\tau^\star$ nearest the vehicle. The speed limit of each mission is drawn from $\{3, 5, 7\}$\,m/s, as in the demonstrations. On the path ($e \le 0.3$\,m) the law is the demonstrations' law ($w = 1$); further off, the expert rejoins first ($w \to 0$, horizontal speed limit falling to 2\,m/s at $e \ge 1$\,m).

\paragraph{Gate.} The policy flies until one of six triggers fires: (1) \emph{obstacle}: the straight look-ahead along the current velocity, over 0.3\,s of reaction plus the braking distance and at least 1.5\,m, sampled every 0.1\,m, comes within 0.5\,m $+$ 0.06\,m per m/s of speed of a trunk; (2) \emph{low}: altitude below 0.8\,m; (3) \emph{off path}: more than 1.5\,m from the path; (4) \emph{heading}: facing more than 75\textdegree{} away from the path; (5) a missed waypoint; (6) \emph{stall}: 5\,s without progress.

\paragraph{Takeover and hand-back.} In a soft takeover, which keeps recording frames, the expert chooses how to fly on from $\tau^\star$: along its plan as is, slowing down along its path with a time-scaled reference, or with a straight stop, by the clearance that a rollout of the delayed closed loop predicts for each option. The rollout replays the setpoints still in flight over a dead time of 0.24\,s and models the vehicle's response as first order with a 0.5\,s lag. A missed waypoint, or a predicted conflict that only a straight stop resolves, leads to a hard takeover: the expert brakes, backs off and re-plans; these frames are not recorded. Control returns to the policy after 2\,s on track.

\paragraph{Rounds.} Each round flies 201 missions over the training worlds; stage~(ii) then trains on the demonstrations and all rounds so far, with DAgger frames weighted $\times3$ ($\times1$ for \ckKw). Round~1 was flown by \bcVtwo, round~2 by \ckRone and round~3 by \ckRtwo; the round-3 recipes \ckNoact, \ckNomotion, \ckSd and \ckKw were trained on rounds 1--3 (Table~\ref{tab:names}).

\section{Full Results}
\label{app:endings}
Table~\ref{tab:endings-full} gives the exact ending counts of the twelve evaluations, Table~\ref{tab:crash-pairs} the paired crash comparisons, and Table~\ref{tab:pairs-shielded} the shielded checkpoints paired with unshielded \ckRone and with \ckRone{} + shield. Tables~\ref{tab:metrics-u} and~\ref{tab:metrics-s} give the secondary metrics, and Table~\ref{tab:density-endings} the shielded endings by density.

\begin{table}[t]
\centering
\caption{Exact ending counts of the six DAgger checkpoints, unshielded (U) and shielded (S), on the same $N = 180$ test missions.}
\label{tab:endings-full}
\small
\begin{tabular}{@{}llcccccc@{}}
\toprule
Checkpoint & & Success & Collision & Ground & Timeout & Camera timeout & Out of bounds \\
\midrule
\ckRone     & U & 38 & 114 & 16 & 5  & 7  & 0 \\
            & S & 94 & 33  & 7  & 42 & 4  & 0 \\
\ckRtwo     & U & 23 & 133 & 11 & 2  & 10 & 1 \\
            & S & 77 & 64  & 5  & 30 & 4  & 0 \\
\ckNoact    & U & 20 & 131 & 17 & 5  & 7  & 0 \\
            & S & 79 & 55  & 10 & 26 & 7  & 3 \\
\ckNomotion & U & 24 & 126 & 13 & 6  & 11 & 0 \\
            & S & 81 & 64  & 13 & 18 & 4  & 0 \\
\ckSd       & U & 28 & 129 & 8  & 6  & 8  & 1 \\
            & S & 82 & 51  & 12 & 29 & 5  & 1 \\
\ckKw       & U & 28 & 136 & 9  & 3  & 4  & 0 \\
            & S & 90 & 39  & 8  & 39 & 2  & 2 \\
\bottomrule
\end{tabular}
\end{table}

\begin{table}[t]
\centering
\caption{Paired crash comparisons, shielded against unshielded ($N = 180$ paired missions): crashes in each condition, discordant pairs and the two-sided exact McNemar $p$.}
\label{tab:crash-pairs}
\small
\begin{tabular}{@{}lccccc@{}}
\toprule
Checkpoint & Crashes U & Crashes S & Only U crashed & Only S crashed & $p$ \\
\midrule
\ckRone     & 130 & 40 & 98  & 8  & $8.08\times10^{-21}$ \\
\ckRtwo     & 144 & 69 & 83  & 8  & $7.58\times10^{-17}$ \\
\ckNoact    & 148 & 65 & 93  & 10 & $5.20\times10^{-18}$ \\
\ckNomotion & 139 & 77 & 78  & 16 & $5.78\times10^{-11}$ \\
\ckSd       & 137 & 63 & 83  & 9  & $3.92\times10^{-16}$ \\
\ckKw       & 145 & 47 & 107 & 9  & $2.00\times10^{-22}$ \\
\bottomrule
\end{tabular}
\end{table}

\begin{table}[t]
\centering
\caption{Each shielded checkpoint paired with unshielded \ckRone and with \ckRone{} + shield ($N = 180$ paired missions; both / only \ckRone{} / only the other / neither, two-sided exact McNemar $p$).}
\label{tab:pairs-shielded}
\small
\begin{tabular}{@{}lcccc@{}}
\toprule
& \multicolumn{2}{c}{Against unshielded \ckRone} & \multicolumn{2}{c}{Against \ckRone{} + shield} \\
\cmidrule(lr){2-3}\cmidrule(lr){4-5}
Shielded checkpoint & counts & $p$ & counts & $p$ \\
\midrule
\ckRone     & 26 / 12 / 68 / 74 & $1.20\times10^{-10}$ & -- & -- \\
\ckRtwo     & 17 / 21 / 60 / 82 & $1.69\times10^{-5}$  & 46 / 48 / 31 / 55 & 0.0712 \\
\ckNoact    & 17 / 21 / 62 / 80 & $7.51\times10^{-6}$  & 50 / 44 / 29 / 57 & 0.101 \\
\ckNomotion & 19 / 19 / 62 / 80 & $1.77\times10^{-6}$  & 52 / 42 / 29 / 57 & 0.154 \\
\ckSd       & 22 / 16 / 60 / 82 & $3.85\times10^{-7}$  & 47 / 47 / 35 / 51 & 0.224 \\
\ckKw       & 22 / 16 / 68 / 74 & $8.54\times10^{-9}$  & 56 / 38 / 34 / 52 & 0.724 \\
\bottomrule
\end{tabular}
\end{table}

\begin{table}[t]
\centering
\caption{Secondary metrics without the shield ($N = 180$ missions per checkpoint). Duration, path, time and path ratios, smallest clearance and timing are medians over missions (time and path ratios over successful missions); waypoints and blind decisions are pooled shares; fallback figures and mean speed are means over missions.}
\label{tab:metrics-u}
\small
\setlength{\tabcolsep}{5pt}
\begin{tabular}{@{}lcccccc@{}}
\toprule
Metric & \ckRone & \ckRtwo & \ckNoact & \ckNomotion & \ckSd & \ckKw \\
\midrule
Waypoints reached                  & 38.5\% & 28.9\% & 27.6\% & 31.3\% & 29.8\% & 31.9\% \\
Duration / path (s, m)             & 38.7 / 46.1 & 28.9 / 35.1 & 30.6 / 33.3 & 32.7 / 43.5 & 30.1 / 41.4 & 35.5 / 38.6 \\
Fallback onsets per mission        & 0.77 & 0.51 & 1.04 & 0.68 & 0.69 & 0.86 \\
Missions with a fallback           & 21.1\% & 21.1\% & 30.6\% & 23.3\% & 26.1\% & 31.1\% \\
Braking or holding (s per mission) & 0.1 & 0.1 & 0.2 & 0.1 & 0.1 & 0.2 \\
Time / expert reference            & 1.15 & 1.00 & 1.26 & 1.05 & 1.10 & 1.10 \\
Path / mission length              & 1.09 & 1.12 & 1.24 & 1.16 & 1.12 & 1.13 \\
Mean speed (m/s)                   & 1.24 & 1.21 & 1.21 & 1.37 & 1.26 & 1.21 \\
Smallest clearance (m)             & 0.35 & 0.35 & 0.35 & 0.35 & 0.35 & 0.35 \\
Blind decisions                    & 26.2\% & 29.5\% & 33.2\% & 31.9\% & 31.3\% & 34.0\% \\
Decision compute p95 (ms)          & 29.4 & 28.8 & 29.1 & 29.4 & 29.4 & 29.6 \\
Decision age p95 (ms)              & 41.1 & 40.4 & 41.3 & 41.0 & 40.6 & 41.3 \\
Longest setpoint gap, median / max (s) & 0.10 / 0.15 & 0.10 / 0.15 & 0.10 / 0.15 & 0.10 / 0.16 & 0.09 / 0.15 & 0.10 / 0.16 \\
\bottomrule
\end{tabular}
\end{table}

\begin{table}[t]
\centering
\caption{Secondary metrics with the shield ($N = 180$ missions per checkpoint; definitions as in Table~\ref{tab:metrics-u}).}
\label{tab:metrics-s}
\small
\setlength{\tabcolsep}{5pt}
\begin{tabular}{@{}lcccccc@{}}
\toprule
Metric & \ckRone & \ckRtwo & \ckNoact & \ckNomotion & \ckSd & \ckKw \\
\midrule
Waypoints reached                  & 76.6\% & 68.4\% & 65.9\% & 62.2\% & 68.9\% & 72.9\% \\
Duration / path (s, m)             & 86.3 / 95.4 & 77.9 / 88.0 & 76.1 / 90.5 & 60.7 / 81.2 & 77.6 / 90.0 & 82.2 / 93.5 \\
Fallback onsets per mission        & 1.53 & 1.13 & 1.39 & 1.04 & 1.32 & 1.25 \\
Missions with a fallback           & 30.0\% & 32.2\% & 34.4\% & 33.9\% & 36.7\% & 34.4\% \\
Braking or holding (s per mission) & 0.3 & 0.2 & 0.2 & 0.1 & 0.2 & 0.2 \\
Time / expert reference            & 1.26 & 1.22 & 1.27 & 1.12 & 1.22 & 1.22 \\
Path / mission length              & 1.13 & 1.14 & 1.19 & 1.18 & 1.19 & 1.16 \\
Mean speed (m/s)                   & 1.18 & 1.17 & 1.16 & 1.36 & 1.20 & 1.20 \\
Smallest clearance (m)             & 0.52 & 0.42 & 0.45 & 0.43 & 0.46 & 0.50 \\
Blind decisions                    & 28.8\% & 32.6\% & 36.4\% & 30.7\% & 35.8\% & 35.8\% \\
Decision compute p95 (ms)          & 33.1 & 32.4 & 32.2 & 32.0 & 32.1 & 32.4 \\
Decision age p95 (ms)              & 44.6 & 43.8 & 43.6 & 43.5 & 43.8 & 43.8 \\
Longest setpoint gap, median / max (s) & 0.10 / 0.18 & 0.10 / 0.16 & 0.10 / 0.15 & 0.10 / 0.13 & 0.10 / 0.17 & 0.10 / 0.16 \\
\bottomrule
\end{tabular}
\end{table}

\begin{table}[t]
\centering
\caption{Shielded endings by density: successes / crashes / timeouts of $N = 60$ missions per density; the remainder of each 60 are camera timeouts or out-of-bounds endings.}
\label{tab:density-endings}
\small
\begin{tabular}{@{}lccc@{}}
\toprule
Checkpoint & Dense & Medium & Sparse \\
\midrule
\ckRone     & 28 / 17 / 15 & 29 / 16 / 15 & 37 / 7 / 12 \\
\ckRtwo     & 21 / 29 / 10 & 32 / 19 / 9  & 24 / 21 / 11 \\
\ckNoact    & 23 / 24 / 12 & 25 / 28 / 6  & 31 / 13 / 8 \\
\ckNomotion & 17 / 31 / 12 & 28 / 30 / 2  & 36 / 16 / 4 \\
\ckSd       & 17 / 31 / 12 & 26 / 21 / 11 & 39 / 11 / 6 \\
\ckKw       & 22 / 20 / 18 & 33 / 12 / 15 & 35 / 15 / 6 \\
\bottomrule
\end{tabular}
\end{table}

\section{Behaviour-Cloning Results}
\label{app:bc}
Table~\ref{tab:bc} gives the endings of the three BC policies on 60 missions (12 of the 36 test worlds, 5 missions each). They were flown before a start-up warm-up fix: the first decision of a fresh process delayed the setpoint stream at the start of the first mission of each world. \bcNoact's camera timeouts happened while four simulators shared the machine. The open-loop errors and selector agreement of \bcVtwo and \bcNoact on expert frames are in Table~\ref{tab:copycat}.

\begin{table}[t]
\centering
\caption{Behaviour-cloning policies in closed loop ($N = 60$ missions each: 12 of the 36 test worlds, 5 missions per world, flown before the warm-up fix). Crash shares are split into collision and ground contact.}
\label{tab:bc}
\small
\begin{tabular}{@{}lcccl@{}}
\toprule
Policy & Stage (i) & Success & Crash (collision / ground) & Other endings \\
\midrule
\bcVone  & non-residual & 0/60 & 83\% (48\% / 35\%) & 10 missions, not broken down \\
\bcVtwo  & residual     & 0/60 & 51.7\% (42\% / 10\%) & timeout 48.3\% \\
\bcNoact & residual     & 0/60 & 70.0\% (22\% / 48\%) & timeout 8.3\%, camera timeout 21.7\% \\
\bottomrule
\end{tabular}
\end{table}

\section{Input-Sensitivity Tests of the Round-3 Recipes}
\label{app:swap}
Table~\ref{tab:swap-full} gives the complete swap tests of \ckSd and \ckKw on the 5000 held-out expert frames. Both still follow the donor's $\bar{a}_t$: their command after the swap is closer to the donor frame's label (0.28 and 0.43\,m/s) than to their own (0.84 and 0.89\,m/s), almost entirely through $S_t$ rather than through the predictor. On the same frames, the expert's commanded direction was blind in 3.7\% of frames and the braking envelope throttled it in 1.4\%; for \ckSd and \ckKw the shares were 5.5\% and 5.1\% blind and 8.5\% and 8.8\% throttled.

\begin{table}[t]
\centering
\caption{Swap tests of \ckSd and \ckKw on $N = 5000$ held-out expert frames: change of the first knot $v_0$ (median / p90, m/s), share of frames whose primitive changed, and the error to the frame's own label and to the donor frame's label (m/s). ``$\bar{a}_t = 0$'': the last command set to zero (first-knot speed afterwards in parentheses). Open loop on the same frames: $|v_0 - v_0^\star|$ median / p90 0.04 / 0.10\,m/s (\ckSd) and 0.03 / 0.07\,m/s (\ckKw); selector agreement 91.8\% and 92.9\%; direction error above $45^\circ$ in 0.6\% and 0.4\% of frames.}
\label{tab:swap-full}
\small
\setlength{\tabcolsep}{4pt}
\begin{tabular}{@{}lcccccccc@{}}
\toprule
& \multicolumn{4}{c}{\ckSd} & \multicolumn{4}{c}{\ckKw} \\
\cmidrule(lr){2-5}\cmidrule(lr){6-9}
Swapped input & change & primitive & own & donor & change & primitive & own & donor \\
\midrule
none                          & 0.00 / 0.00 & 0.0\%  & 0.04 & 1.14 & 0.00 / 0.00 & 0.0\%  & 0.03 & 1.14 \\
image $z_t$                   & 0.04 / 0.13 & 12.2\% & 0.04 & 1.14 & 0.03 / 0.08 & 10.4\% & 0.03 & 1.15 \\
goal $g_t$                    & 0.03 / 0.40 & 16.9\% & 0.05 & 1.09 & 0.02 / 0.09 & 9.0\%  & 0.03 & 1.14 \\
velocity $v^{\mathcal{H}}_t$  & 0.16 / 1.54 & 12.9\% & 0.18 & 0.98 & 0.49 / 1.85 & 28.9\% & 0.50 & 0.97 \\
attitude $n^{\mathcal{B}}_t$  & 0.05 / 0.89 & 17.3\% & 0.07 & 1.16 & 0.04 / 0.57 & 12.3\% & 0.05 & 1.16 \\
body rates                    & 0.05 / 0.87 & 20.8\% & 0.07 & 1.16 & 0.02 / 0.27 & 11.2\% & 0.04 & 1.15 \\
last command $\bar{a}_t$      & 0.83 / 2.21 & 30.1\% & 0.84 & 0.28 & 0.88 / 2.27 & 43.1\% & 0.89 & 0.43 \\
\quad $\bar{a}_t$ in $S_t$ only & 0.74 / 2.05 & 30.1\% & 0.75 & 0.34 & 0.88 / 2.26 & 43.1\% & 0.89 & 0.43 \\
\quad $\bar{a}_t$ to $P_\omega$ only & 0.02 / 0.12 & 0.0\% & 0.05 & 1.11 & 0.00 / 0.01 & 0.0\% & 0.03 & 1.14 \\
$\bar{a}_t = 0$               & 0.84 / 2.19 & 22.0\% & 0.86 & (1.31) & 0.83 / 2.06 & 40.1\% & 0.83 & (1.47) \\
all of $S_t$                  & 1.09 / 2.89 & 40.6\% & 1.09 & 0.05 & 1.14 / 2.96 & 44.0\% & 1.13 & 0.03 \\
\bottomrule
\end{tabular}
\end{table}

\end{document}